\documentclass[onefignum,onetabnum]{siamonline220329}

\crefname{section}{Section}{Sections}
\crefname{subsection}{Subsection}{Subsections}

\usepackage{bm}
\usepackage{dsfont}
\usepackage[labelformat=brace]{subcaption}

\usepackage{lipsum}
\usepackage{amsfonts}
\usepackage{graphicx}
\usepackage{epstopdf}
\usepackage{algorithmic}
\ifpdf
\DeclareGraphicsExtensions{.eps,.pdf,.png,.jpg}
\else
\DeclareGraphicsExtensions{.eps}
\fi

\usepackage{enumitem}
\setlist[enumerate]{leftmargin=.5in}
\setlist[itemize]{leftmargin=.5in}

\newsiamremark{remark}{Remark}
\newsiamremark{hypothesis}{Hypothesis}
\crefname{hypothesis}{Hypothesis}{Hypotheses}
\newsiamthm{claim}{Claim}

\newcommand{\centered}[1]{\setlength{\tabcolsep}{0mm}\begin{tabular}{l} #1 
\end{tabular}}

\headers{Efficient Parallel Algorithms for Inpainting}{N. K\"amper, V. 
Chizhov, and J. Weickert}

\title{Efficient Parallel Algorithms for Inpainting-Based 
	Representations of 4K Images \\ - Part I: Homogeneous Diffusion Inpainting
    \thanks{Part of this work has been published as a conference 
    paper~\cite{KW22}.
    \funding{This work has received funding from the European 
        Research Council (ERC) under the European Union's Horizon 2020 
        research and innovation programme (grant agreement no. 741215, ERC 
        Advanced Grant IN\-CO\-VID)}
    }
}

\author{Niklas K\"amper \thanks{Mathematical Image Analysis Group, Faculty of 
Mathematics and Computer Science, Campus E1.7, Saarland University, 66041 
Saarbr\"ucken, 
Germany (\email{kaemper@mia.uni-saarland.de}, 
\email{chizhov@mia.uni-saarland.de}, \email{weickert@mia.uni-saarland.de})}
	\and Vassillen Chizhov\footnotemark[2]
	\and Joachim Weickert\footnotemark[2]}

\usepackage{amsopn}

\begin{document}

\maketitle              


\begin{abstract}
In recent years inpainting-based compression methods have been shown to 
be a viable alternative to classical codecs such as JPEG and JPEG2000.
Unlike transform-based codecs, which store coefficients in the transform 
domain, inpainting-based approaches store a small subset of the original 
image pixels and reconstruct the image from those by using a suitable 
inpainting operator.
A good candidate for such an inpainting operator is homogeneous diffusion 
inpainting, as it is simple, theoretically well-motivated, and can achieve 
good reconstruction quality for optimized data. However, a major challenge 
has been to design fast solvers for homogeneous diffusion inpainting that 
scale to 4K image resolution ($3840 \times 2160$ pixels) and are real-time 
capable. We overcome this with a careful adaptation and fusion of two of the 
most efficient concept from numerical analysis: multigrid and domain 
decomposition. 
Our domain decomposition algorithm efficiently utilizes GPU parallelism by 
solving inpainting problems on small overlapping blocks.  Unlike simple block 
decomposition strategies such as the ones in JPEG, our approach yields block 
artifact-free reconstructions. Furthermore, embedding domain decomposition 
in a full multigrid scheme provides global interactions and allows us to 
achieve optimal convergence by reducing both low- and high-frequency errors 
at the same rate.
We are able to achieve 4K color image reconstruction at more than $60$ frames 
per second even from very sparse data - something which has been previously 
unfeasible. 
\end{abstract}




\section{Introduction}
\label{sec:intro}

The classical image inpainting task deals with the reconstruction of missing 
parts of partially corrupted 
images~\cite{ALM10,BSCB00,EL99a,GL14,MM98a,SSZK23,Sc15}.
However, it has also been successfully used in image compression
methods~\cite{AG94,BHK10,Ca88,DMMH96,GWWB08,GLG12,JS23,LSWL07,PHNH16,
PKW17,RPSH20,SPME14,SCSA04,WZSG09,XSW10,ZD11} as an alternative to classical
transform-based approaches.
In the encoding phase, 
these methods store a small optimized subset of the image data.  
The decoding phase then consists of reconstructing the missing parts of the 
image from the stored data by an inpainting process. 
Sparse nonlinear diffusion-based inpainting for image compression was 
introduced by Gali\'c et al.~\cite{GWWB05} in 2005, later improved by 
Schmaltz et al.~\cite{SPME14}, and extended to color images by 
Peter et al.~\cite{PKW17}.
They showed that inpainting-based methods can qualitatively outperform 
the widely used JPEG~\cite{PM92} and JPEG2000~\cite{TM02} for images with 
small to medium amounts of texture.

Surprisingly, already with simple linear homogeneous diffusion~\cite{Ii62} 
one can obtain very good inpainting results, if the locations as well as the 
values of the known pixels are carefully 
optimized~\cite{BBBW08,BLPP17,BHR21,CRP14,HSW13,MHWT12}.
For piecewise constant images, such as cartoon images, depth maps or flow 
fields, one can achieve state-of-the-art results~\cite{GLG12,HMWP13,LSJO12,MBWF11}, and some of these
methods~\cite{JPW20,JPW21} even outperform HEVC~\cite{SOHW12}. 
Its advantage, in contrast to nonlinear inpainting methods, is its 
simplicity, since it is parameter-free and its finite difference 
discretization yields a sparse pentadiagonal linear system of size 
$N \times N$, where $N$ is the number of pixels in the image. 

However, computing the solution of such large linear systems 
can be quite time-consuming. Nowadays, 4K color images of size $3840 \times 
2160$ pixels are a standard, resulting in systems with approximately $8$ 
million equations and the same number of unknowns for each color channel, and 
either 30 or 60 frames per second are usually required for video playback. 
Thus, it would be desirable to achieve real-time decoding for 4K 
color images with at least 30 or 60 frames per second. 
Even though there has been substantial research to accelerate 
inpainting-based compression 
\cite{APKMWH21, APW16, CW21, HPW15, KN22, KSFR07, MBWF11, PSMM15}, 
achieving 4K color inpainting at 60 frames per second for very sparse data 
has been previously unfeasible.
Current approaches are at least one order of magnitude slower. 
However, most of their numerical solvers have been tailored towards 
sequential or mildly parallel architectures, such as CPUs. 
They do not exploit the potential of highly parallel architectures such 
as GPUs, which are widely available these days.

We exploit this potential with our approach by carefully adapting and merging 
two of the most efficient concepts from numerical analysis, that have been 
rarely used in image analysis so far: domain decomposition~\cite{DJN15, TW05} 
and multigrid~\cite{Br77,BHM00, Hac85, TOS01, Wes92}.
Domain decomposition methods~\cite{DJN15, TW05} decompose the image domain 
into multiple subdomains and solve linear systems on each subdomain in 
parallel, instead of working directly on the large global linear system. 

They benefit from highly parallel architectures, due to the decomposition 
into multiple small subdomains. 
By providing mechanisms for communication across subdomain boundaries and
iterating this process, one can achieve convergence to the solution of the 
global problem. This results in a reconstruction without any artifacts at the 
block boundaries, unlike simple block decompositions such as in JPEG. 
However, while domain decomposition methods are very efficient at decreasing 
high frequency modes of the error, they are not able to decrease the lower 
frequencies at the same rate. 
Multigrid methods~\cite{Br77,BHM00, Hac85, TOS01, Wes92} allow 
for a more efficient interaction between subdomains that are far 
apart. They construct a multiresolution representation of the problem 
and decimate the different frequencies of the error at the different 
resolutions.
This results in a similar rate of decrease of both the low- and 
high-frequency components of the error, and also leads to an optimal linear 
scaling behavior. 


\subsection{Our Contribution}
\label{sec:contribution}

The goal of our paper is to develop a fast solver for homogeneous diffusion 
inpainting, in order to make 4K image and video compression with homogeneous 
diffusion practical. We do so by devising a new powerful algorithmic 
approach that fuses ideas from two very efficient concepts in numerical 
analysis:  full multigrid \cite{Br77,BHM00,Hac85,TOS01,Wes92} and domain 
decomposition methods~\cite{DJN15, TW05}. 

In our conference paper contribution on real-time 4K image 
inpainting~\cite{KW22} we have successfully employed a multilevel domain 
decomposition method to solve the homogeneous diffusion inpainting 
equation in real-time for 4K color images with 30 frames per second on a 
contemporary GPU. We have shown that domain decomposition methods are well 
suited for homogeneous diffusion inpainting, because in practice the 
influence area of each known pixel is localized.

We expand on our conference paper by extending our multilevel 
optimized restricted additive Schwarz (ORAS) method for homogeneous 
diffusion inpainting to a full multigrid approach with even more resolution 
layers. We improve the visual quality of our multigrid approach by a new 
initial downsampling method that adapts better to the image and the 
inpainting data. 
The full multigrid ORAS solver is more robust with 
respect to the mask density, which enables real-time performance even 
for very low densities. Furthermore, we are able to reach the current 
standard in video coding of 60 frames per second at 4K resolution 
if at least 2\% of the image data is known.

This paper is the first part of a companion paper on homogeneous diffusion 
inpainting. In the second part, we focus on the encoding, which consists of 
optimizing the inpainting data.


\subsection{Related Work}
\label{sec:related}

In the following we discuss prior work on accelerating diffusion-based 
inpainting. In contrast to previous methods, we exploit the highly parallel 
nature of current GPUs and do not rely on any sort of temporal coherence of subsequent video frames. 
We integrate multigrid ideas, but complement them with a highly parallel 
domain decomposition method.


\paragraph{Domain Decomposition in Image Processing}
While domain decomposition methods have not been used for sparse image 
inpainting, there exist a few domain decomposition approaches for image 
processing tasks, such as denoising~\cite{CHC19, Ko07a, XTW10}, optic flow 
computation~\cite{KSBW05} or deblurring~\cite{XCQ14}. 


\paragraph{Multigrid Methods} 
Multigrid methods~\cite{Br77,BHM00, Hac85, TOS01, Wes92} belong to the most 
efficient solvers for linear and nonlinear systems. They solve problems on 
multiple resolutions simultaneously, in order to achieve uniform convergence 
for both high- and low-frequency components of the error. Both K\"ostler et 
al.~\cite{KSFR07} and Mainberger et al.~\cite{MBWF11} use them for homogeneous 
diffusion inpainting on mildly parallel architectures such as multicore 
CPUs~\cite{MBWF11} or the \textit{Playstation 3}~\cite{KSFR07}. 
For the smoothing operations on each level of the multigrid method, both 
approaches use a simple Gauss-Seidel iteration~\cite{Sa03}. 
K\"ostler et al.~\cite{KSFR07} also considered nonlinear anisotropic 
diffusion with a coarse-to-fine multilevel method.
While our approach also employs a (full) multigrid method, we improve the 
downsampling by adapting it to the inpainting operator, and we use a more 
sophisticated smoother, i.e.\ domain decomposition, that better utilizes 
GPU parallelism.


\paragraph{Green's Functions}
Another approach for accelerated inpainting goes back to 
Hoffmann et al.~\cite{HPW15}. It is based on discrete Green's functions 
\cite{AWA19, BL58, CY00}. They can be used to decompose the inpainting 
problem into pixel-wise contributions from which the reconstruction 
can be assembled as a linear combination. 
This leads to an alternative system matrix of size $|K| \times |K|$ 
where $|K|$ is the number of known pixels, which determines the 
coefficients in the linear combination. 
This approach has the advantage that its runtime depends only on 
the number of mask pixels instead of the overall number of pixels. 
For very sparse inpainting data it can, thus, outperform multigrid 
methods. However, its system matrix is dense, and its size grows 
quadratically with the number of mask pixels. Thus, for 4K images 
and typical mask densities between 1\% and 5\% this approach is 
infeasible due to memory and runtime constraints. 
Kalmoun and Nasser's~\cite{KN22} approach is based on 
continuous Green's functions used in the method of fundamental 
solutions~\cite{Ka98a,KA64}. They use a GMRES~\cite{Sa03} 
solver with a fast multipole method~\cite{GR87a}. 
The closed-form of the Green's functions coupled with the fast multipole 
method allow them to avoid storing the system matrix circumventing the
memory constraints. Nevertheless, the inpaintings are still not real-time 
even at low resolutions such as $256\times 256$.


\paragraph{Finite Elements}
Chizhov and Weickert~\cite{CW21} replace the standard finite difference 
discretization of homogeneous diffusion inpainting with adaptive finite 
element approximations. This leads to faster inpaintings, since one 
replaces the fine regular pixel grid by a coarser adaptive triangulation 
resulting in a smaller system matrix and fewer unknowns.
On the other hand, the mesh construction and corresponding sparse matrix 
assembly results in a non-negligible overhead. The latter two steps are also 
hard to parallelize, which makes it less suitable for a GPU implementation.


\paragraph{Video Coding}
In the context of video compression, Peter et al.~\cite{PSMM15} proposed a 
method for nonlinear anisotropic diffusion inpainting. It relies on fast 
explicit finite difference schemes~\cite{WGSB16}, which are well-suited for 
parallelization and benefit strongly from a good initialization available 
from the previous frame. With this advantage, they were able to achieve 
real-time decoding of $640 \times 480$ videos on an 
\textit{Nvidia GeForce GTX 460} GPU. In our work we do not utilize such 
temporal coherence which allows us to be more general. However, this also 
suggests that our methods can be made even faster when applied to videos.

Two other real-time capable video codecs that exploit temporal coherence 
go back to Andris et al.~\cite{APKMWH21, APW16}. They combine global 
homogeneous diffusion inpainting of keyframes with optic flow based 
prediction for the frames in between. In~\cite{APKMWH21} they achieve 
real-time performance for FullHD color videos on an
\textit{Intel Xeon CPU W3565@3.20GHz}, by solving the inpainting with 
a multilevel conjugate gradient method~\cite{BD96}. 
	
	
\subsection{Paper Structure}
\label{sec:organisation}
	
We review the basics of homogeneous diffusion inpainting in 
\cref{sec:inpainting}. In \cref{sec:oras} we describe our domain 
decomposition approach for homogeneous diffusion inpainting. In order to 
accelerate the domain decomposition inpainting, specifically the convergence 
of the low-frequency modes of the error, we discuss how we embed it in a full 
multigrid framework in \cref{sec:multigrid}. 
We evaluate our inpainting method experimentally in 
\cref{sec:experiments}. Finally, we conclude our paper and provide an outlook 
on future work in \cref{sec:conclusions}.

	
\section{Homogeneous Diffusion Inpainting}
\label{sec:inpainting}

In this section we briefly review homogeneous diffusion 
inpainting~\cite{Ca88}, which is our inpainting method of choice and is used 
for all results in this paper. Homogeneous diffusion inpainting is a simple 
and very popular method for inpainting-based compression. It is 
parameter-free, allows to establish a comprehensive data selection theory, 
and may lead to particularly fast algorithms. It has been shown that, if 
the inpainting data is carefully chosen, we can achieve very good 
results~\cite{HMWP13,JPW20,MBWF11,MHWT12}. 


\subsection{Continuous Formulation}
\label{sec:inpainting-continuous}

Let us consider a continuous grayscale image $f: \Omega \to \mathbb{R}$ 
defined on a rectangular image domain $ \Omega \subset \mathbb{R}^2$. 
We store only a fraction of the full image data on a small subset 
$K \subset \Omega$. 
Homogeneous diffusion inpainting aims at restoring $f$ in the inpainting 
domain $\Omega \setminus K$ based only on the stored data $f|_K$. More 
specifically, the reconstruction $u$ is obtained by using  the known data 
$f|_{K}$ as Dirichlet boundary conditions, while solving the Laplace equation 
on $\Omega\setminus K$ with reflecting boundary conditions on the image 
boundary $\partial \Omega$:
\begin{alignat}{2}
	\label{eq:inp-continuous-1}
	-\Delta u (\bm{x}) &= 0, &\quad &\bm{x} \in \Omega
	\setminus K, \\
	\label{eq:inp-continuous-2}
	u (\bm{x}) &= f (\bm{x}), &\quad &\bm{x} \in K, \\
	\label{eq:inp-continuous-3}
	\partial_{\bm{n}} u (\bm{x}) &= 0, &\quad &\bm{x} 
	\in 
	\partial \Omega,
\end{alignat}
where $\partial_{\bm{n}} u$ denotes the derivative of $u$ in outer normal 
direction $\bm{n}$. For convenience, we may combine 
\cref{eq:inp-continuous-1,eq:inp-continuous-2} into a single equation by 
using an indicator function $c = \mathds{1}_K$, which is 1 on the known 
data domain $K$ and 0 outside of it:
\begin{alignat}{2}
\label{eq:inp-final-1}
	c(\bm{x})u(\bm{x}) + 
	(1-c(\bm{x}))(-\Delta)u(\bm{x}) &= 
	c(\bm{x})f(\bm{x})  &\quad &\text{on } \Omega, \\
\label{eq:inp-final-2}
		\partial_{\bm{n}} u (\bm{x}) &= 0 &\quad &\text{on }
	\partial \Omega.
\end{alignat}


\subsection{Discrete Formulation}
\label{sec:inpainting-discrete}

To apply homogeneous diffusion inpainting on a digital image $\bm{f} \in 
\mathbb{R}^N$ with $N$ pixels, we have to discretize 
\cref{eq:inp-final-1,eq:inp-final-2}. The discretization with finite 
differences leads to  a linear system of equations~\cite{MBWF11,MM05}.
We term the approximation of the indicator function $\mathds{1}_K \approx 
\bm{c}\in\{0,1\}^N$ the \emph{inpainting mask}. Furthermore, we discretize 
the negated Laplacian $-\Delta \approx \bm{L} \in \mathbb{R}^{N\times N}$ with 
reflecting boundary conditions, using the standard 5-point stencil 
discretization. Its values are given by
\begin{equation}
L_{i,j} = 
\begin{cases}
-\frac{1}{h^2} &\text{if } j\in \mathcal{N}(i),\\
\frac{\lvert \mathcal{N}(i) \rvert}{h^2} &\text{if } i=j, \\
0 &\text{else},
\end{cases}
\end{equation}
where $\mathcal{N}(i)$ are the direct (non-diagonal) neighbors of pixel $i$, 
$|\mathcal{N}(i)|$ is their count, and $h$ is the distance between pixels. 
Then the discrete formulation of \cref{eq:inp-final-1,eq:inp-final-2} is 
given by
\begin{equation}
	\label{eq:inp-discrete-lsi}
	\underbrace{(\bm{C}+(\bm{I}-\bm{C}) \bm{L})}_{=:\bm{A}} \bm{u} 
	= \underbrace{\bm{C} \bm{f}}_{=:\bm{b}},
\end{equation}	
where $\bm{I} \in \mathbb{R}^{N\times N}$ is the identity matrix, 
and $\bm{C} := \text{diag}(\bm{c}) \in \mathbb{R}^{N\times N}$ is a 
diagonal matrix with the components of the inpainting mask on the diagonal.


\paragraph{Practical Considerations}
For an RGB color image, the inpainting is performed on each of the three 
channels separately. Thus, we get three linear systems of this type with 
different right-hand sides.
While $\bm{A} \bm{u} = \bm{b}$ is not symmetric, it can be reduced to 
a symmetric and positive-definite system by removing the mask 
pixels~\cite{MBWF11}. Since the extended system (\ref{eq:inp-discrete-lsi}) 
acts exactly the same as the reduced one at all unknown pixels, we can solve 
it with a conjugate gradient method (CG)~\cite{HS51}.
For highly parallel hardware such as GPUs, solving the global homogeneous 
diffusion inpainting system with a CG method is, however, not very efficient 
compared to domain decomposition methods.
The reason for this lies on the one hand in the global scalar products 
which entail two global synchronizations in each iteration, but on the other 
hand also in the extensive use of global GPU memory, which is quite slow. 
Domain decomposition methods, instead, require only a single global
scalar product per iteration and can use faster local GPU memory for the 
local problems on the subdomains. This makes them much more efficient on 
highly parallel GPUs. For these reasons we discuss domain decomposition 
methods in the following section and specifically our method of choice:
the optimized restricted additive Schwarz method (ORAS)~\cite{SGT07}.  


\section{Optimized Restricted Additive Schwarz}
\label{sec:oras}
\begin{figure}[tb]
	\centering
	\includegraphics[width=4cm]{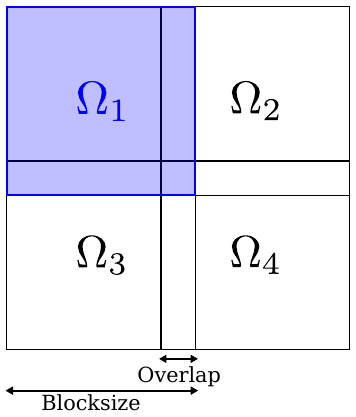}
	\caption{\textbf{Domain Decomposition Example.} The domain is 
            divided into four overlapping subdomains. The subdomain 
            $\Omega_1$ is highlighted in blue.}
	\label{fig:subdivision}
\end{figure}\textbf{}
The \textit{restricted additive Schwarz (RAS)} method~\cite{CS99}, in its 
continuous formulation, is an iterative technique for solving boundary value 
problems, such as 
\cref{eq:inp-continuous-1,eq:inp-continuous-2,eq:inp-continuous-3}.
It is one of the simplest domain decomposition methods~\cite{DJN15, TW05} and 
is well suited for parallelization. As a first step, the image domain 
$\Omega$ is divided into $k\in \mathbb{N}$ overlapping subdomains 
$\Omega_1,...,\Omega_k \subset \Omega$, such that 
$\cup_{i=1}^k\Omega_i=\Omega$. An example of such an overlapping division 
into four subdomains is shown in \cref{fig:subdivision}. 
%
\begin{algorithm}[htb]
	\caption{RAS for Homogeneous Diffusion Inpainting: 
                    Continuous Formulation} 
	\label{alg:ras-continuous} 
	\begin{enumerate}
		\item
		Compute the global residual $r^n: \Omega \to \mathbb{R}$:
		\begin{equation}
			r^n(\bm{x}) = 
			\begin{cases}
				0 + \Delta u^n(\bm{x}) & \quad 
				\text{for } \bm{x} \in \Omega \setminus K, \\
				0  & \quad \text{for } \bm{x} \in K.
			\end{cases} 
		\end{equation}
		\item
		For $i \in \{1, ..., k\}$ solve the following PDE for 
            the local correction $v_i^n$:
		\begin{alignat}{2}
			-\Delta v_i^n (\bm{x}) &= r^n(\bm{x}) &\quad 
			&\bm{x} \in \Omega_i \setminus K, \\
			v_i^n (\bm{x}) &= r^n(\bm{x}) &\quad 
			&\bm{x} \in \Omega_i \cap K, \\
			\partial_{\bm{n}} v_i^n (\bm{x}) &= 0 &\quad 
			&\bm{x} \in \partial \Omega_i \cap \partial \Omega, \\
			\label{eq:ras-continous-subdomain-boundaries}
			v_i^n (\bm{x}) &= 0 &\quad 
			&\bm{x} \in 
			\partial \Omega_i \setminus \partial \Omega.
		\end{alignat}
		\item 
		Update $u^n$ with the extensions of the weighted 
            local corrections $v_i^n$:
		\begin{equation}
			u^{n+1}(\bm{x}) = u^n(\bm{x}) + \sum_{i=1}^{k} 
			E_i(\chi_i(\bm{x}) v_i^n(\bm{x})).
		\end{equation}
	\end{enumerate}
\end{algorithm}
%
In the continuous setting, the RAS method is given by iterating 
\cref{alg:ras-continuous}.  
Starting with an initial approximation $u^0: \Omega \to \mathbb{R}$ that
fulfills the interpolation condition $u^0|_K = f|_K$ at the known locations,
in each iteration $n$, we try to correct the current approximation $u^n$
by locally computing correction terms $v_i: \Omega_i \to \mathbb{R}$ to the 
global error $e^n = u - u^n$ for every subdomain. We obtain these local 
corrections by solving local versions of the global PDE, but with the global 
residual $r^n$ as the right-hand side. At the subdomain boundaries which are 
not part of the global image boundary, the so-called inner subdomain 
boundaries, we need additional boundary conditions. There we impose 
homogeneous Dirichlet conditions. Together with the homogeneous Neumann 
conditions at the image boundaries and the Dirichlet conditions at the 
mask points, we get local PDEs with conditions over three different sets: 
$\partial\Omega_i\cap\partial\Omega$, 
$\partial\Omega_i\setminus \partial\Omega$, and $\Omega_i\cap K$.
We extend the local corrections $v_i$ from functions on $\Omega_i$ to 
functions on $\Omega$, with an extension operator 
$E_i (v_i): \Omega \to \mathbb{R}$ that sets them to zero outside of 
$\Omega_i$. This allows us to add them to the global solution $u^n$.
However, if we simply add the corrections from all subdomains to the global 
solution, we would add multiple corrections in the overlapping regions. 
This would result in divergence; see e.g.~\cite{EG03,Ga08}.
In order to achieve convergence, we have to weight the local corrections in 
the overlapping regions before adding them to the global solution, by defining 
partition of unity functions $\chi_i: \Omega_i \to \mathbb{R}$, such that
\begin{equation}
	w = \sum_{i=1}^{k} E_i (\chi_i w\lvert_{\Omega_i})
        \label{eq:partition_of_unity_condition}
\end{equation}
holds for any function $w:\Omega \to \mathbb{R}$. This ensures that at every 
point the weights always add up to $1$. As we iterate this method, $u^n$ 
converges towards the solution $u$ of the global PDE, as long as the 
subdomains overlap. The convergence speed of RAS depends on the size of the 
overlap as well as the overall number of subdomains. 

By replacing the inner subdomain boundary conditions with mixed Robin 
boundary conditions, we can optimize the communication between neighboring 
subdomains. This improves the convergence speed of the method. The resulting 
algorithm is then called the \textit{optimized restricted additive Schwarz 
(ORAS)} method~\cite{SGT07}. This means that we replace the Dirichlet boundary 
conditions \cref{eq:ras-continous-subdomain-boundaries} with mixed boundary 
conditions,
\begin{equation}
\partial_{\bm{n}} v_i^n (\bm{x}) + \alpha \cdot v_i^n 
(\bm{x}) = 0 \quad \bm{x} \in \partial \Omega_i \setminus 
\partial \Omega,
\end{equation}
where $\alpha > 0$ is a weighting factor between the Dirichlet 
and the Neumann condition. 


\paragraph{Discrete Optimized Restricted Additive Schwarz}
In the discrete case we have to solve a sparse linear system of equations 
\cref{eq:inp-discrete-lsi}. To this end, the discrete  image domain $\Omega$ 
with $N$ pixels is also divided into $k$ overlapping domains 
$\Omega_1, ..., \Omega_k \subset \Omega$ such that 
$\cup_{i=1}^k\Omega_i=\Omega$. 
We define restriction matrices $\bm{R}_i\in \mathbb{R}^{|\Omega_i| \times N}$, 
where $|\Omega_i|$ is the number of pixels in subdomain $\Omega_i$, that 
restrict global vectors to the local domains by extracting the values from 
the subdomain $\Omega_i$ and ignoring all values outside of it. 
Their transposes $\bm{R}^\top_i$ are then extension matrices analogues to 
$E_i$ from the continuous case, that extend local vectors from the subdomain 
$\Omega_i$ to the global domain $\Omega$.

For the discrete counterpart of the partition of unity functions $\chi_i$, 
we define diagonal nonnegative matrices $\bm{D}_i \in 
\mathbb{R}^{\lvert\Omega_i\rvert \times \lvert\Omega_i\rvert}$, such that
\begin{equation}
\label{eq:partition-of-unity-discrete}
\bm{I} = \sum_{i=1}^{k}  \bm{R}_i^\top \bm{D}_i \,
\bm{R}_i,
\end{equation}
where $\bm{I} \in\mathbb{R}^{N\times N}$ is the identity matrix of size 
$N \times N$. The latter implies the discrete counterpart 
$\bm{w} = \sum_{i=1}^k \bm{R}_i^{\top}\bm{D}_i\,\bm{R}_i\,\bm{w}$ 
of the condition \cref{eq:partition_of_unity_condition}.

The local corrections $\bm{v}_i^n$ are computed by solving small linear 
systems, where the system matrices $\bm{A}_i $ are given by restricting the 
global system matrix $\bm{A}$ to the subdomains $\Omega_i$ and imposing mixed 
Robin boundary conditions at the inner subdomain boundaries.
Starting with an initial guess $\bm{u}^0$ the resulting discrete ORAS method 
is given by iterating \cref{alg:oras-discrete}.
\begin{algorithm}[htb] 
	\caption{ORAS for Homogeneous Diffusion Inpainting: 
                Discrete Formulation} 
	\label{alg:oras-discrete} 
	\begin{enumerate}
		\item
		Compute the global residual $\bm{r}^n \in \mathbb{R}^N$:
		\begin{equation}
		\bm{r}^n = \bm{b} - \bm{A} \bm{u}^n
		\end{equation}
		\item
		For $i \in \{1, ..., k\}$ solve for a local correction 
		$\bm{v}_i^n$:
		\begin{equation}
		\bm{A}_i 
		\bm{v}_i^n = \bm{R}_i \bm{r}^n
		\end{equation}
		\item 
		Update $\bm{u}^n$ with the weighted and extended local 
		corrections $\bm{v}_i^n$:
		\begin{equation}
		\bm{u}^{n+1} = \bm{u}^n + \sum_{i=1}^{k} 
		\bm{R}_i^\top \bm{D_i} \bm{v}_i^n
		\end{equation}
		
	\end{enumerate}
	
\end{algorithm}

\subsection{Implementation Details}

\begin{figure}[tb]
	\centering
	\centerline{\includegraphics[width=1.0\linewidth]
		{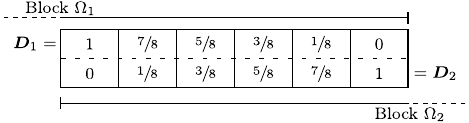}}
	
	\caption{\textbf{Visualization of the Overlap Weights.} 
		The weights $\bm{D}_1$ and $\bm{D}_2$ in the 
		overlap of two blocks $\Omega_1$ and $\Omega_2$ 
		are chosen, such that they always add up to 1 at each pixel.}
	\label{fig:block-overlap-weights}
\end{figure}

Even though domain decomposition methods are typically used as 
preconditioners for Krylov methods such as CG, in the current work we use 
them directly. 
Due to the approximately local influence of each mask pixel, the global 
coupling from Krylov methods is not necessary for us. As global Krylov 
methods need several synchronizations per iteration and rely on global 
GPU memory, using ORAS as a direct method is more efficient for our GPU 
implementation. In our ORAS inpainting method we decompose the image domain 
into multiple overlapping blocks of size $32\times 32$ pixels with an 
overlap of $6$ pixels. These values are optimized for our 
\textit{Nvidia GeForce GTX 1080 Ti} GPU, which results in $36852$ local 
problems in each iteration of the ORAS method for a 4K color image. 
The local problems are solved using a simple conjugate gradient 
algorithm~\cite{Sa03}. We stop the local CG iterations on a block when 
the squared 2-norm of the residual has been reduced to a fixed fraction 
of the squared global residual norm. Compared to a stopping criterion 
based on the local relative residual norm, this leads to a more uniform 
convergence, since all blocks have the same squared residual norm after 
each ORAS iteration. This also avoids using unnecessarily many CG 
iterations in blocks that are already sufficiently converged. 
For the weight matrices $\bm{D}_i$, we use separable weights in 
$x$- and $y$-direction, which means we only have to ensure that the partition 
of unity holds for an overlap between two blocks in a single direction. 
\cref{fig:block-overlap-weights} visualizes our weighting. For the outermost 
pixels we have to use a weight of $0$ to ensure convergence~\cite{SGT07}, 
which leads to a weight of $1$ for the innermost pixel in the overlapping 
region. In between, we use a linear weighting, such that both weights add 
up to $1$.

\section{Embedding ORAS in Multigrid}
\label{sec:multigrid}

Simple domain decomposition methods such as \newline ORAS have one 
drawback: They only allow communication between neighboring blocks, so it 
takes many iterations to spread information over large distances. 
Thus, high frequencies of the error are reduced much faster than low 
frequencies. This leads to a fast convergence rate at the beginning, which 
slows down significantly after the first few iterations. As a remedy, domain 
decomposition methods are usually used together with a simple two-level scheme 
\cite{DJN15, TW05}.
Besides the local subdomain corrections, it computes an additional correction 
on a very coarse grid to couple all subdomains together. This helps to better 
attenuate the low-frequency error components. As the influence area of each 
mask pixel in homogeneous diffusion inpainting is approximately localized, 
such a global coupling of all subdomains is not really necessary. Instead, a 
more fine-grained coupling between subdomains is more useful. 

Multigrid methods~\cite{Br77,BHM00, Hac85, TOS01, Wes92} can offer such a 
coupling between subdomains. They compute useful correction steps on coarser 
levels to improve the convergence of the low-frequency error components. In 
the next section we describe a simple two level multigrid method.

\subsection{Two-Grid Cycle}

Iterative solvers such as Jacobi~\cite{Sa03}, Gauss-Seidel~\cite{Sa03}, or 
the ORAS method~\cite{SGT07} reduce the high frequency modes of the error 
efficiently, but they are much slower at decreasing the low frequencies. 
Multigrid addresses this by transferring the problem to a coarser grid where 
the low frequencies appear as higher frequencies which can be reduced more 
efficiently by the very same solvers. We first study the two grid formulation 
of multigrid which consist of iterating what's known as a 
\textit{two-grid cycle}. 
Our two grids are the original fine grid with grid spacing $h$, and a coarser 
grid with spacing $H>h$. In our implementation, we set $H=2h$.

We start the cycle $k+1$ with an approximation $\bm{u}_h^{k}$ of the solution 
$\bm{u}_h$ of the problem $\bm{A}_h\bm{u}_h = \bm{b}_h$. Here $\bm{A}_h$ is 
the original system matrix and $\bm{b}_h$ is the right-hand side (both on the 
original fine grid). If we knew the true error $\bm{u}_h-\bm{u}_h^k$ we could 
correct our approximation to get the true solution: 
$\bm{u}_h^k + (\bm{u}_h-\bm{u}^k_h) = \bm{u}_h$. Consequently, our goal is to 
find an approximation of the error efficiently, by using the aforementioned 
idea of decimating different frequencies of the error over different grids. 
The error $\bm{u}_h-\bm{u}_h^{k}$ typically contains both high- and 
low-frequency components. We reduce the high frequencies by performing a few 
iterations $\vartheta_\text{pre}$ with a \textit{smoother}: an iterative 
solver that dampens the high frequencies efficiently - e.g.\ damped Jacobi, 
Gauss-Seidel, or in our case ORAS. This results in the approximation of the 
solution 
\begin{equation}
\bm{u}_h^{k+1/3} = 
\text{pre-smooth}
(\bm{A}_h,\bm{b}_h,\bm{u}_h^{k},\vartheta_\text{pre}).
\end{equation}
Provided we have used enough iterations $\vartheta_\text{pre}$, the error 
$\bm{e}_h^k = \bm{b}_h-\bm{u}^{k+1/3}_h$ will have only negligible 
high-frequency components. We can then represent it almost perfectly on the 
coarser grid. The transfer between the grids can be formalized using a 
restriction matrix $\bm{R}^H_h$ and a prolongation matrix $\bm{P}^h_H$. 
The two are designed so that sufficiently low frequency components 
are reproduced exactly, while higher frequency components are smoothed 
out to avoid aliasing.

A problem of the above formulation is that we do not know the error 
in practice, so we cannot transfer it to the coarser grid. 
However, we know the residual and its relation to the error:
\begin{equation}
    \bm{r}^k_h = \bm{b}_h-\bm{A}_h\bm{u}_h^{k+1/3} = 
    \bm{A}_h\bm{u}_h - \bm{A}_h\bm{u}_h^{k+1/3} = 
    \bm{A}_h\left(\bm{u}_h-\bm{u}_h^{k+1/3}\right) = 
    \bm{A}_h\bm{e}_h^k.
\end{equation}
On the coarse grid this reads $\bm{A}_H\bm{e}_H^k=\bm{r}_H^k$, where 
$\bm{A}_H$ is an analogue of $\bm{A}_h$ on the lower level. To compute 
$\bm{A}_H$, we discretize the inpainting problem on the coarser grid (see 
\cref{sec:lower_resolution_problems_construction}). 
We get the coarse residual $\bm{r}^k_H$ by transferring the fine grid 
residual $\bm{r}^k_h$ to the coarse grid:
\begin{equation}
    \bm{r}_H^k = \bm{R}^H_h \bm{r}^k_h.
\end{equation}
In the two-grid cycle we then solve the coarse grid problem for the error 
exactly:
\begin{equation}
    \label{eq:coarse_grid_solve}
    \bm{A}_H\bm{e}_H^k = \bm{r}^k_H.
\end{equation}
We note that the matrix $\bm{A}_H$ is twice smaller in each dimension 
compared to $\bm{A}_h$, thus the computational cost is also reduced.
We can then transfer $\bm{e}_H^k$ to the fine grid by using the 
prolongation matrix $\tilde{\bm{e}}^k_h = \bm{P}^h_H\bm{e}^k_H$ 
and use it to correct our approximation of the solution:
\begin{equation}
    \bm{u}_h^{k+2/3} = \bm{u}^{k+1/3} + \tilde{\bm{e}}^k_h = 
    \bm{u}^{k+1/3} + \bm{P}^h_H\bm{e}^k_H.
\end{equation}
A subsequent post-smoothing with $\vartheta_\text{post}$ iterations is then 
applied to smooth high frequency error components potentially introduced by 
the correction step:
\begin{equation}
    \bm{u}^{k+1}_h = 
    \text{post-smooth}
    (\bm{A}_h, \bm{b}_h, \bm{u}_h^{k+2/3}, \vartheta_\text{post}).
\end{equation}
All of these steps together result in \Cref{alg:2-grid-cycle}.
%
\begin{algorithm} 
	\caption{Two-Grid Cycle} 
	\label{alg:2-grid-cycle} 
	\begin{enumerate}
		\item
		Pre-smooth: $\bm{u}_h^{k+1/3} = \text{ORAS}(\bm{A}_h, 
		\bm{b}_h,\bm{u}_h^{k}, \vartheta_\text{pre})$
		\item 
		Compute Residual: $\bm{r}^k_h = \bm{b}_h - 
		\bm{A}_h \, \bm{u}_h^{k+1/3}$
		\item 
		Restriction: $\bm{r}^k_H = \bm{R}_h^H \, \bm{r}^k_h$
		\item 
		Coarse Solve: $\bm{A}_H \, \bm{e}^k_H = \bm{r}^k_H$
		\item 
		Prolongation + Correction: $\bm{u}_h^{k+2/3} = 
		\bm{u}_h^{k+1/3} + 
		\bm{P}_H^h \, \bm{e}^k_H$
		\item 
		Post-smooth: $\bm{u}_h^{k+1} = 
		\text{ORAS}(\bm{A}_h, 
		\bm{b}_h,\bm{u}_h^{k+2/3}, \vartheta_\text{post})$
		
	\end{enumerate}
\end{algorithm}
%
While solving $\bm{A}_H \bm{x}_H = \bm{b}_H$ is generally computationally 
cheaper than solving $\bm{A}_h\bm{x}_h = \bm{b}_h$, for high resolution 
images even the coarse problem is expensive.
Thus, one often considers recursively applying this process over a sequence 
of grids with decreasing resolution until a grid of a sufficiently low 
resolution is reached. 
The latter results in an almost optimal workload of $O(N\log N)$ for 
computing the solution. Another issue is how to choose a good initial guess 
$\bm{u}^0_h$. 
It turns out that a good quality guess can be constructed efficiently by 
starting from the coarsest level. The recursive approach with this initial 
guess has an optimal workload of $O(N)$. Both of these ideas are discussed 
in the following subsection.

\subsection{V-Cycles and Full Multigrid}

Instead of using only two resolution levels, we can extend the two-grid cycle 
to multiple levels, which allows to efficiently dampen different frequencies 
of the error at different resolutions. The solve on the coarse level can be 
replaced by another correction step on an even coarser level. This process 
can be iterated recursively until we reach a level that is coarse enough for 
the lowest possible frequencies in our problem.
The hierarchical application of the two-grid cycle is called a 
\textit{V-cycle}.    

\begin{figure}[tb]
	\centering
	\includegraphics[scale=0.85] 
	{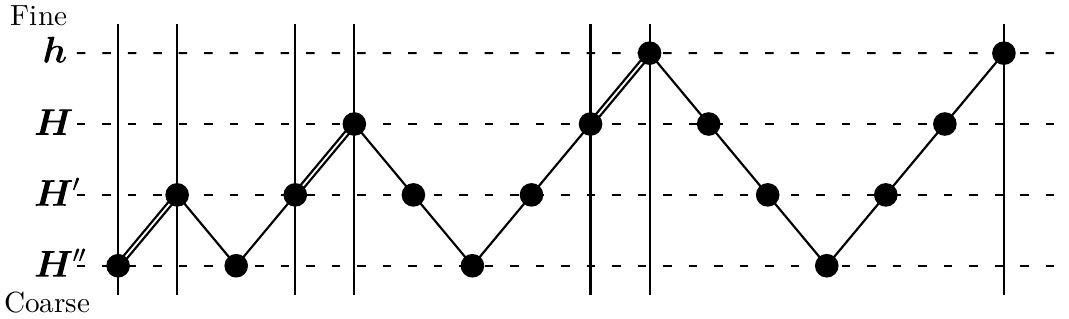}
	\caption{\textbf{Full Multigrid Scheme.} Example with four resolution 
	layers and a single V-cycle for each level. The doubled lines represent 
        the FMG prolongations to initialize the V-cycle for the next finer 
        level.}
	\label{fig:multigrid}
\end{figure}

\begin{figure}[H]
	\centering
	\includegraphics[scale=0.85] 
	{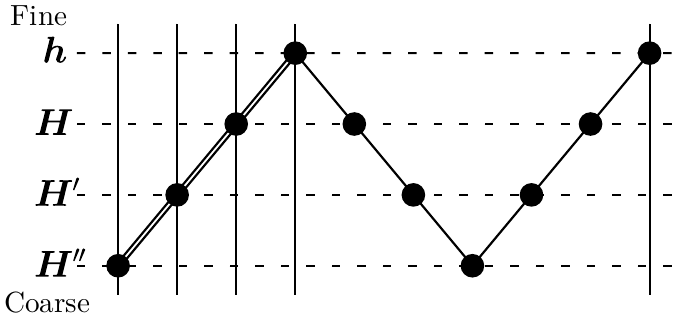}
	\caption{\textbf{Reduced Full Multigrid Scheme.} 
        The initial guess is constructed in a coarse-to-fine 
        manner, also known as one-way or cascadic multigrid. 
        Then we continue with additional V-cycle correction steps 
        (a single V-cycle is visualized above).
        }
	\label{fig:multigrid_reduced}
\end{figure}

Moreover, we can speed up the computation even further by providing a good 
initialization. Such an initialization can be obtained by computing an 
inpainting solution on a coarser grid and prolongating it to the fine 
resolution. This can be extended to a complete coarse-to-fine initialization 
strategy. Starting from a very coarse resolution grid, we successively refine 
the problem, where inpainting solutions from coarser levels are interpolated 
and used as an initialization for finer ones. At each level a V-cycle is used 
to solve the resulting linear system. This coarse-to-fine method with the 
error correction steps is called \textit{full multigrid (FMG)}. 
\cref{fig:multigrid} shows an illustration how the different grids are 
traversed in such an FMG scheme. 
However, restriction and prolongation between different grids can be 
costly in terms of runtime on parallel hardware such as GPUs, as they consist 
mainly of global memory operations. To reduce the amount of restriction and 
prolongation operations without losing the advantages of FMG, we propose to 
skip the V-cycles on the lower levels and just use a single pre-smoothing 
iteration instead. This reduced FMG scheme, is depicted in 
\cref{fig:multigrid_reduced}. 
For homogeneous diffusion inpainting, this reduced FMG scheme is sufficient 
for a good convergence and improves the overall runtime compared to the 
non-reduced scheme.  

\subsection{Lower Resolution Problems Construction}
\label{sec:lower_resolution_problems_construction}

Full multigrid requires constructing a problem $\bm{A}_H\bm{u}_H = \bm{b}_H$ 
on the coarser grid that is similar enough to the problem 
$\bm{A}_h\bm{u}_h = \bm{b}_h$ on the finer grid.
To achieve this, we use a downsampling ratio of two in each direction so that 
the difference between the two problems is as small as possible.
We note that in the multigrid literature the standard considered boundary is  
rectangular, and thus its connectivity and geometry remains the same under 
downsampling. In our case, only the reflecting Neumann boundary is on the 
rectangle corresponding to the image boundary. The Dirichlet boundary given 
by the mask $\bm{c}$ can be arbitrarily complex and usually consists of 
scattered pixels with largely varying density over the image domain. This 
means that both the geometry and connectivity of our Dirichlet boundary 
greatly changes with downsampling. 
This is important, as it is known that grid transfer operations in multigrid 
have to be adapted near the boundary for optimal performance~\cite{Wes92}.

\begin{figure}[tb]
	\begin{subfigure}[b]{0.24\linewidth}
		\centering
		\centerline{\includegraphics[width=1.0\linewidth]
			{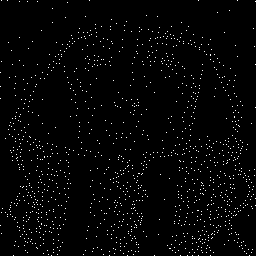}}
		\subcaption{$256 \times 256$ }
	\end{subfigure}
	\hfill
	\begin{subfigure}[b]{0.24\linewidth}
		\centering
		\centerline{\includegraphics[width=1.0\linewidth]
			{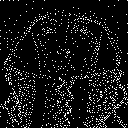}}
		\subcaption{$128 \times 128$ }
	\end{subfigure}
	\hfill
	\begin{subfigure}[b]{0.24\linewidth}
		\centering
		\centerline{\includegraphics[width=1.0\linewidth]
			{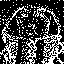}}
		\subcaption{$64 \times 64$ }
	\end{subfigure}
	\hfill
	\begin{subfigure}[b]{0.24\linewidth}
		\centering
		\centerline{\includegraphics[width=1.0\linewidth]
			{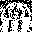}}
		\subcaption{$32 \times 32$ }
	\end{subfigure}
	\caption{\textbf{Mask Restriction Example  on \textit{trui} 
                with a 2\% Mask Density}.
                With each resolution level the mask density increases 
                while the connectivity of the mask changes. }
	\label{fig:mask_restriction}
\end{figure}

\paragraph{Mask Downsampling}
In order to construct coarser versions of the mask $\bm{c}$, we use a 
specific dyadic downsampling: A pixel on the coarser level becomes a mask 
pixel if any one of its corresponding pixels on the finer level is a mask 
pixel (max pooling as opposed to average pooling). For an illustration of 
this process, see \Cref{fig:mask_restriction}. 
This is similar in spirit to what is done in~\cite{MST10}, but there the 
authors deal with simpler boundaries. We note that the density of the 
boundary set increases as we downsample in this manner, which generally 
makes solutions on the coarser grid more efficient. At the same time, 
however, on coarser levels the boundary's geometry and even connectivity 
changes, such that our coarsest problem discretization is quite far 
removed from the initial fine grid problem. 
This explains for instance why a multigrid preconditioned conjugate gradient 
solver is not a good fit for our problem, or why downsampling with an 
aggressive ratio, as done in standard multilevel domain decomposition methods, 
is not optimal for our problem.

\begin{figure}[tb]
	\begin{subfigure}[b]{0.49\linewidth}
		\centering
     	\includegraphics[width=0.94\linewidth]
		{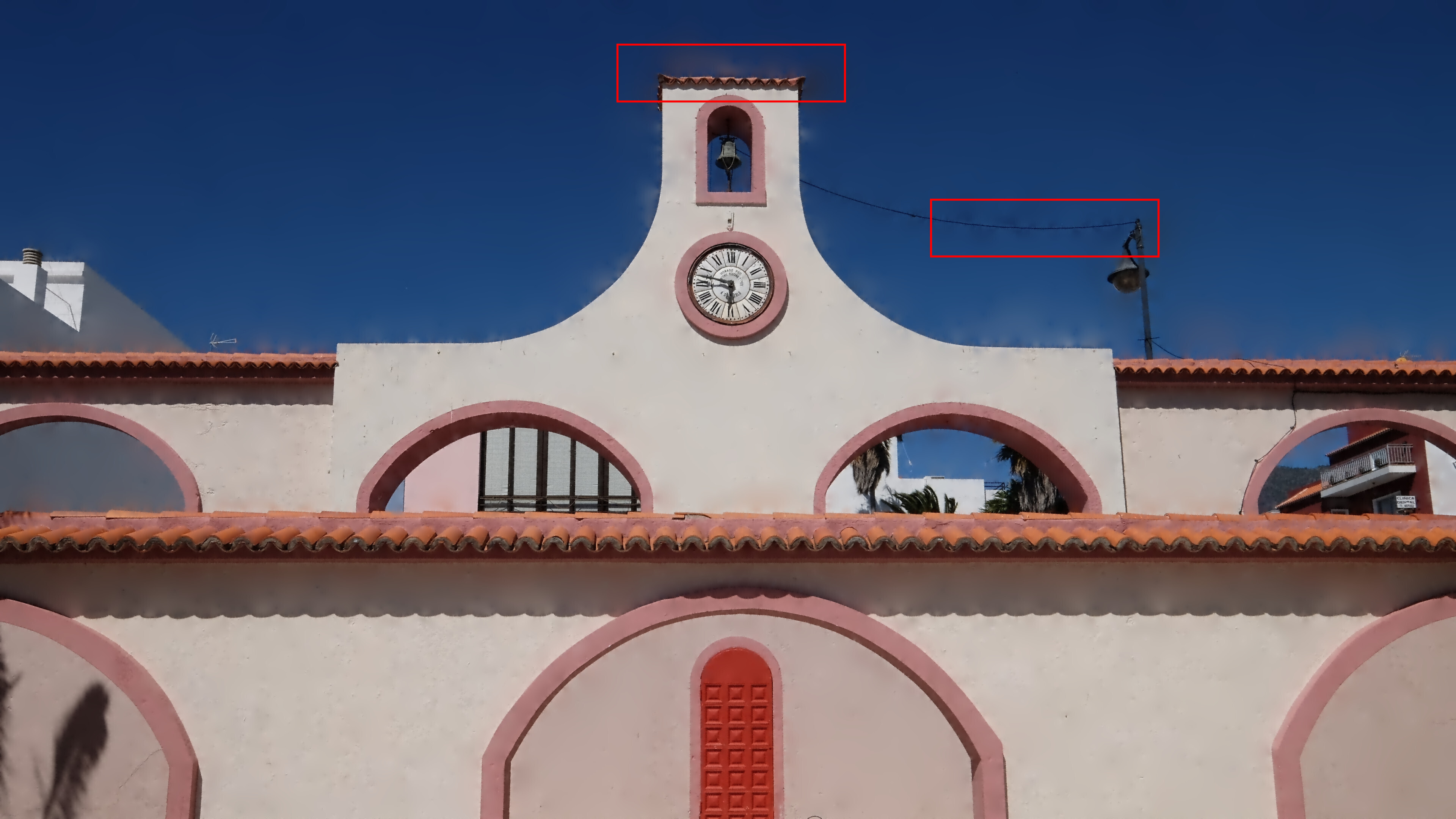}
		\\[0.015\linewidth]
		\setlength{\tabcolsep}{0.01125\linewidth}
		\begin{tabular}{cc}
            \includegraphics[width=0.459\linewidth]
			{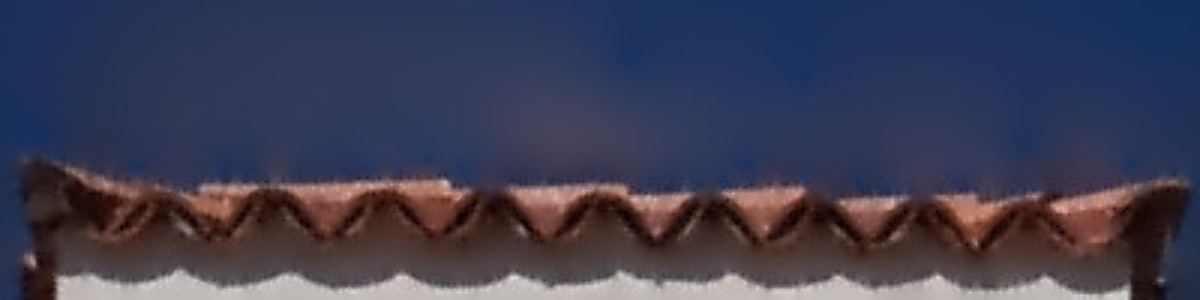} &
            \includegraphics[width=0.459\linewidth]
			{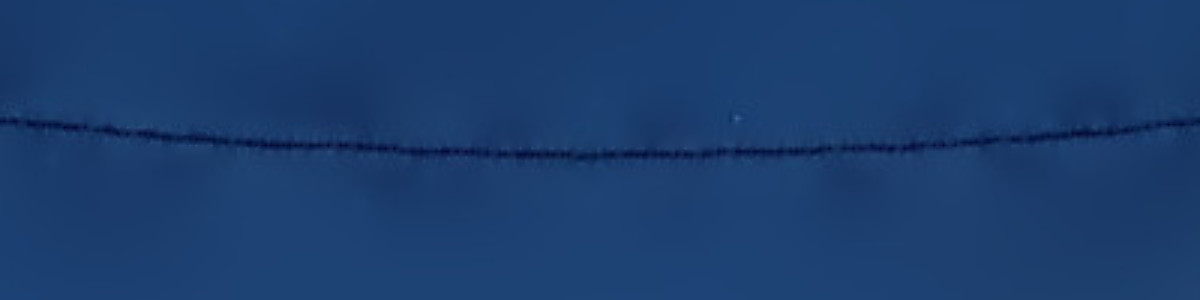}
        \end{tabular}
		\subcaption{na{\"i}ve downsampling}
        \label{fig:rhs_restriction_leakage_comparison_a}
	\end{subfigure}
	\hspace{-5mm}
	\begin{subfigure}[b]{0.49\linewidth}
		\centering
		\includegraphics[width=0.94\linewidth]
		{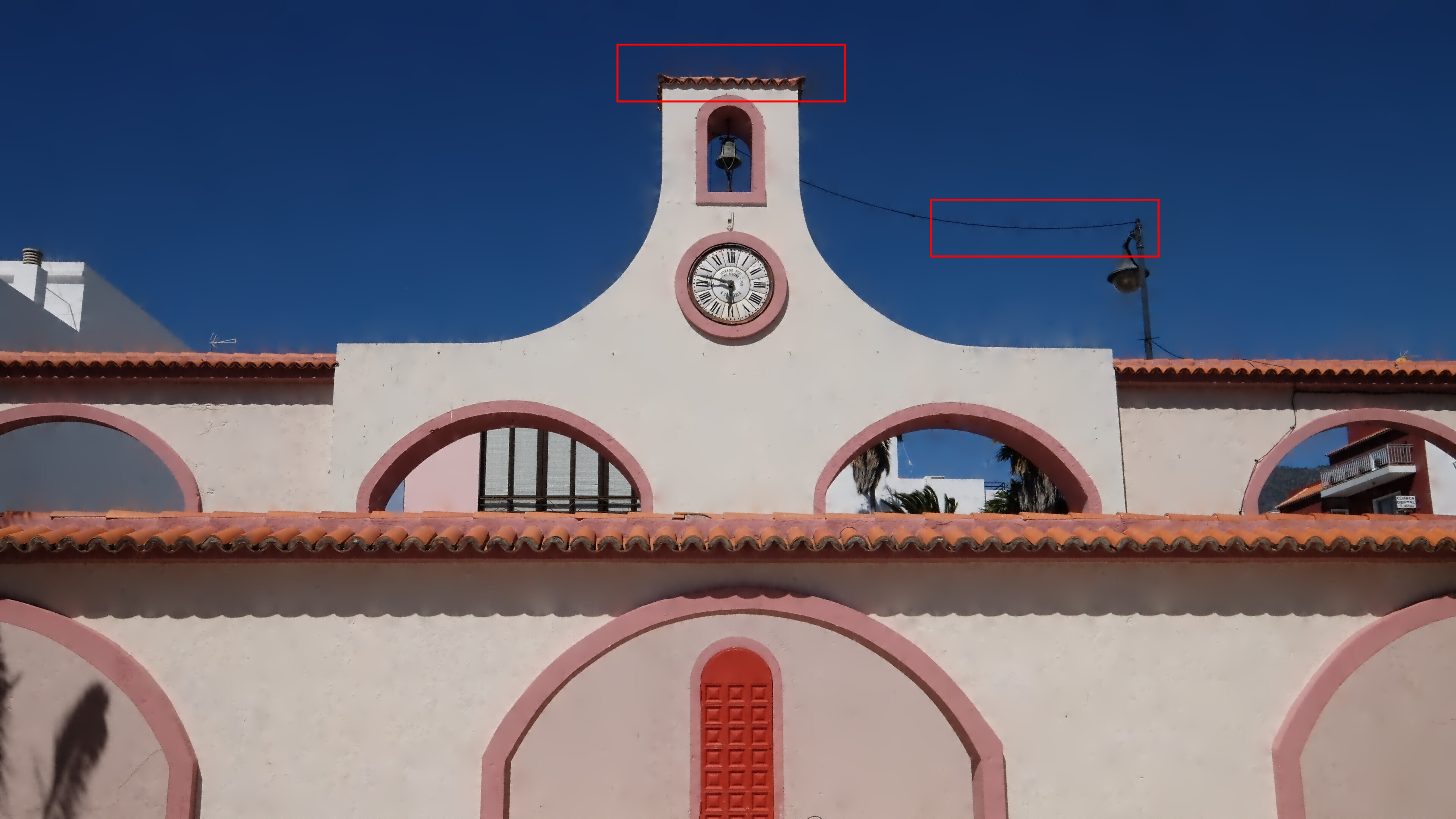}
		\\[0.015\linewidth]
		\setlength{\tabcolsep}{0.01125\linewidth}
		
        \begin{tabular}{cc}
            \includegraphics[width=0.459\linewidth]
			{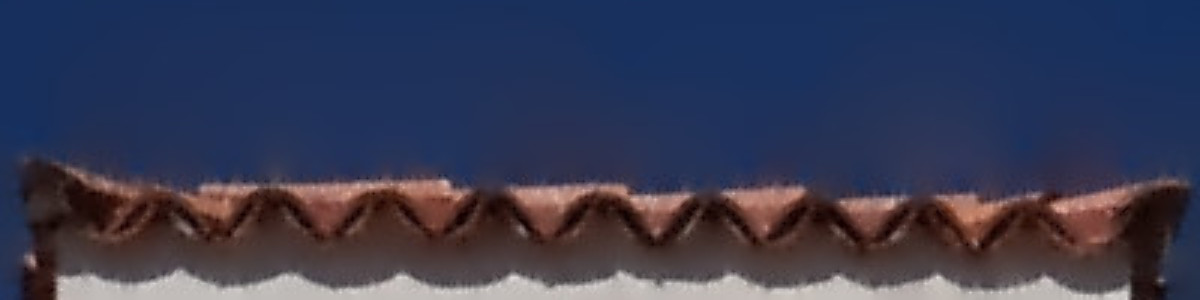} &
            \includegraphics[width=0.459\linewidth]
			{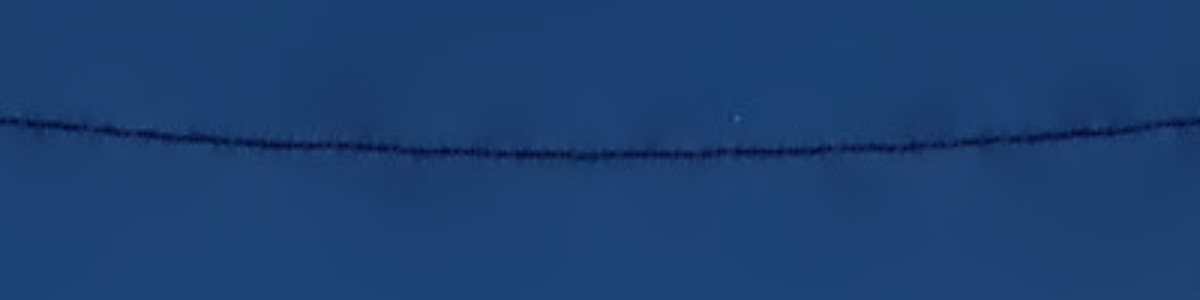}
        \end{tabular}
		\subcaption{modified downsampling}
  \label{fig:rhs_restriction_leakage_comparison_b}
	\end{subfigure}
	\caption{\textbf{Downsampling Leakage Example.}  
 Illustration of the leakage caused by na{\"i}ve dyadic downsampling 
 for a 2\% mask using only one-way multigrid.}
	\label{fig:rhs_restriction_leakage_comparison}
\end{figure}

\paragraph{Na{\"i}ve Mask Values Downsampling}
Besides for the mask, we also need to find lower resolution counterparts 
$\bm{b}_H=\bm{C}_H\bm{f}_H$ for the mask values $\bm{b}_h = \bm{C}_h \, 
\bm{f}_h$. 
The na{\"i}ve downsampling  averages the values of all mask pixels in a 
$2\times2$ neighborhood on the finer level. If there are no mask pixels in 
the neighborhood, the corresponding coarse level pixel itself is not 
considered as a mask pixel and thus the right-hand side should be set to $0$.
As an example, the na{\"i}ve coarse level right-hand side $b_{H}^{2,2}$ for 
the coarse level pixel $c_H^{2,2}$ is given by
\begin{equation}
    {\renewcommand{\arraystretch}{1.2}\begin{tabular}{|c|c|}
    \hline
    $c_h^{3,3}$ & $c_h^{4,3}$ \\
    \hline
    $c_h^{3,4}$ & $c_h^{4,4}$ \\
    \hline
    \end{tabular}}\,\,, \quad 
    {\renewcommand{\arraystretch}{1.2}\begin{tabular}{|c|c|}
    \hline
    $b_h^{3,3}$ & $b_h^{4,3}$ \\
    \hline
    $b_h^{3,4}$ & $b_h^{4,4}$ \\
    \hline
    \end{tabular}}\,\,, \quad
    b_H^{2,2} = \frac{c^{3,3}_h b_h^{3,3} 
             + c^{3,4}_h b_h^{3,4} 
             + c^{4,3}_h b_h^{4,3} 
             + c^{4,4}_h b_h^{4,4}}{
             \max\left(1,c^{3,3}_h+c^{4,3}_h+c^{3,4}_h+c^{4,4}_h\right)}.
\end{equation}
As one may expect with such a complex boundary, it turns out that the above 
simple averaging and downscaling is not ideal. 
A na{\"i}ve average mixes as many as four mask values from the finer level 
to get the value of the corresponding mask pixel on the coarser level. 
However, a key observation specific to homogeneous diffusion inpainting is 
that the influence of a mask pixel is localized to regions enclosed by other 
mask pixels. 
This means that a na{\"i}ve downsampling can cause mask pixels, that had a 
localized effect on the finer level, to have their influence leak over 
boundaries on the coarser level: note the pronounced color bleeding near 
edges in~\Cref{fig:rhs_restriction_leakage_comparison_a}. 
To rectify this, we modify the downsampling procedure for the mask values 
by allowing mask pixels to suppress the contribution of their neighbors. 

\paragraph{Modified Mask Values Downsampling}
If the four neighbors of a mask pixel are also mask pixels, then its value 
should not be averaged on the lower level, since its influence is suppressed 
by its neighbors. Note that the neighbors can be either fine scale or coarse 
scale mask pixels, since either will suppress the influence of the pixel 
provided that they are its direct neighbors.
As an example we consider the fine mask pixel $c_h^{3,3}$, then its direct 
neighbors are: its fine grid neighbors $c_h^{3,4}$ and $c_h^{4,3}$, and its 
coarse grid neighbors $c_H^{1,2}$ and $c_H^{2,1}$. 
An illustration of this is presented below, where the coarse pixel 
$c_H^{2,2}$ is split into its fine scale constituents $c_h^{3,3}$, 
$c_h^{3,4}$, $c_h^{4,3}$, $c_h^{4,4}$.

\begin{equation}
    {\renewcommand{\arraystretch}{1.2}
\begin{tabular}{|c|@{}c@{}|c|}
    \hline
    $\ast$ & $c_H^{2,1}$ & $\ast$ \\
    \hline
    $c_H^{1,2}$ & 
    \begin{tabular}{c|c}
    $c_h^{3,3}$ & $c_h^{4,3}$ \\
    \hline
    $c_h^{3,4}$ & $c_h^{4,4}$ \\
    \end{tabular}
    & $c_H^{3,2}$ \\
    \hline
    $\ast$ & $c_H^{2,3}$ & $\ast$ \\
    \hline
\end{tabular}}
\end{equation}
To model this suppression effect, we may assign each fine grid pixel a weight 
in the averaging based on the number of its non-mask pixel neighbors. 
This yields the following modified downsampling procedure for $b_H^{2,2}$:
\begin{gather}
\begin{alignedat}{3}
    w_h^{3,3} &= 
    c_h^{3,3}\left(4-c_H^{1,2}-c_H^{2,1}-c_h^{4,3}-c_h^{3,4}\right), 
    &\quad& 
    w_h^{3,4} &= 
    c_h^{3,4}\left(4-c_H^{1,2}-c_h^{3,3}-c_h^{4,3}-c_H^{2,3}\right), \\[1mm]
    w_h^{4,3} &= 
    c_h^{4,3}\left(4-c_h^{3,3}-c_H^{2,1}-c_H^{3,2}-c_h^{3,4}\right),
    &\quad&
    w_h^{4,4} &= 
    c_h^{4,4}\left(4-c_h^{3,3}-c_h^{3,3}-c_H^{3,2}-c_H^{2,3}\right),
\end{alignedat} \\[2mm]
    b^{2,2}_H = \frac{w_h^{3,3}b_h^{3,3} 
              + w_h^{3,4}b_h^{3,4}
              + w_h^{4,3}b_h^{4,3}
              + w_h^{4,4}b_h^{4,4}}
              {\max\left(1,w_h^{3,3}+w_h^{3,4}+w_h^{4,3}+w_h^{4,4}\right)}.
\end{gather}
The effect of this modification is illustrated in 
\Cref{fig:rhs_restriction_leakage_comparison_b}, which exhibits much less 
color bleeding than \Cref{fig:rhs_restriction_leakage_comparison_a}.
While the latter could be rectified with more V-cycles or complex smoothers, 
the modified downsampling allows to diminish such artifacts at a very low 
computational cost.

\subsection{Implementation Details}

For our multigrid ORAS inpainting method, we use a single V-cycle on the 
finest level, and we set the pre- and post-smoothing iterations to a single 
ORAS iteration. We choose the number of levels such that the coarsest level 
consists of a single ORAS block of size $32\times32$ pixels. 
Since we reduce the resolution by two in each direction with each level, we 
get a total of $8$ resolution levels for a 4K image.

For the restriction outside of the mask pixels we employ a simple averaging.
The prolongation to a finer grid is implemented with bilinear interpolation, 
except at the mask pixels from the fine grid where we have to fulfill the 
interpolation conditions.
The prolongated coarse grid solutions, used in the initial cascadic part, 
are set to the corresponding values from the right-hand side at the mask 
pixels such that the interpolation condition 
$\bm{C}_h \, \bm{P}_H^h \bm{u}_H = \bm{C}_h\bm{f}_h$ is satisfied.
We set the prolongated coarse grid corrections to zero at the mask pixels, 
since the residual is always zero there.   


\section{Experiments}
\label{sec:experiments}	

\begin{figure}[p]
	\setlength{\tabcolsep}{1mm}
	\begin{tabular}{ccccc}
		
		& \small original image& \small inpainting mask & \small inpainted 
		image & \\
		
		\centered{\rotatebox{90}{\small \textit{lofsdalen}}} &
		\centered{\includegraphics[width=0.3\textwidth]
		{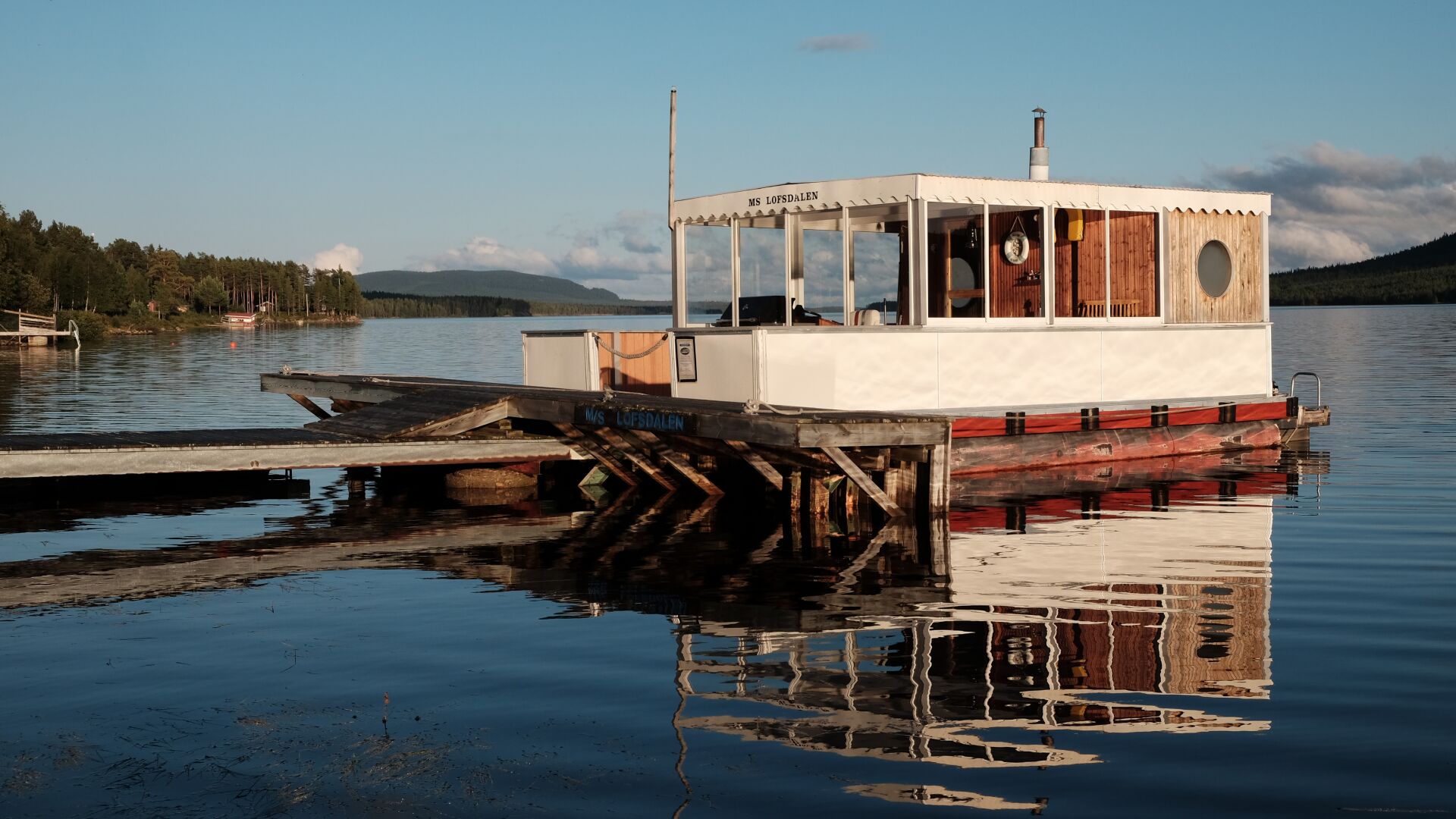}} & 
		\centered{\includegraphics[width=0.3\textwidth]
		{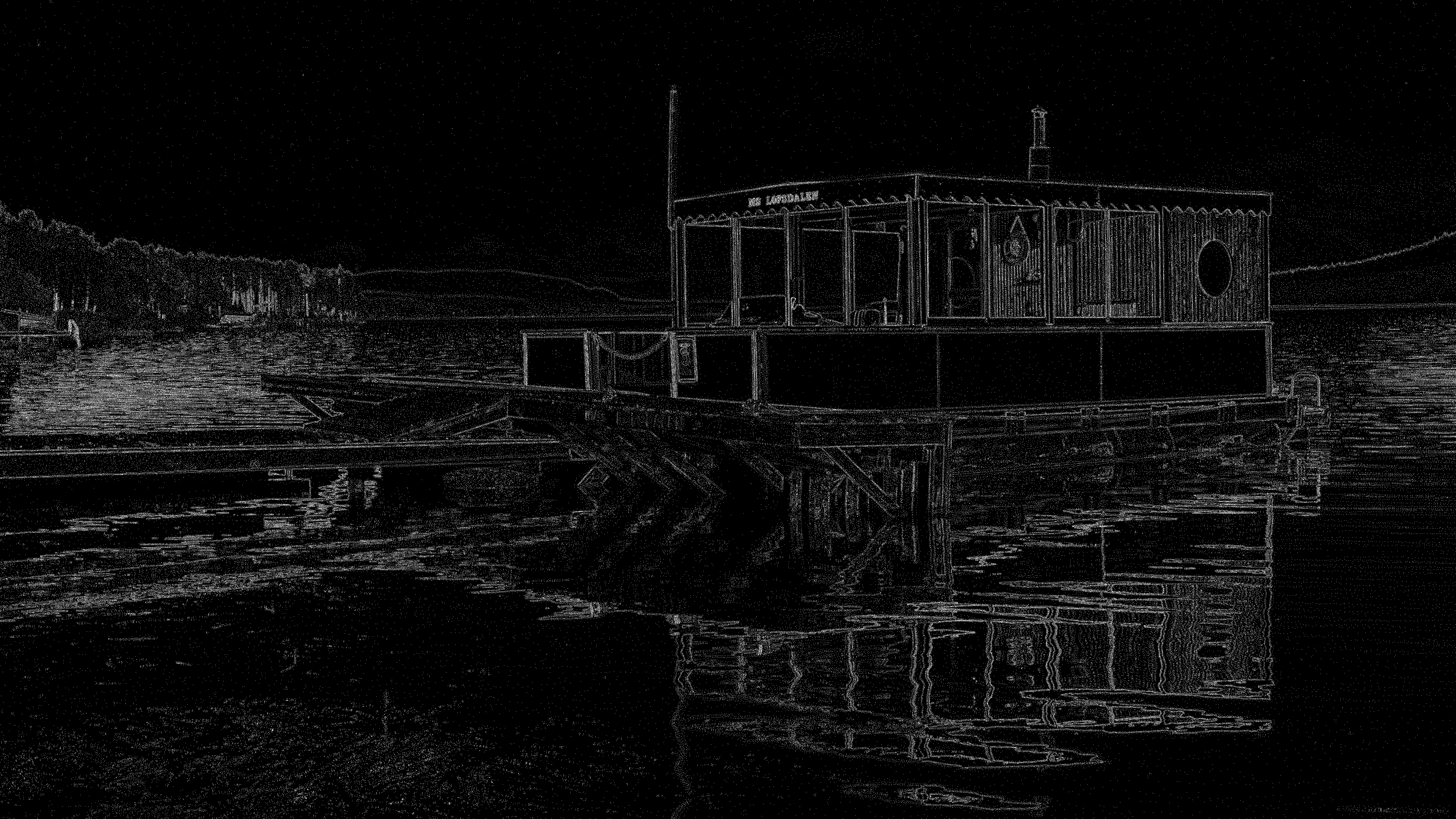}} &
		\centered{\includegraphics[width=0.3\textwidth]
	{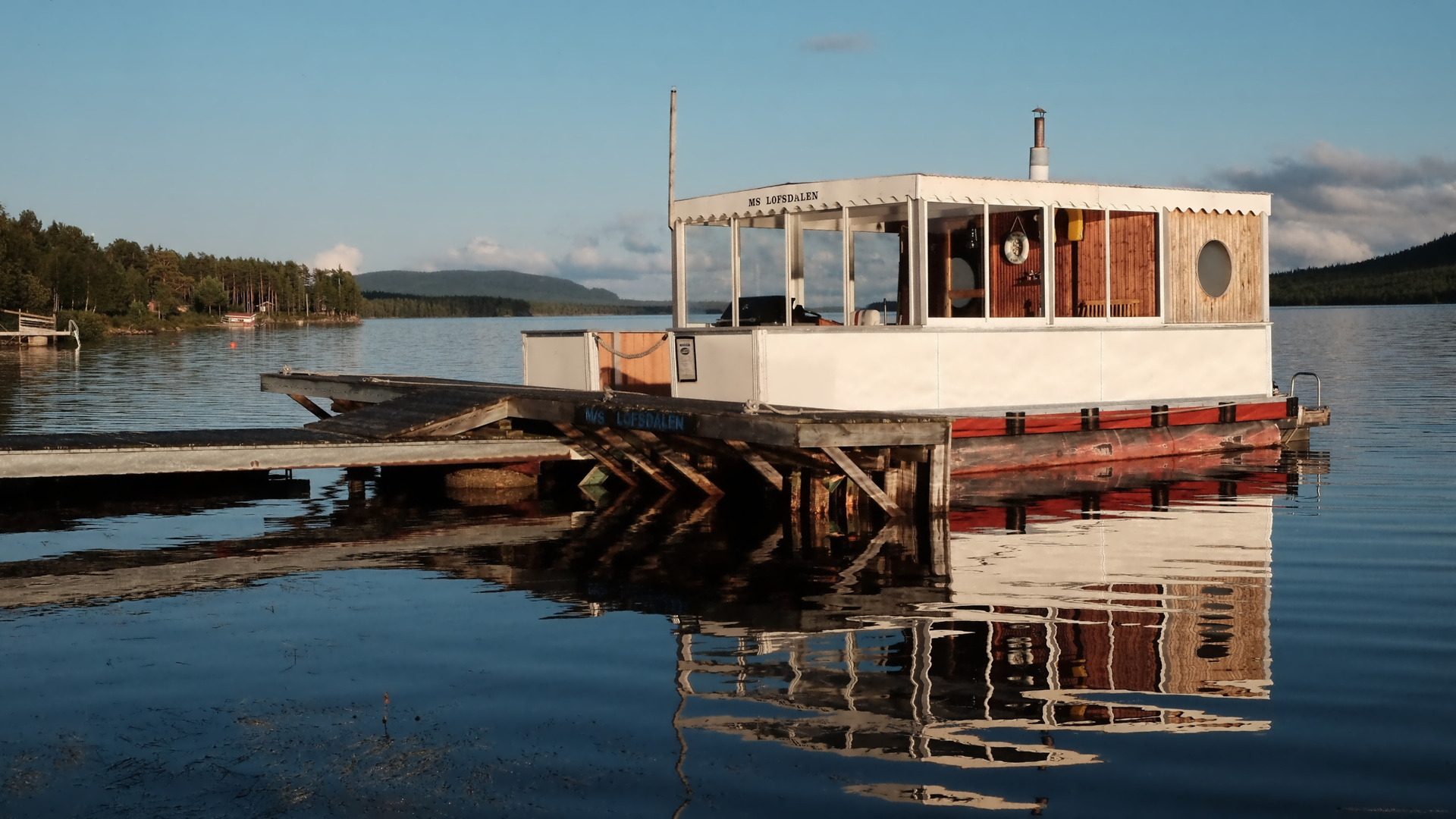}} & 
		\centered{\rotatebox{90}{\small PSNR: 34.31}}\vspace{0.2mm}\\ 
		
		\centered{\rotatebox{90}{\small \textit{shed}}} &
		\centered{\includegraphics[width=0.3\textwidth]
		{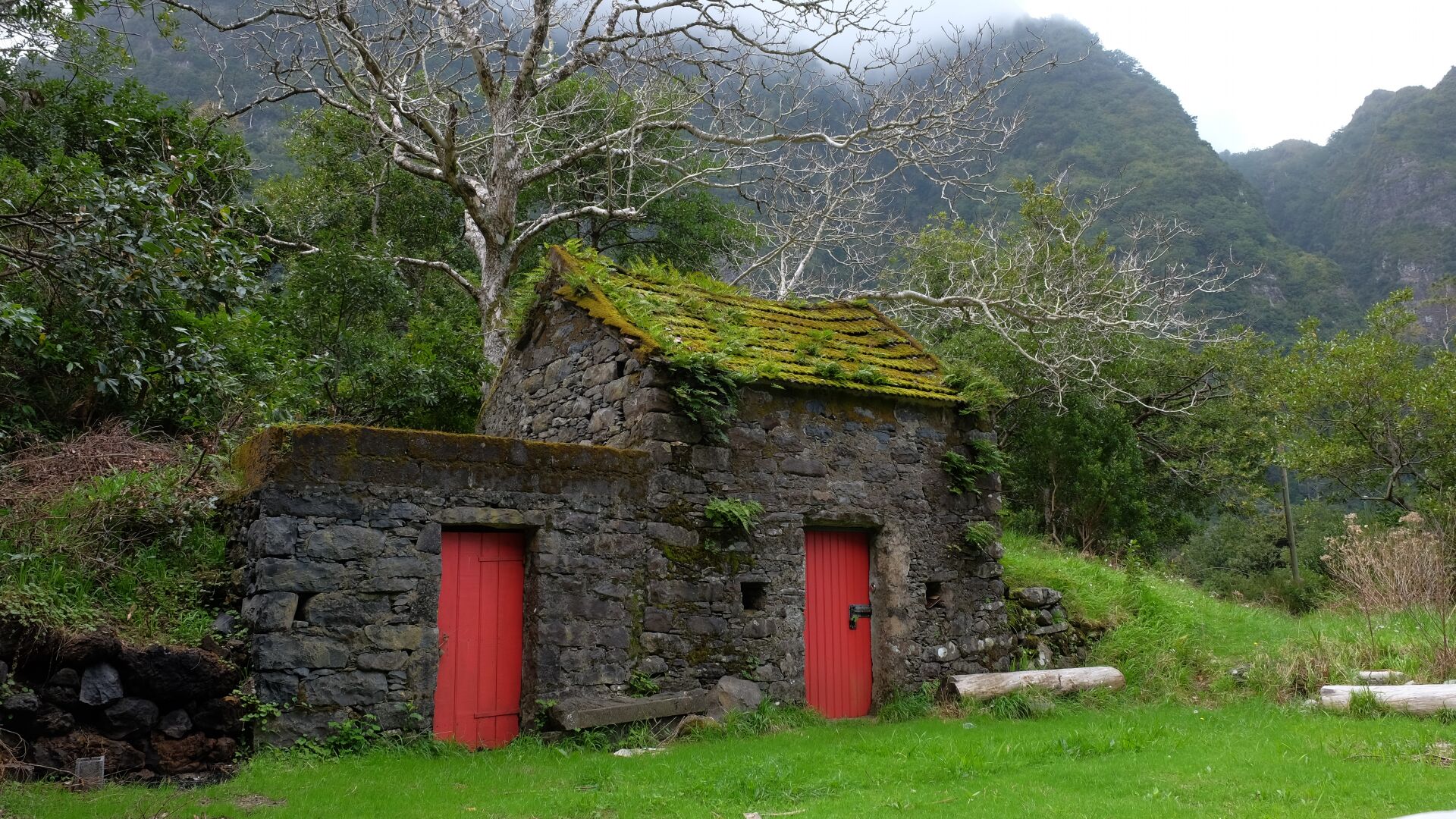}} &
		\centered{\includegraphics[width=0.3\textwidth]
		{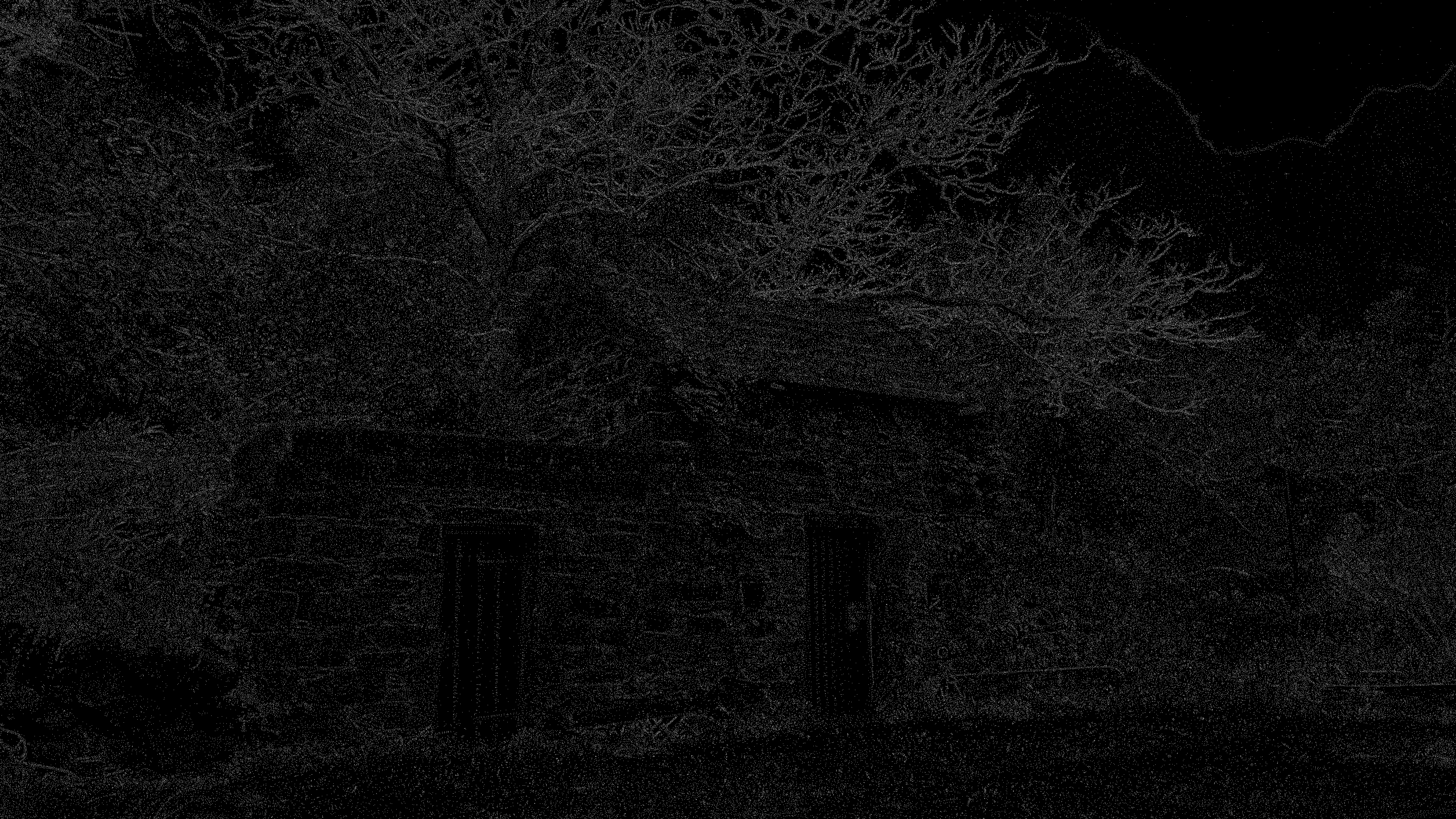}}  & 
		\centered{\includegraphics[width=0.3\textwidth]
		{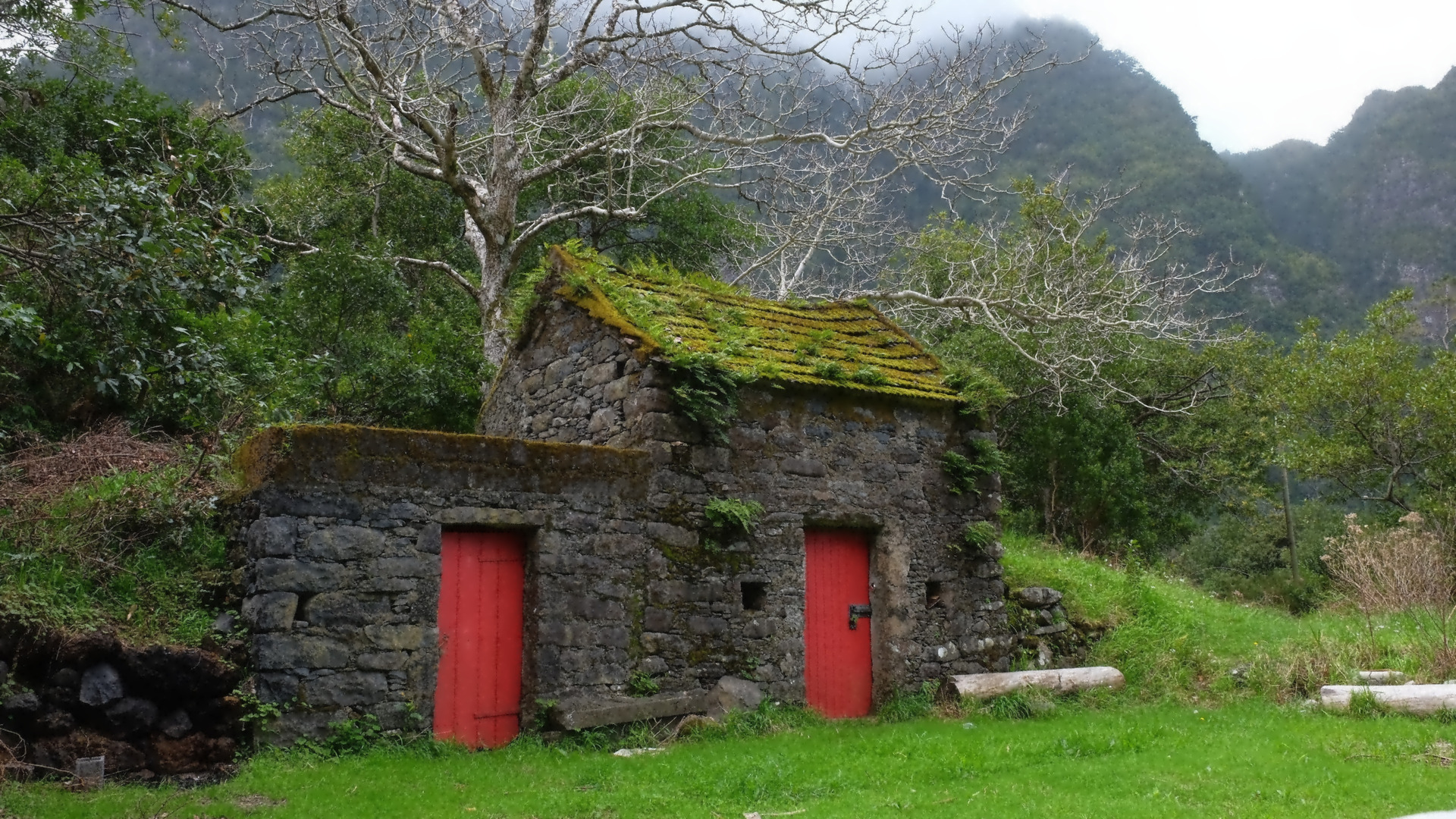}} &
		\centered{\rotatebox{90}{\small PSNR: 25.36}}\vspace{0.2mm}\\ 
		
		\centered{\rotatebox{90}{\small \textit{rogen}}} &
		\centered{\includegraphics[width=0.3\textwidth]
		{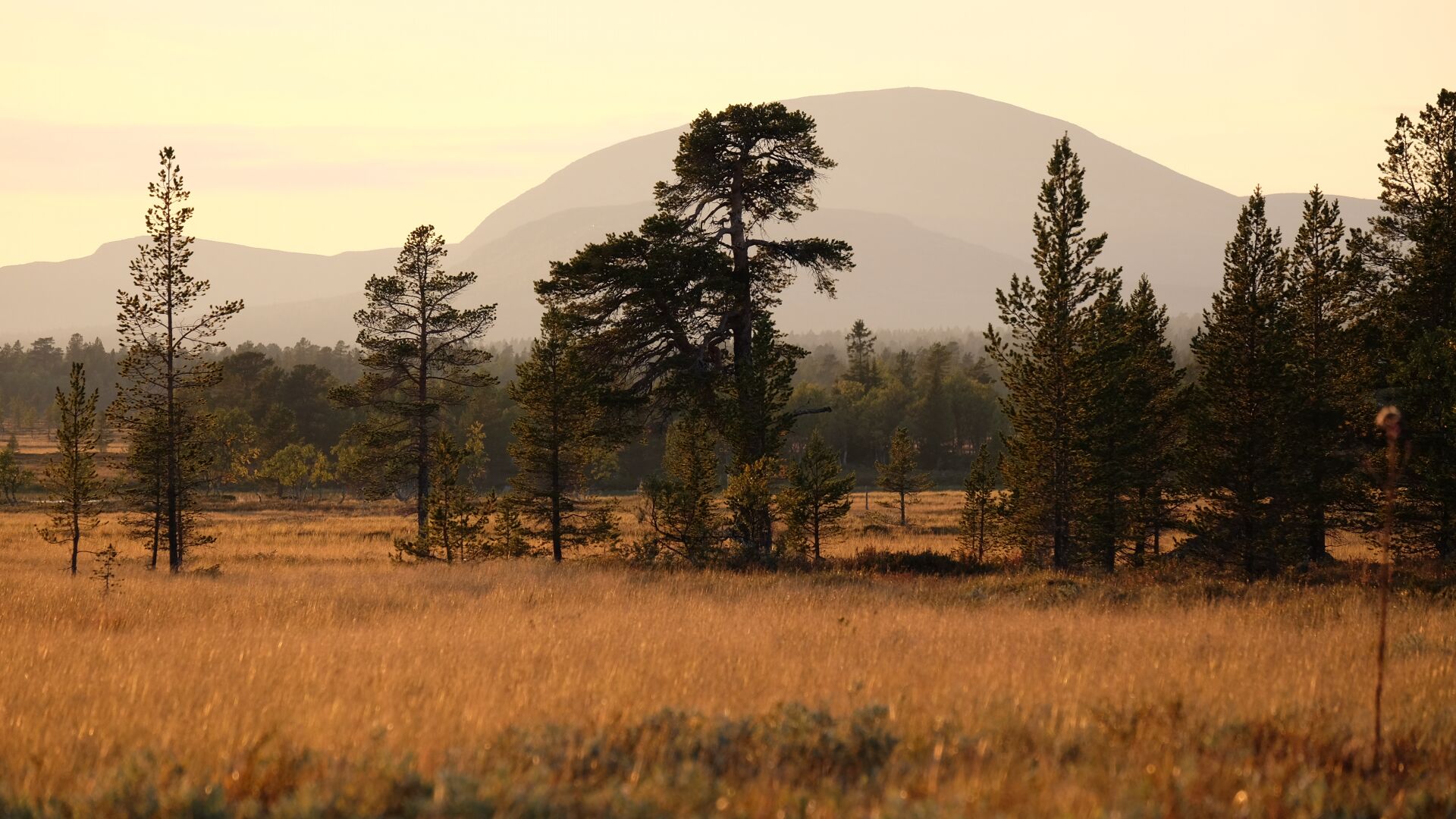}} & 
		\centered{\includegraphics[width=0.3\textwidth]
		{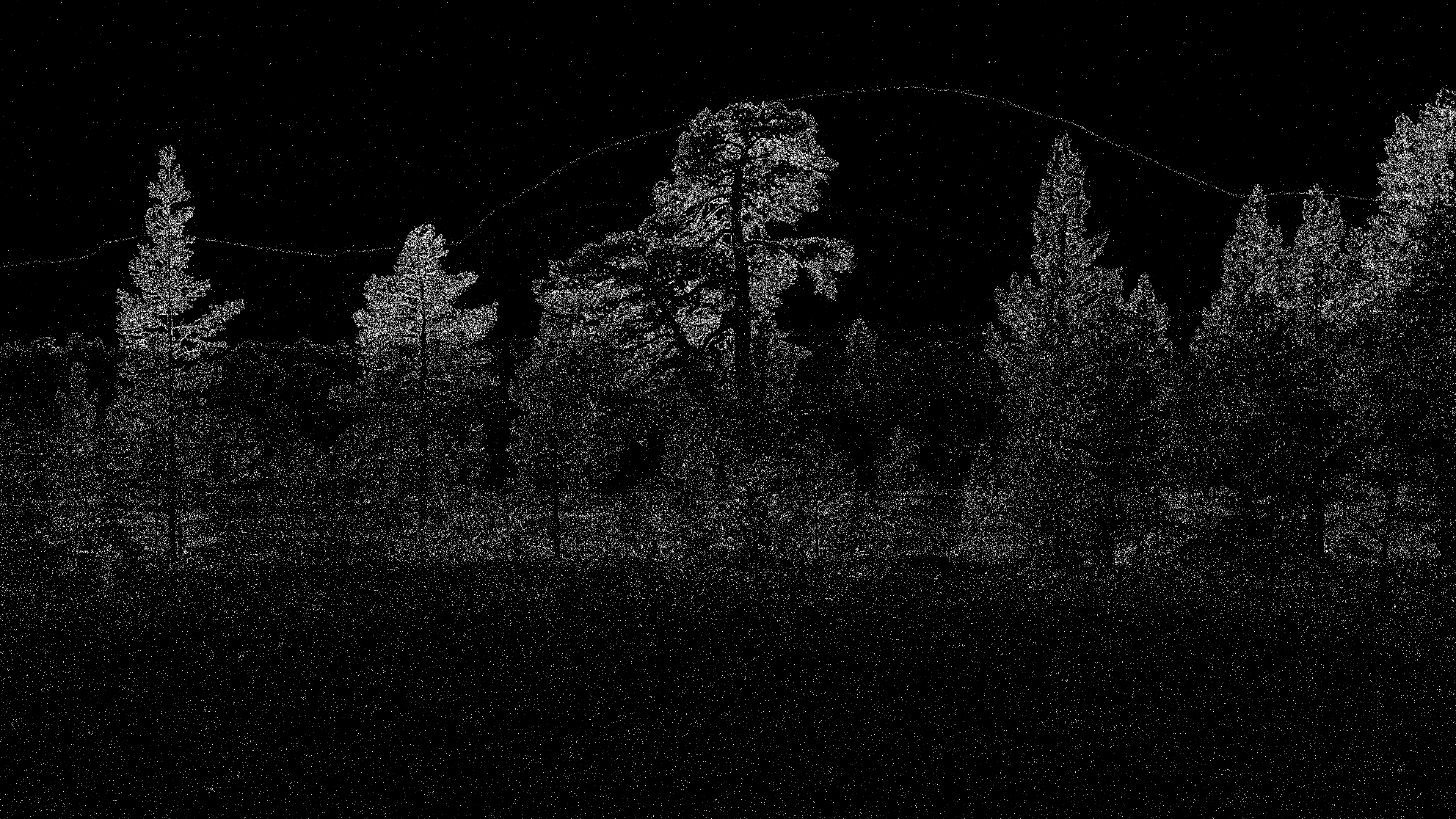}} & 
		\centered{\includegraphics[width=0.3\textwidth]
		{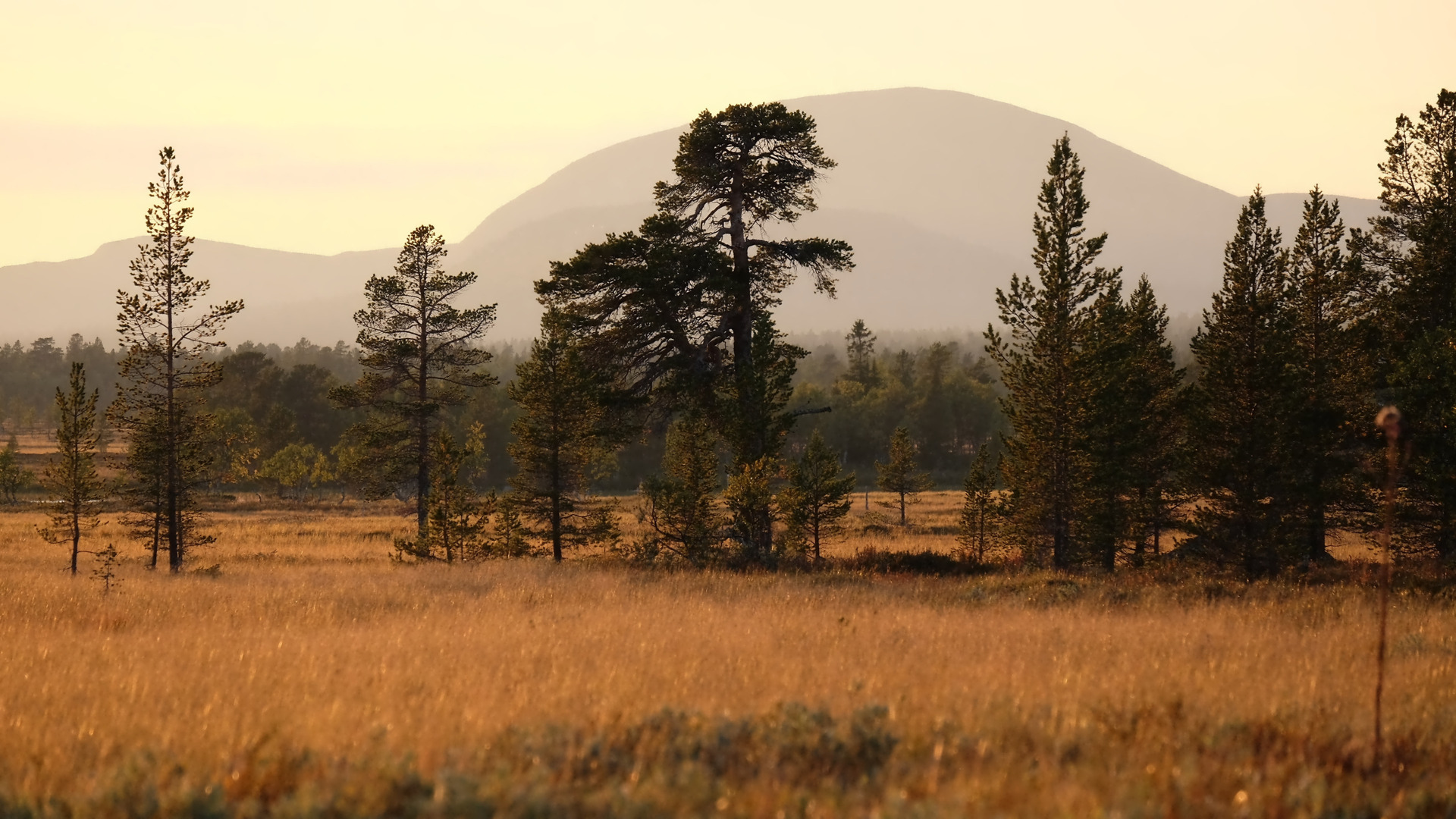}} &
		\centered{\rotatebox{90}{\small PSNR: 35.59}}\vspace{0.2mm}\\  
		
		\centered{\rotatebox{90}{\small \textit{saaremaa}}} &
		\centered{\includegraphics[width=0.3\textwidth]
		{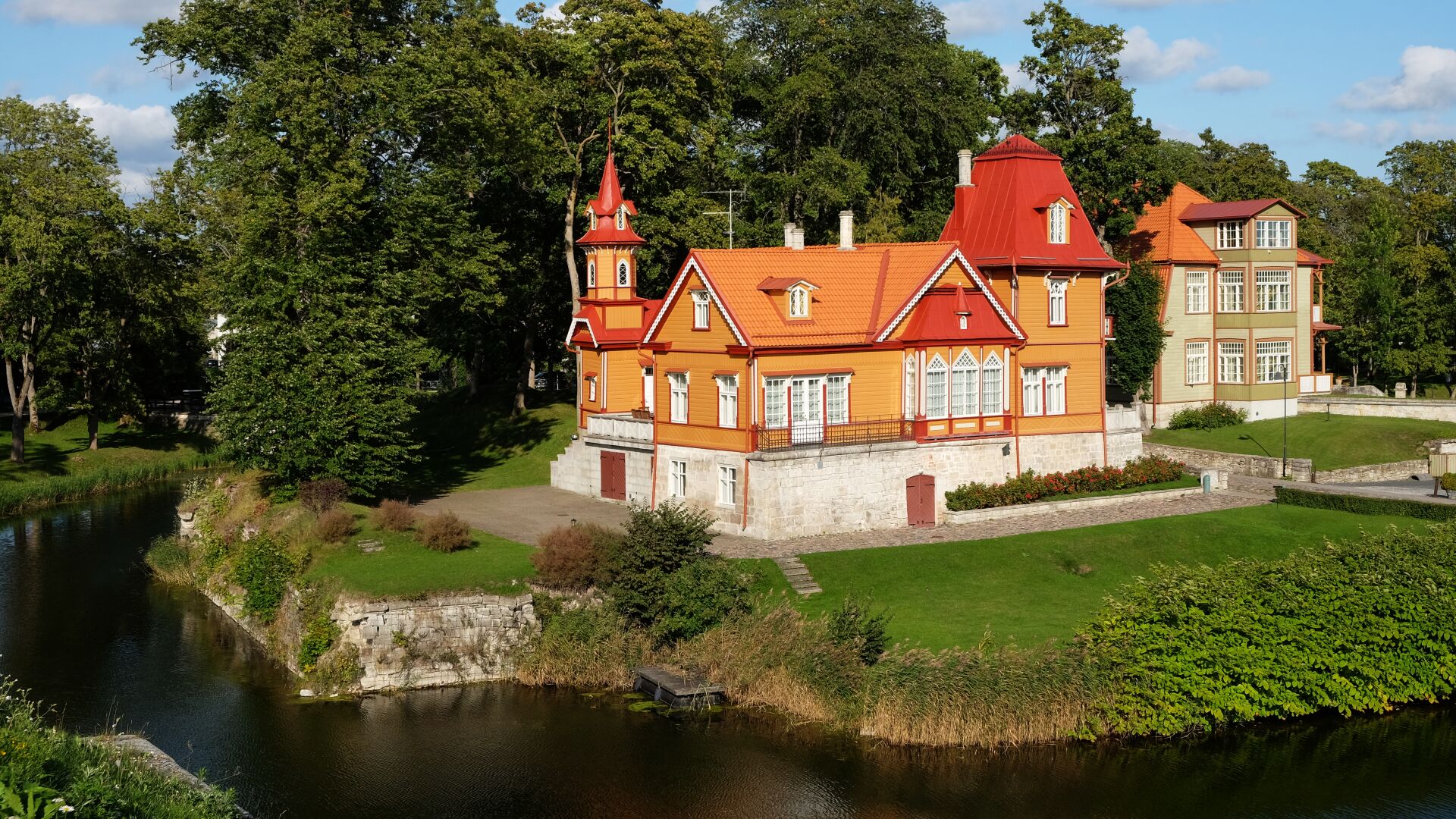}} & 
		\centered{\includegraphics[width=0.3\textwidth]
		{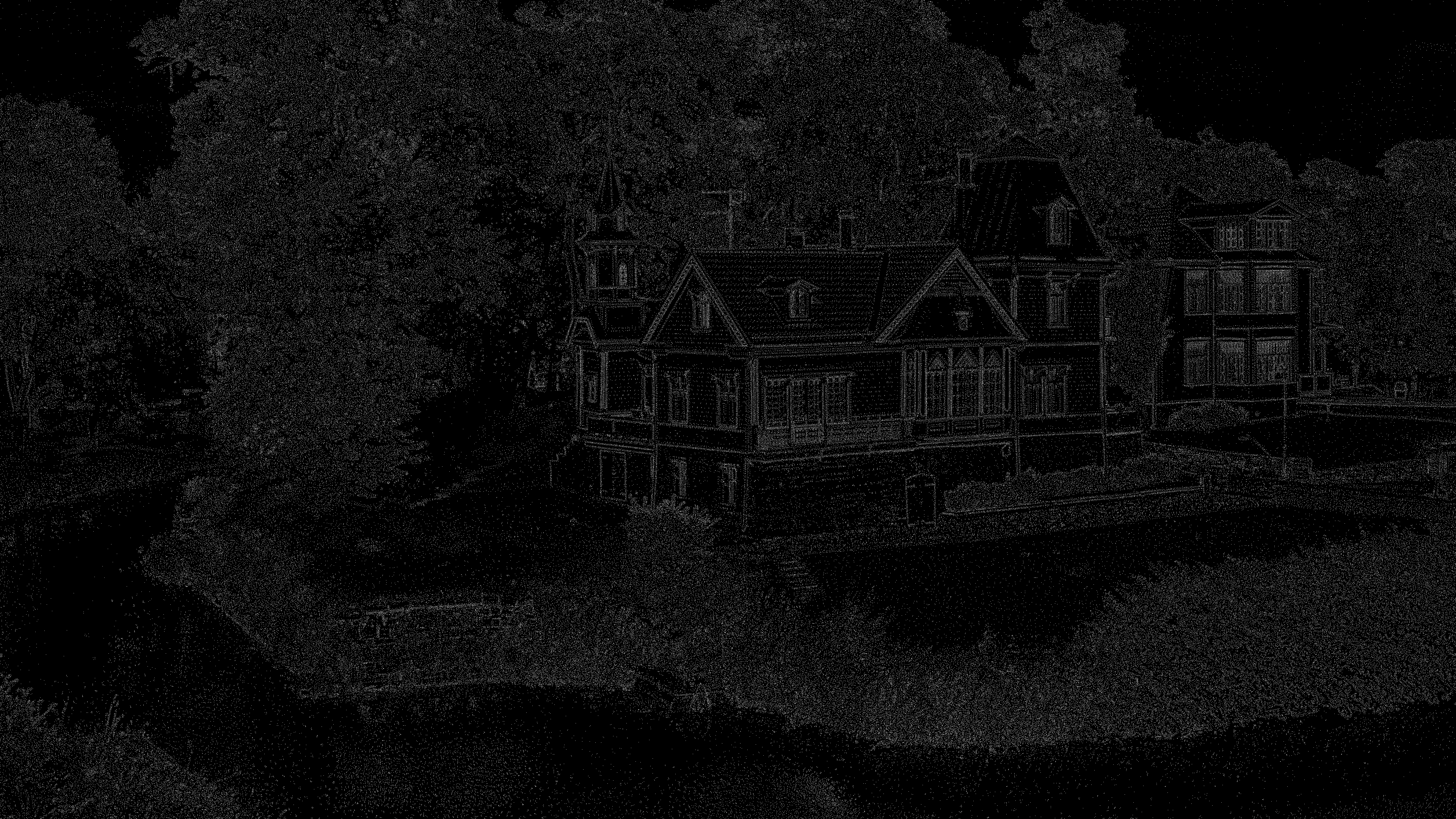}} & 
		\centered{\includegraphics[width=0.3\textwidth]
		{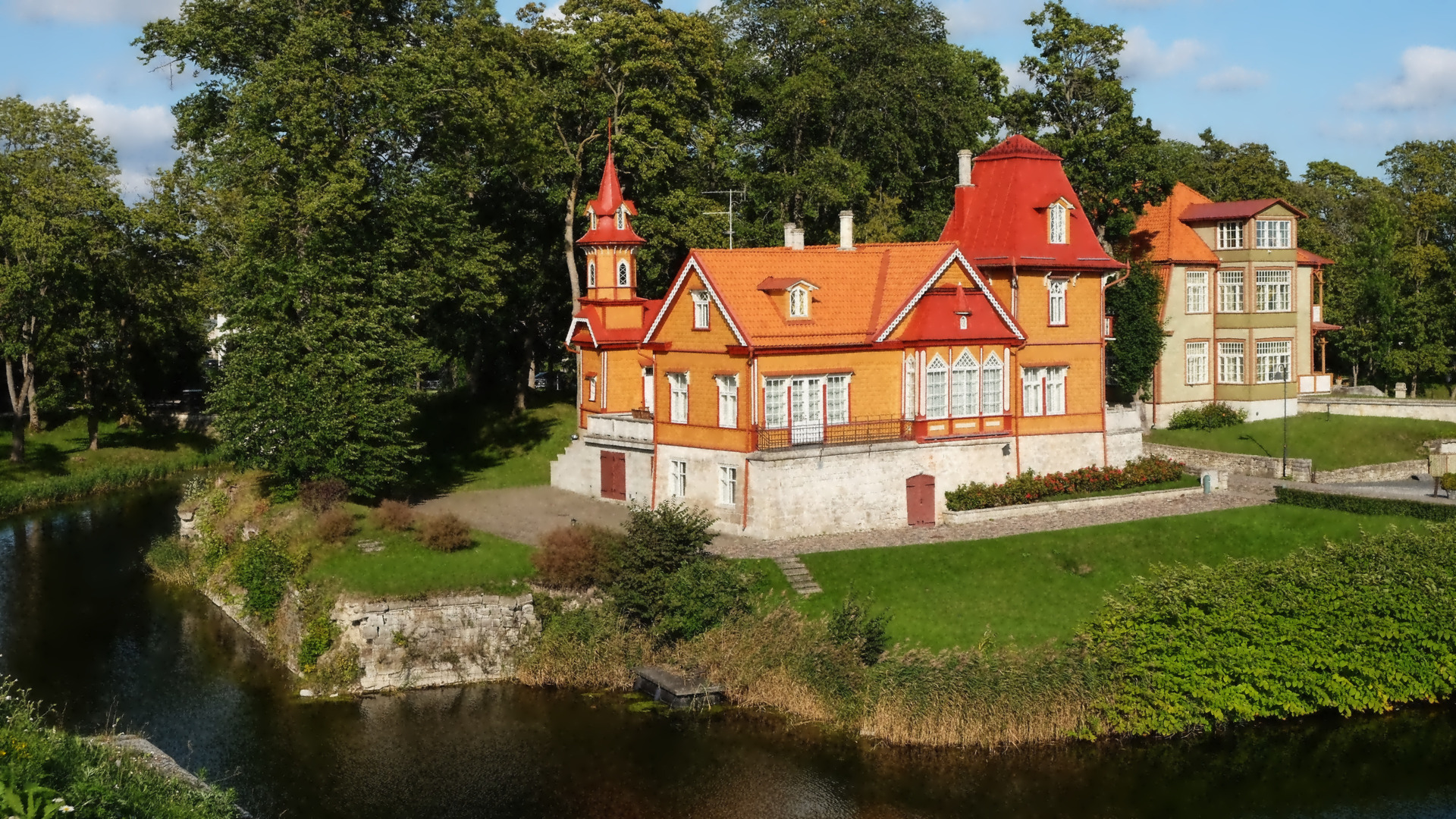}} &
		\centered{\rotatebox{90}{\small PSNR: 25.27}}\vspace{0.2mm}\\ 
		
		\centered{\rotatebox{90}{\small \textit{fuerteventura}}} &
		\centered{\includegraphics[width=0.3\textwidth]
		{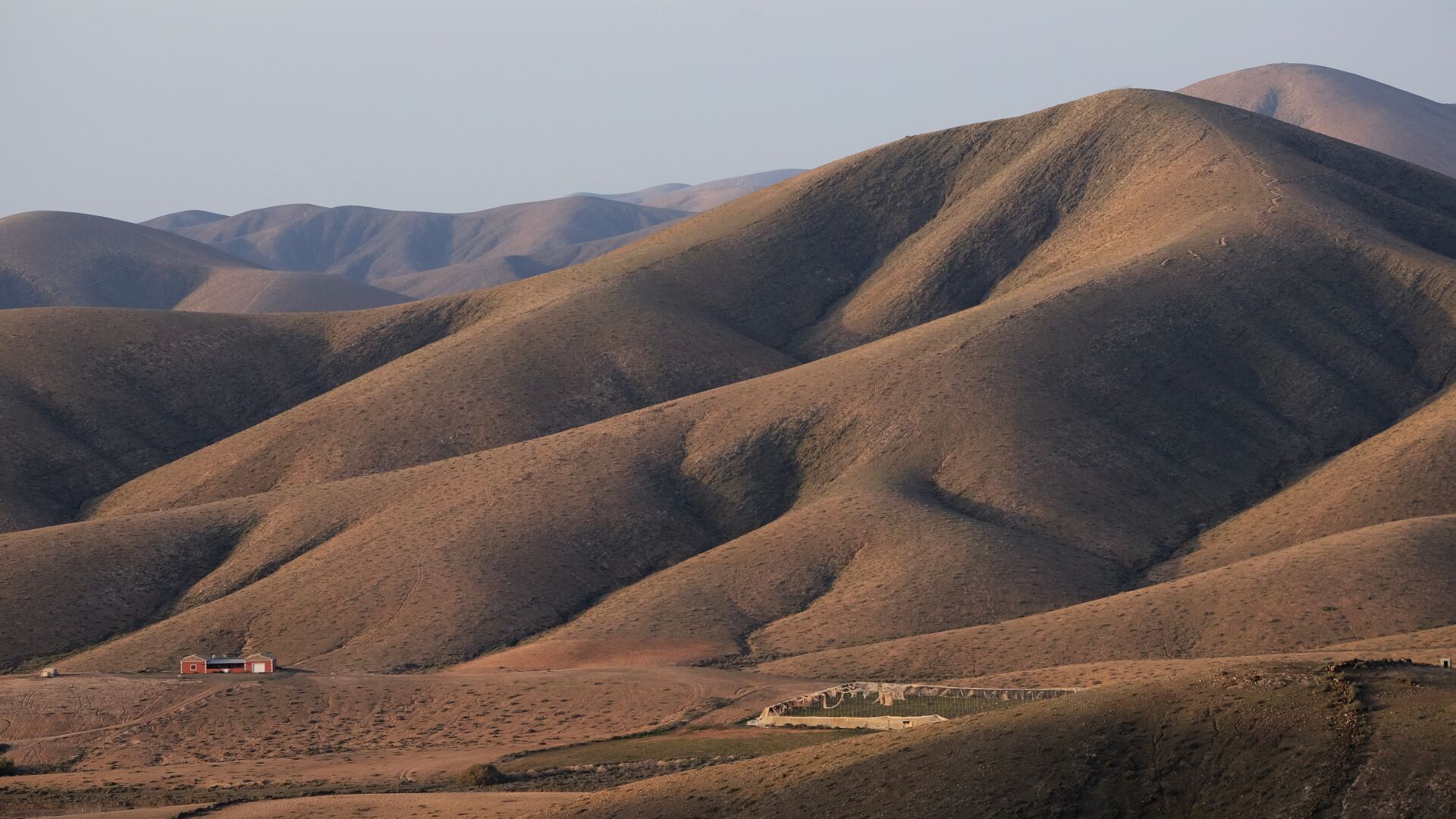}} & 
		\centered{\includegraphics[width=0.3\textwidth]
		{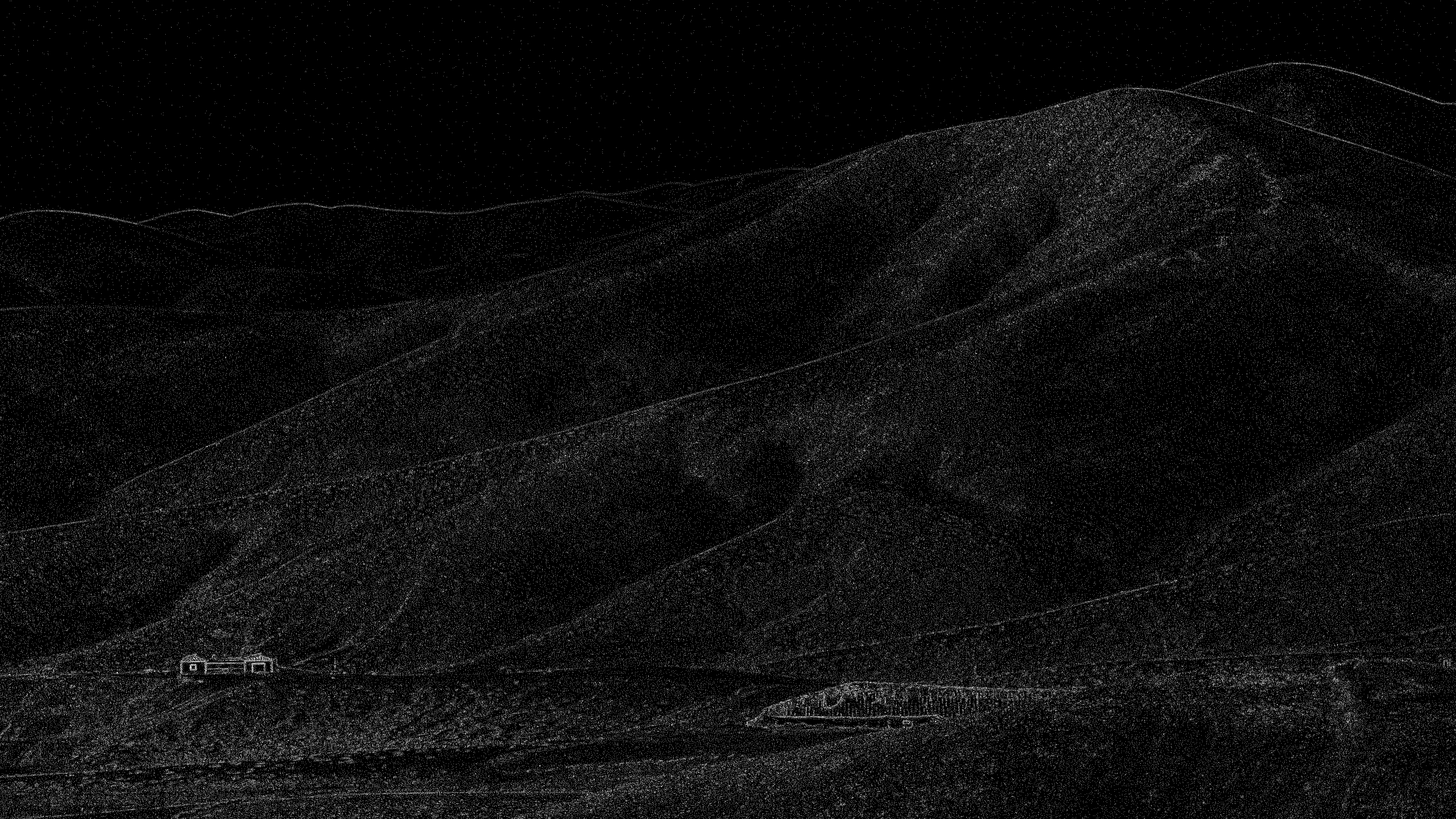}} & 
		\centered{\includegraphics[width=0.3\textwidth]
	{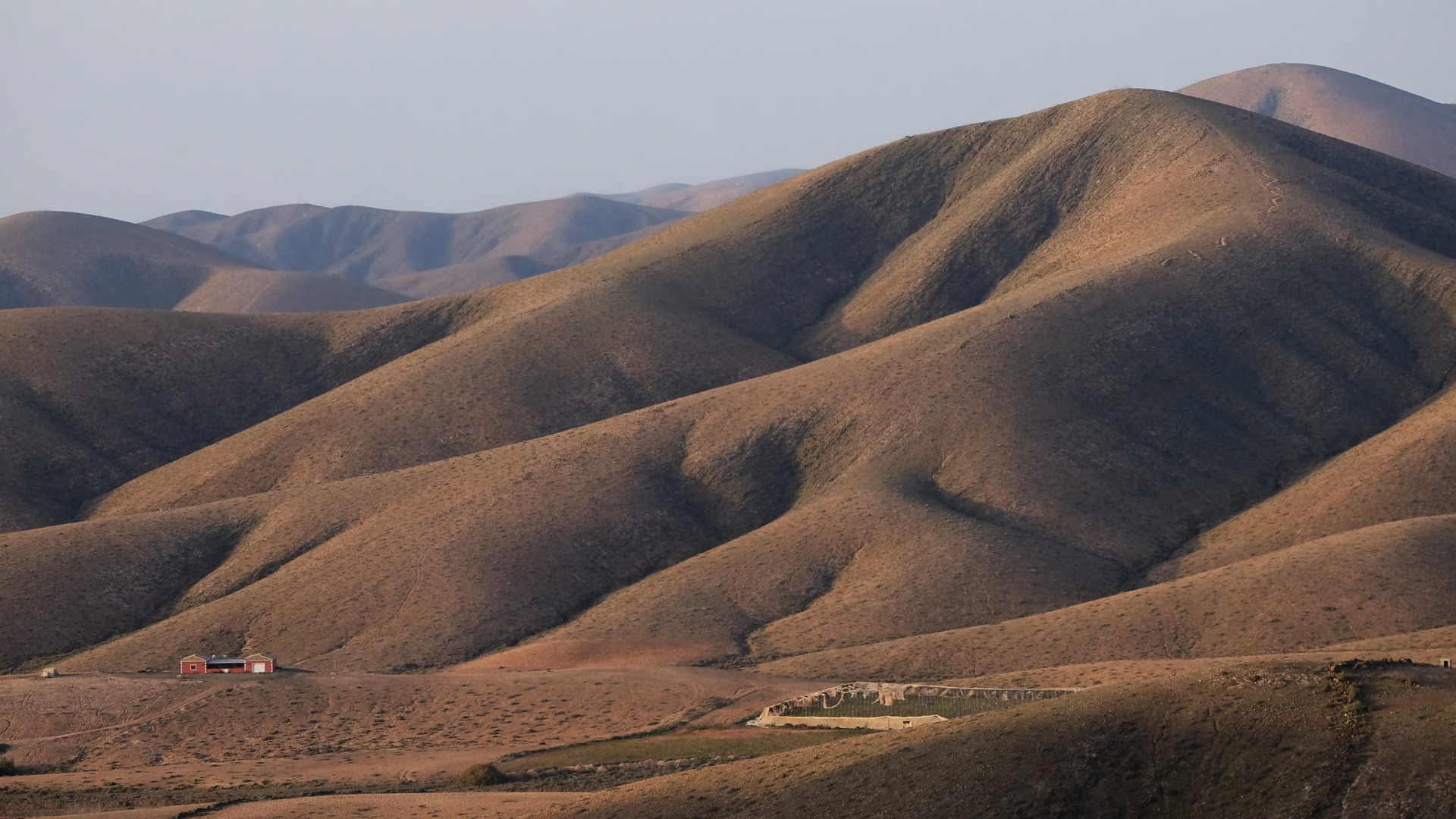}} &
		\centered{\rotatebox{90}{\small PSNR: 35.45}}\vspace{0.2mm}\\ 
		
		\centered{\rotatebox{90}{\small \textit{elpaso}}} &
		\centered{\includegraphics[width=0.3\textwidth]
		{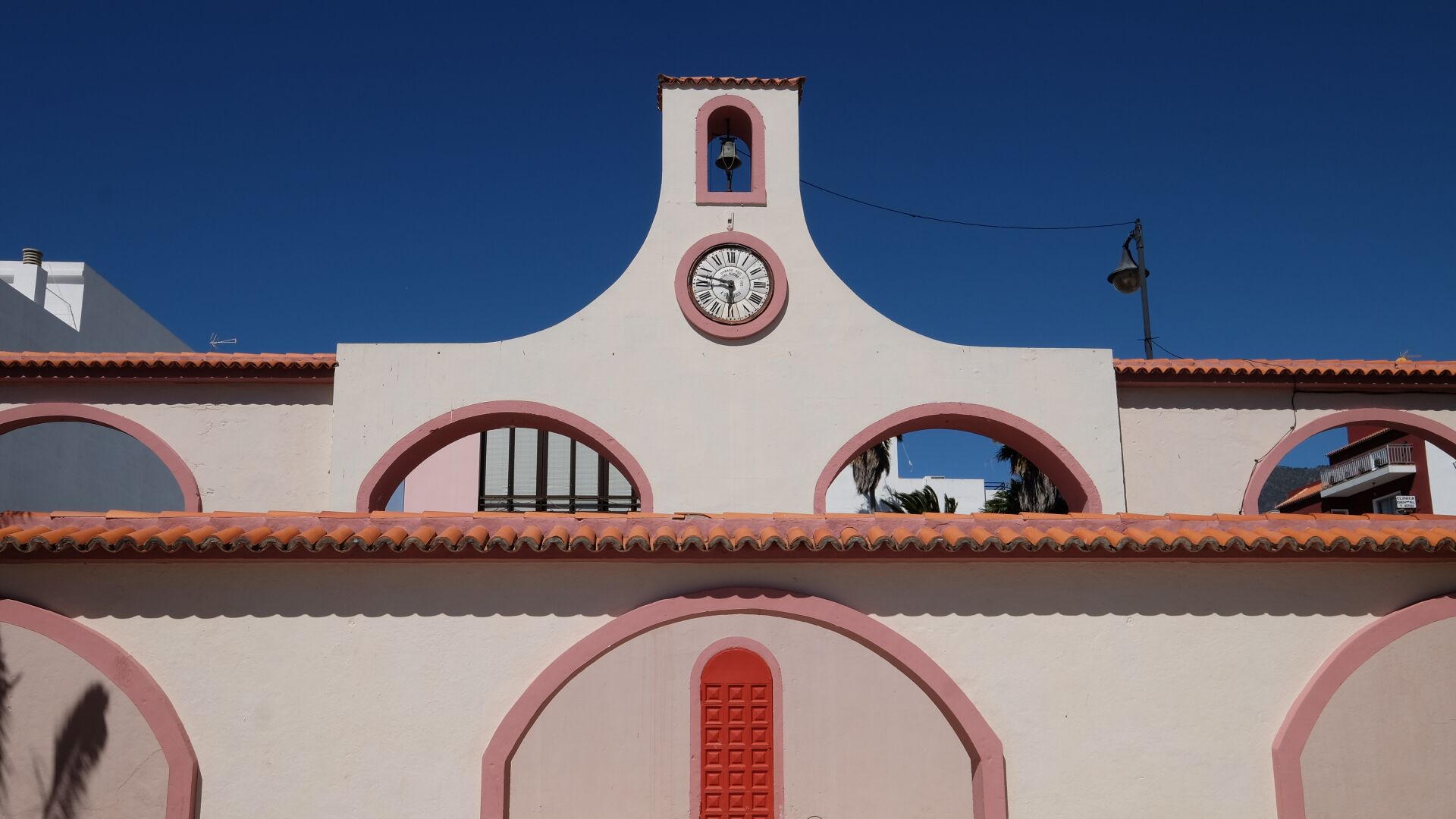}} & 
		\centered{\includegraphics[width=0.3\textwidth]
		{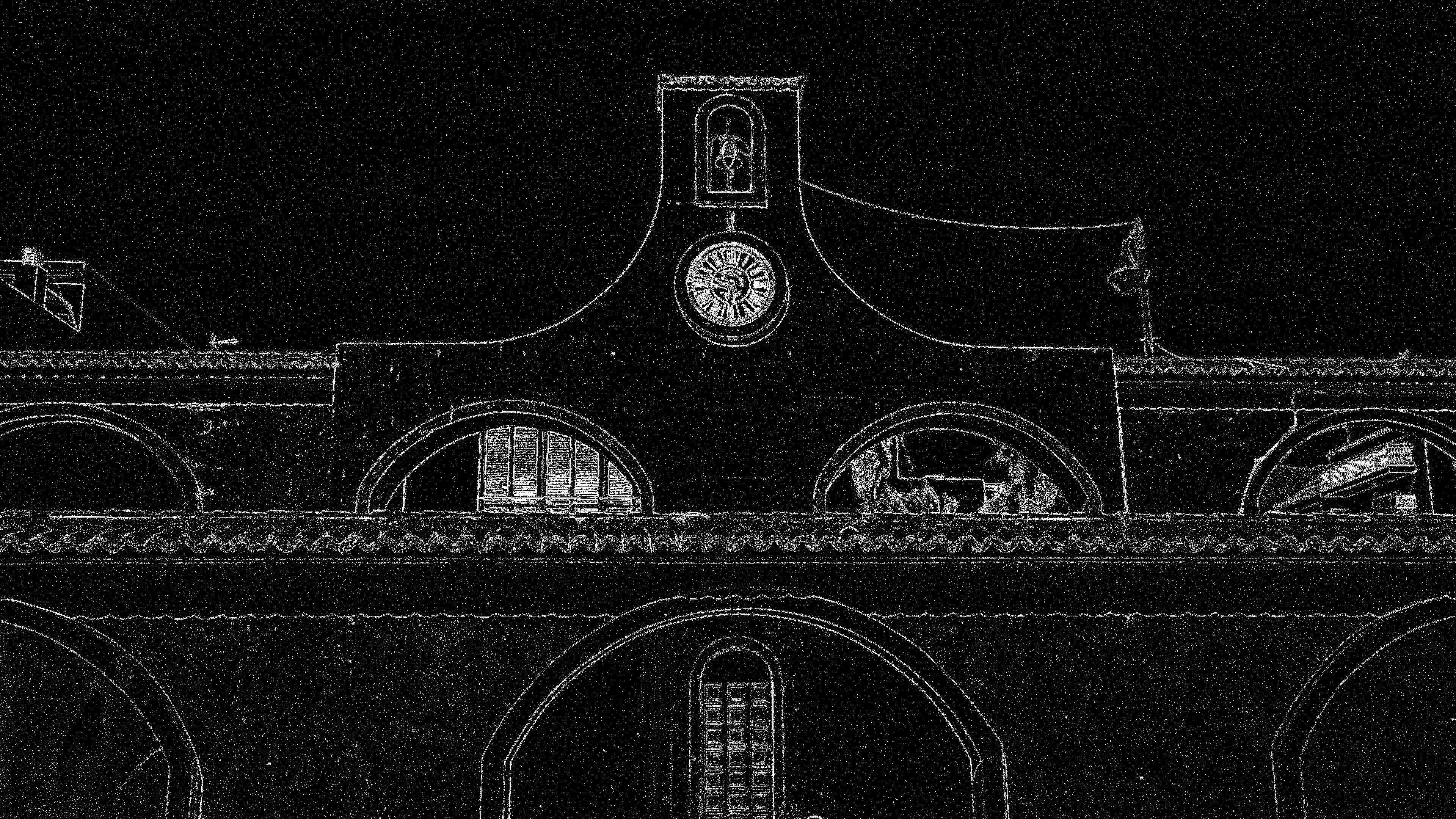}} & 
		\centered{\includegraphics[width=0.3\textwidth]
		{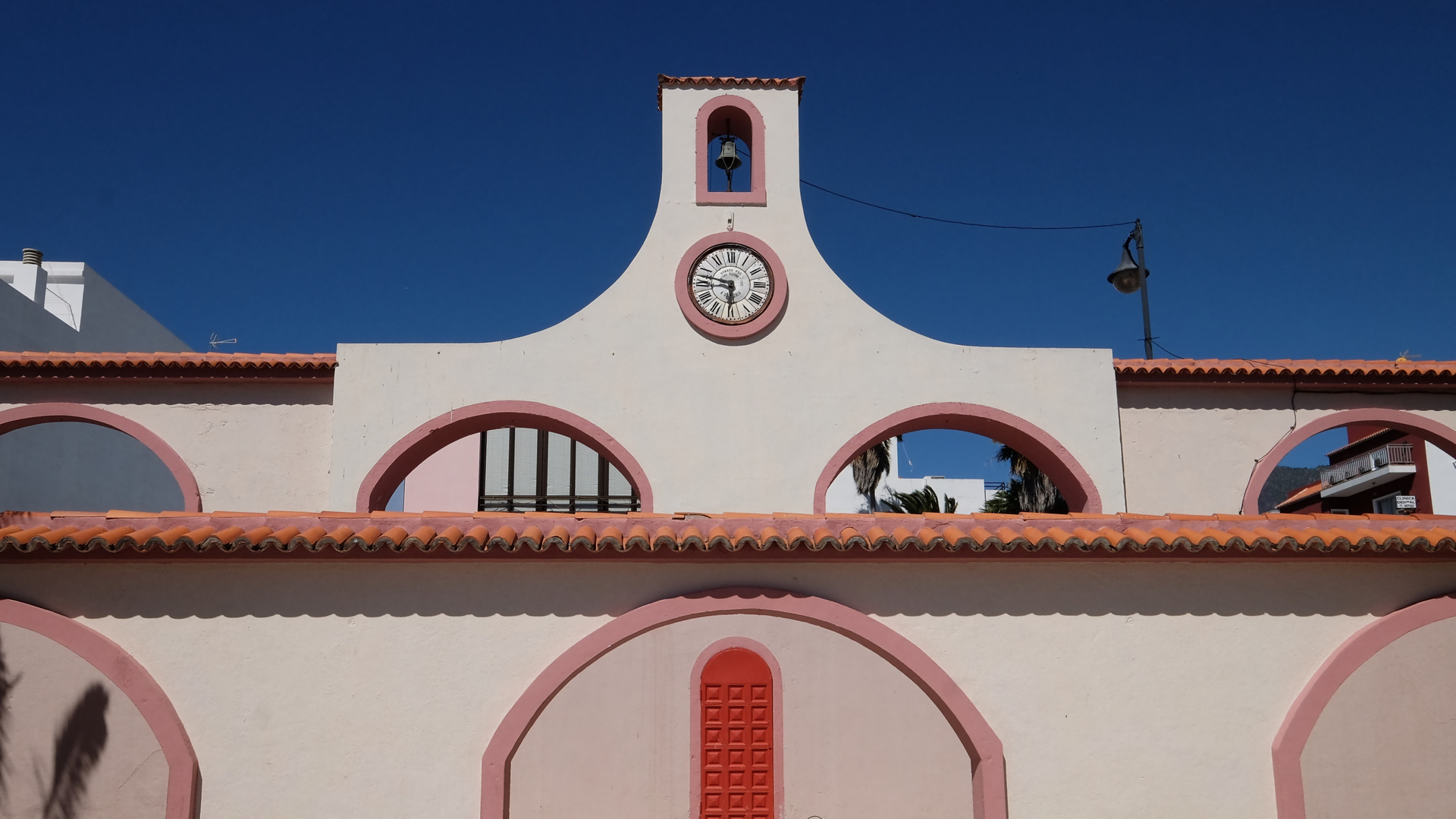}} &
		\centered{\rotatebox{90}{\small PSNR: 44.22}}\vspace{0.2mm}\\ 
				
	\end{tabular}

	\caption{\textbf{Sparse Inpainting with 5 \% Known Data}. Images 1 to 6 of our test dataset of size $3840 \times 2160$ with an 5\% optimized 
	inpainting mask and the corresponding inpainting. Photos by J. Weickert.}
	\label{fig:ds-inpainting-1}
\end{figure}

\begin{figure}[p]
	\setlength{\tabcolsep}{1mm}
	\begin{tabular}{ccccc}

		& \small original image & \small inpainting 
		mask & \small inpainted 
		image 
		& \\
		
		\centered{\rotatebox{90}{\small \textit{sunset}}} &
		\centered{\includegraphics[width=0.3\textwidth]
		{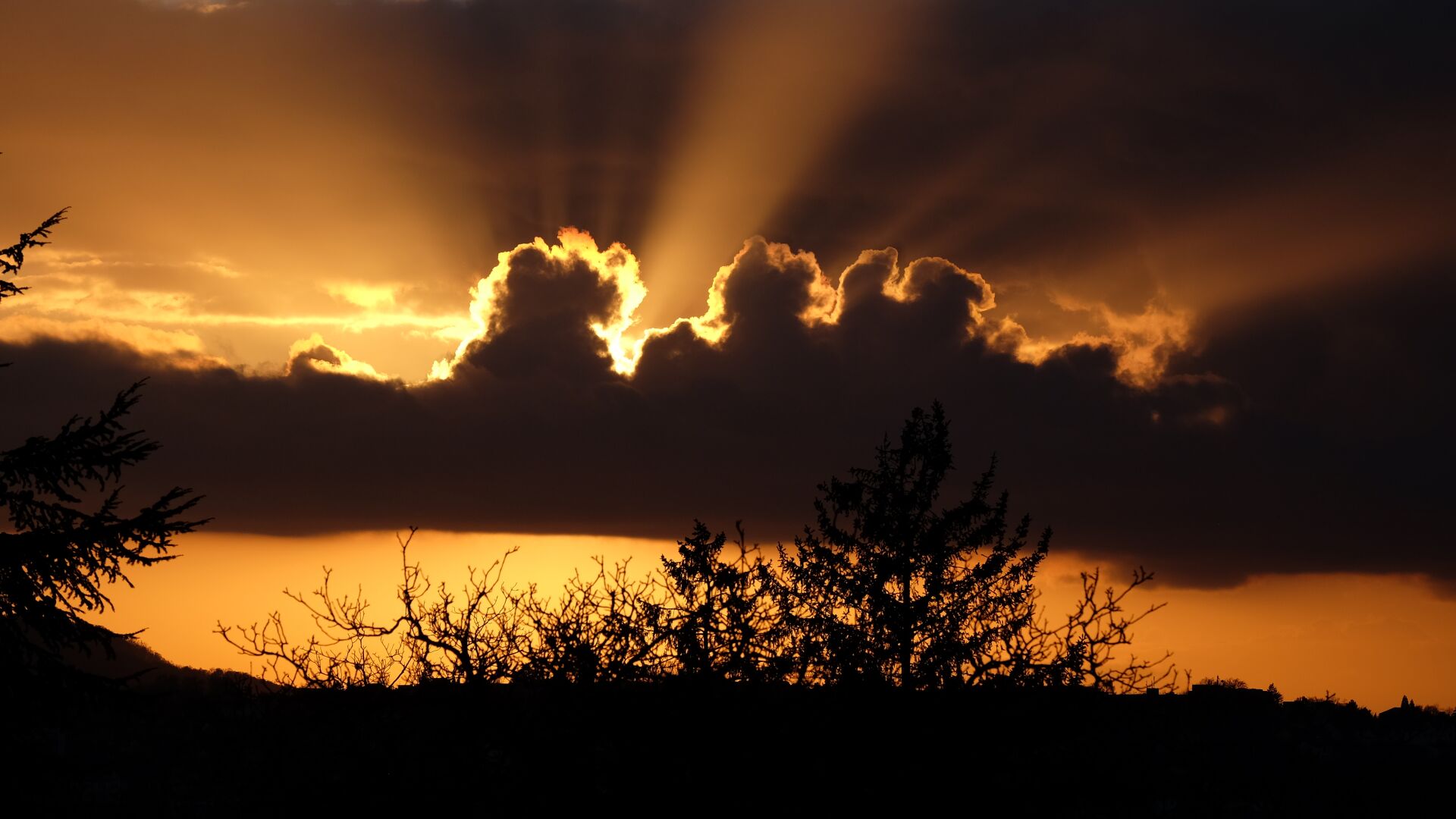}} & 
		\centered{\includegraphics[width=0.3\textwidth]
		{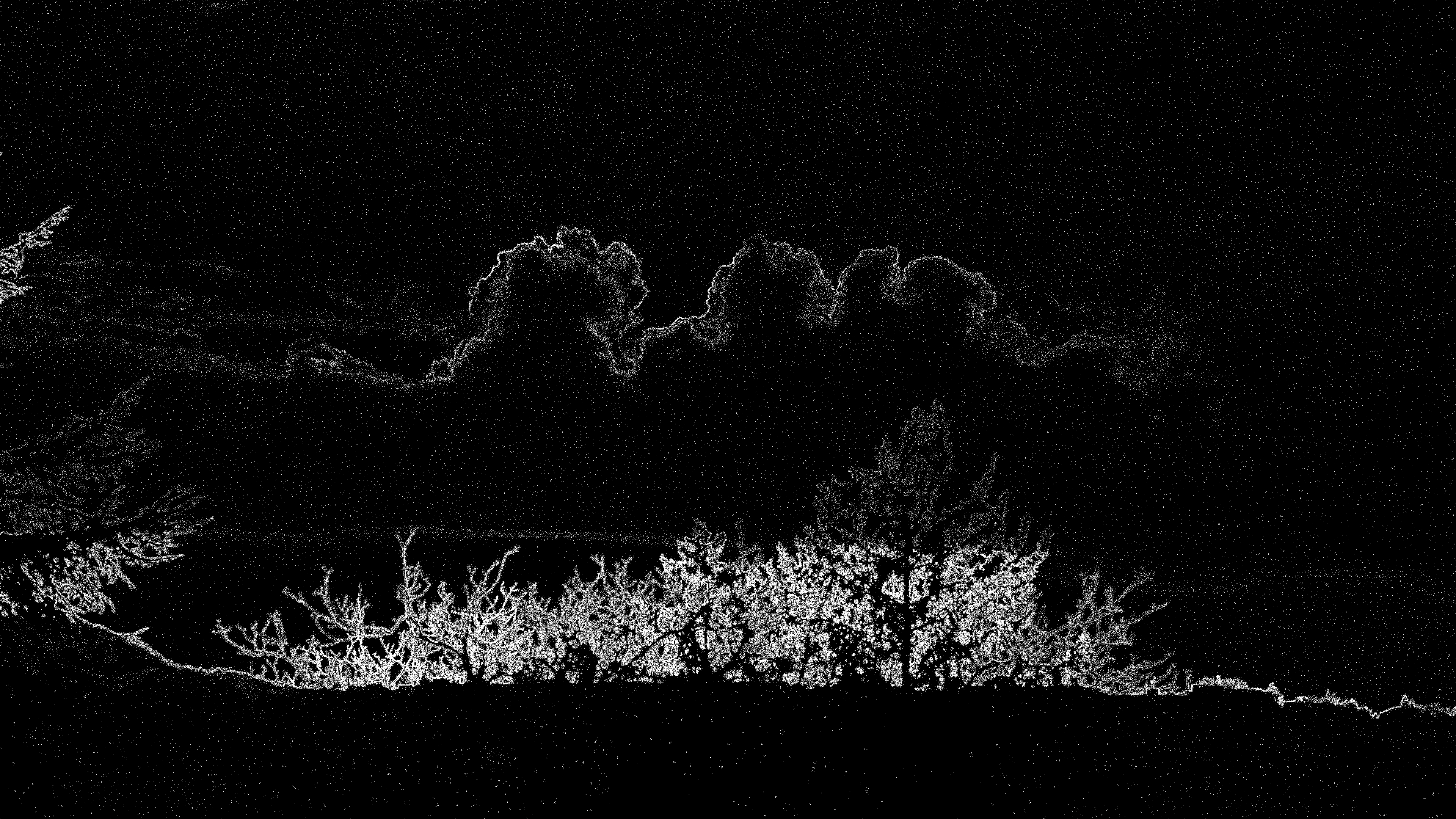}} &
		\centered{\includegraphics[width=0.3\textwidth]
		{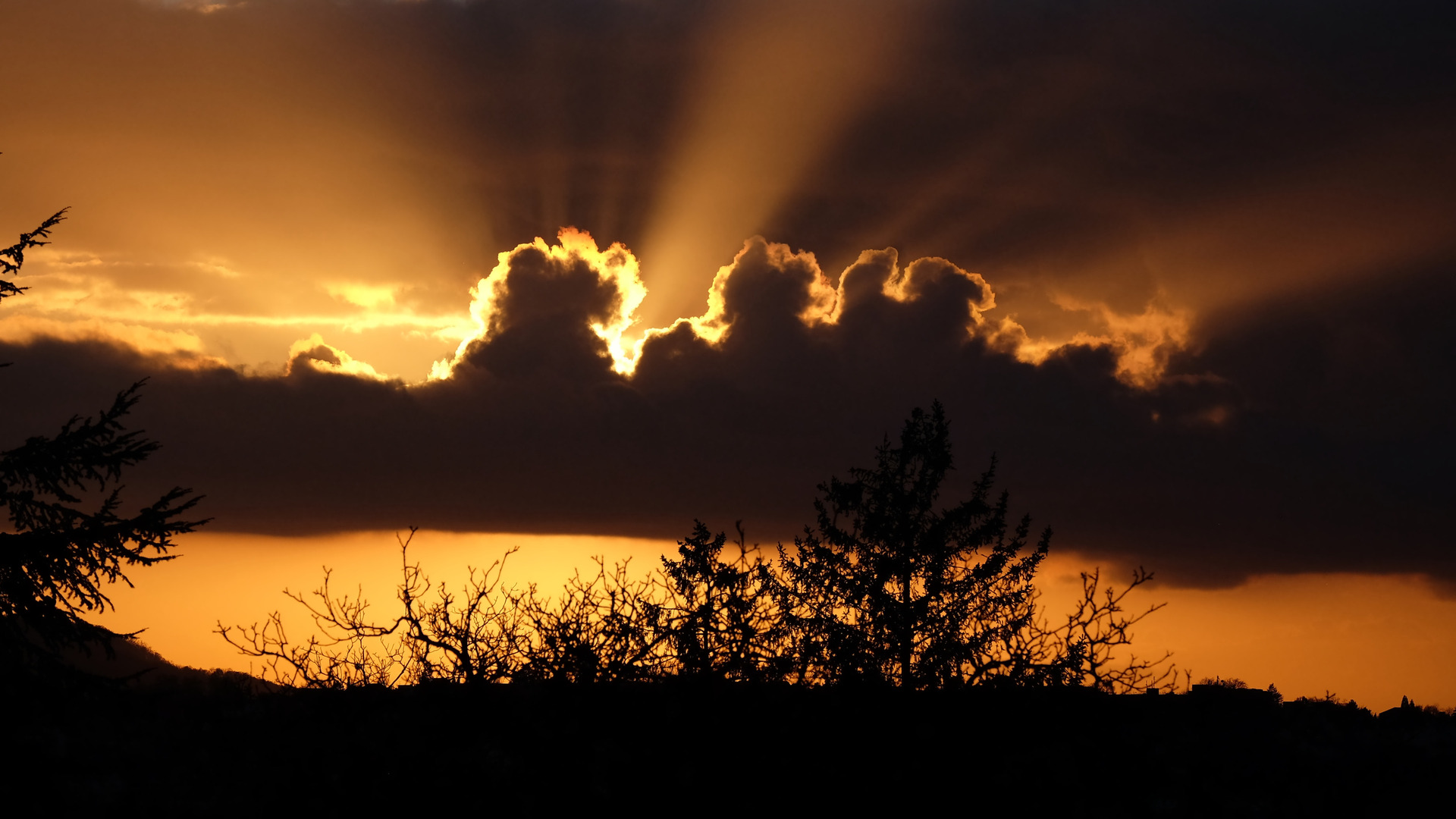}} & 
		\centered{\rotatebox{90}{\small PSNR: 45.62}}\vspace{0.2mm}\\  
		
		\centered{\rotatebox{90}{\small \textit{boats}}} &
		\centered{\includegraphics[width=0.3\textwidth]
		{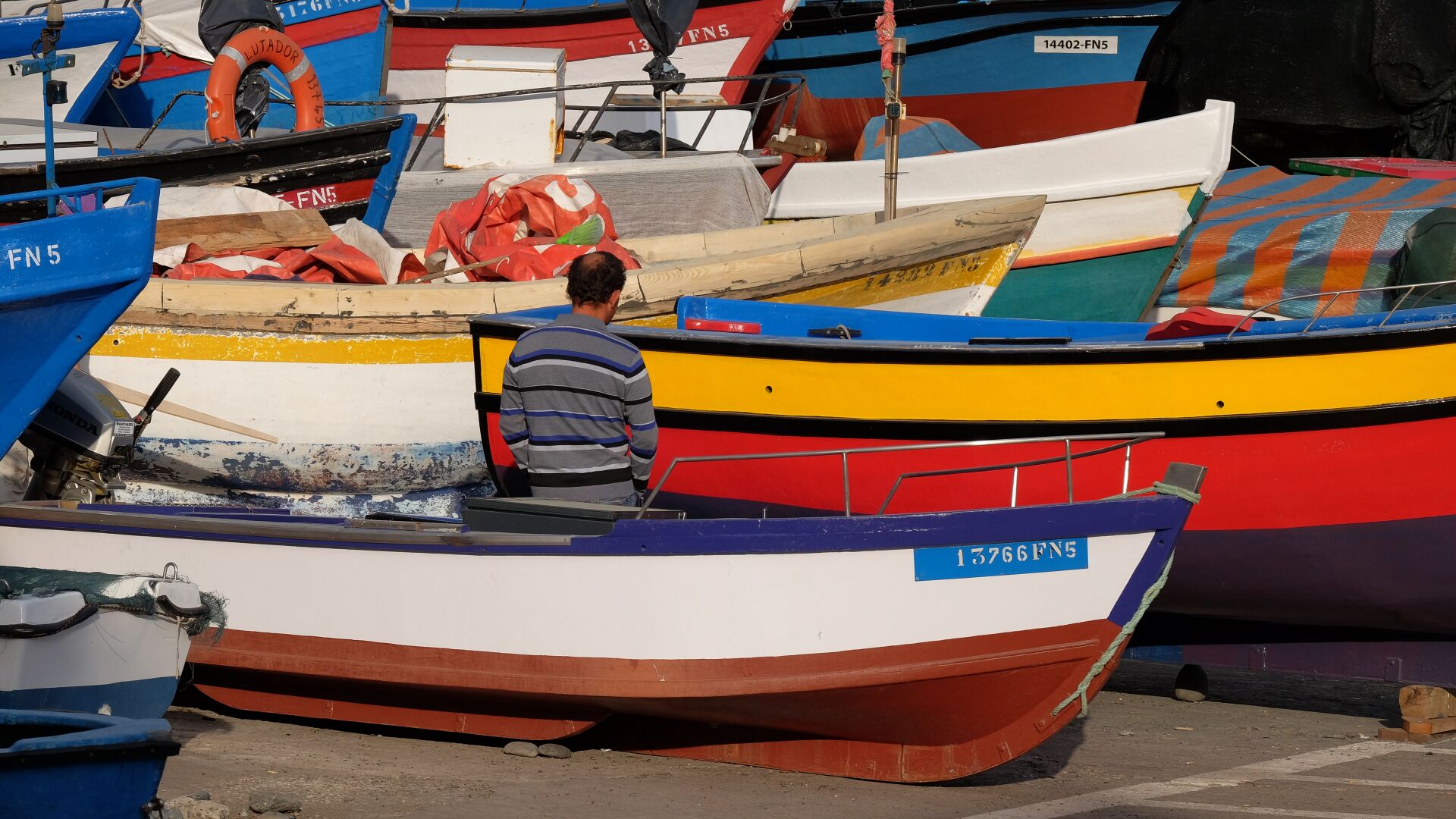}} &
		\centered{\includegraphics[width=0.3\textwidth]
		{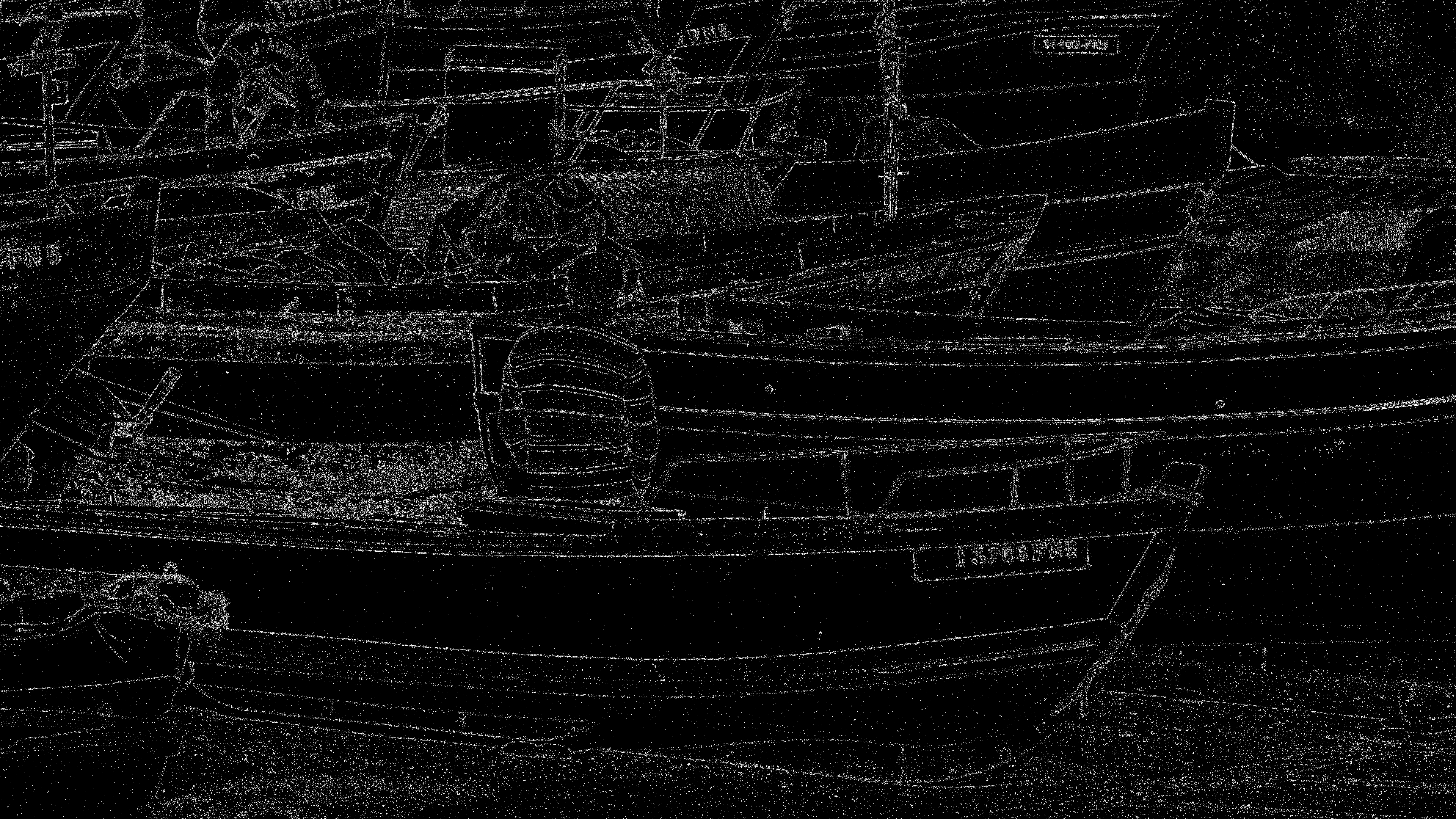}}  & 
		\centered{\includegraphics[width=0.3\textwidth]
		{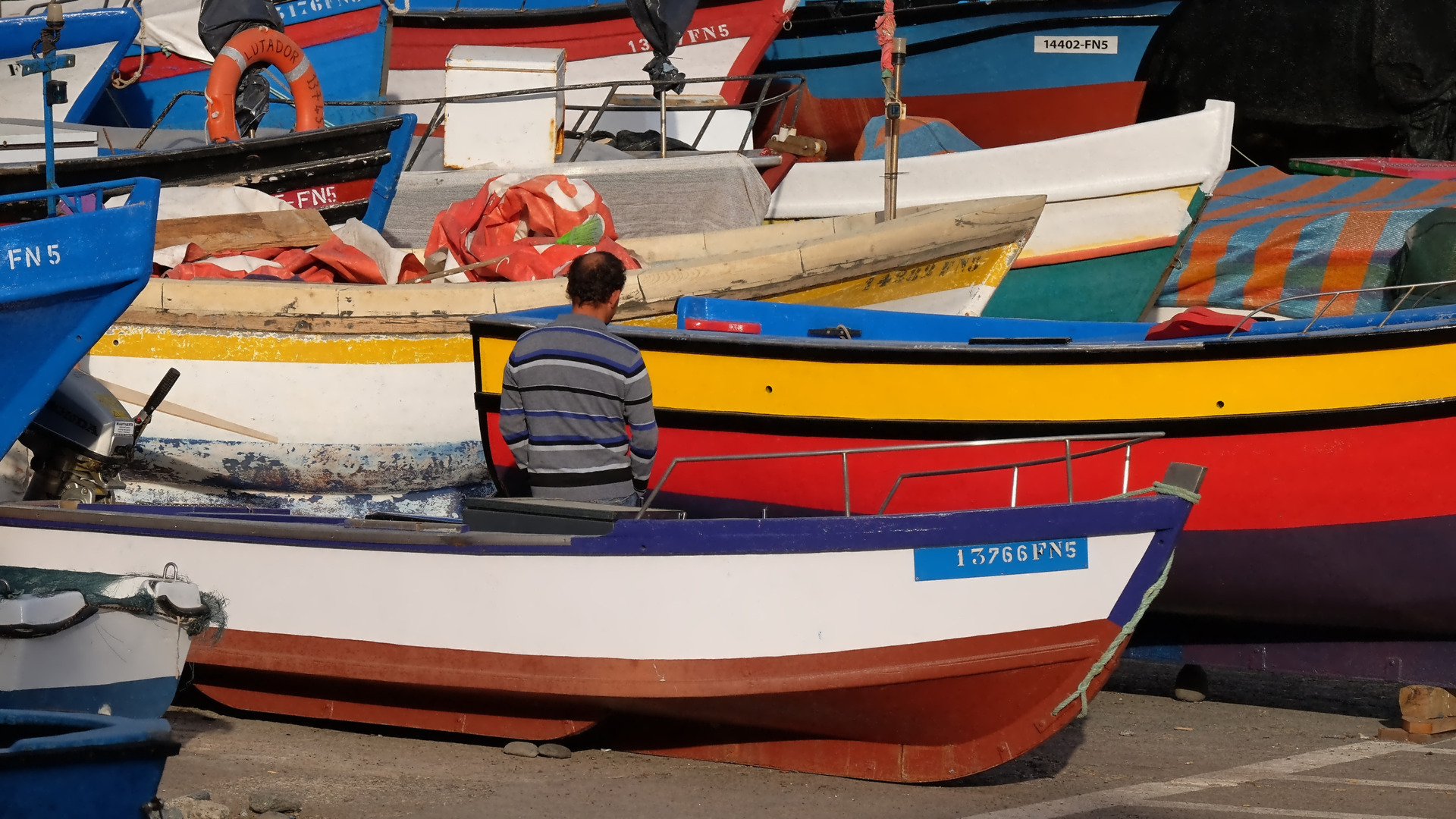}} &
		\centered{\rotatebox{90}{\small PSNR: 33.37}}\vspace{0.2mm}\\  
		
		\centered{\rotatebox{90}{\small \textit{pico}}} &
		\centered{\includegraphics[width=0.3\textwidth]
		{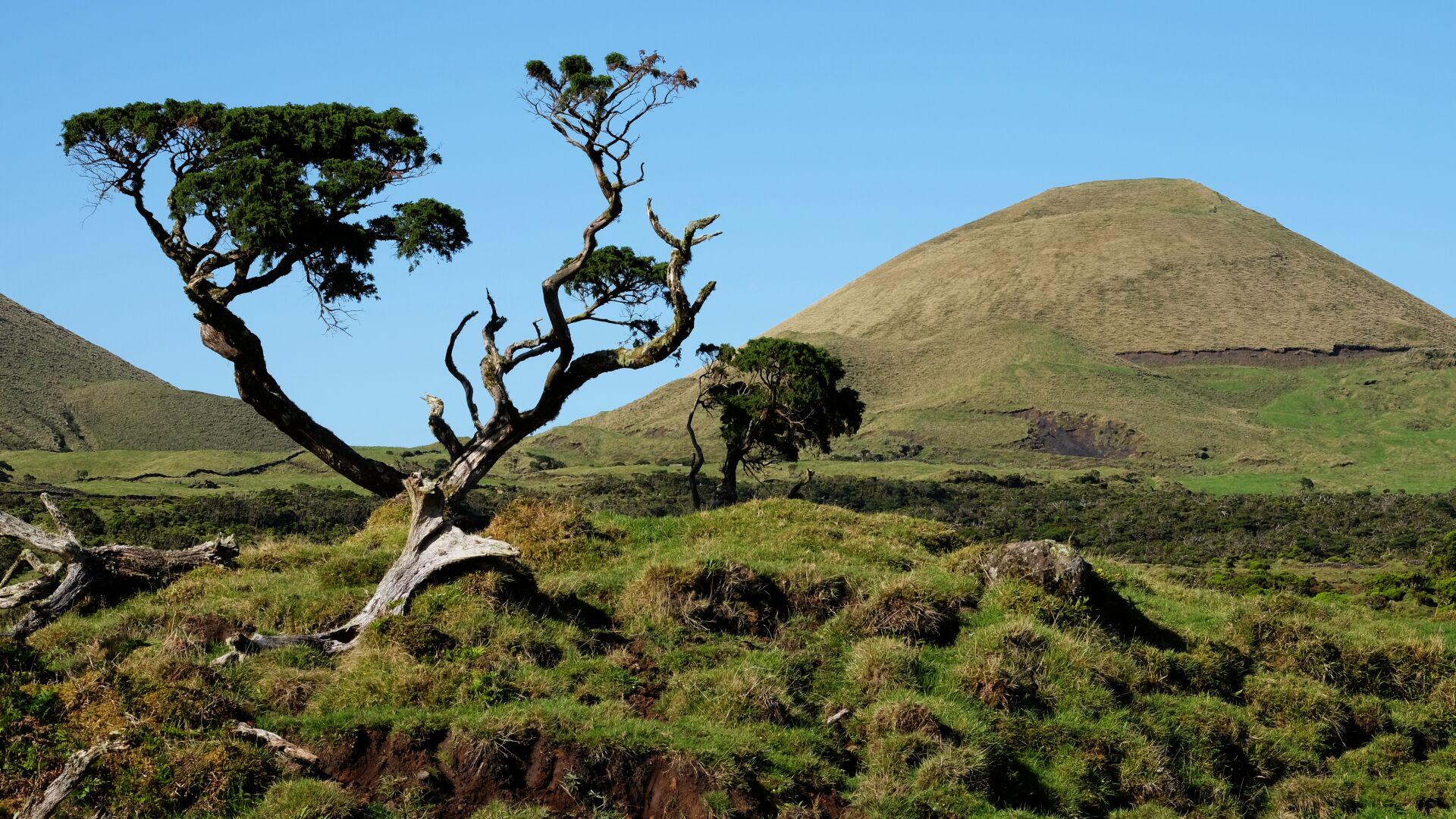}} & 
		\centered{\includegraphics[width=0.3\textwidth]
		{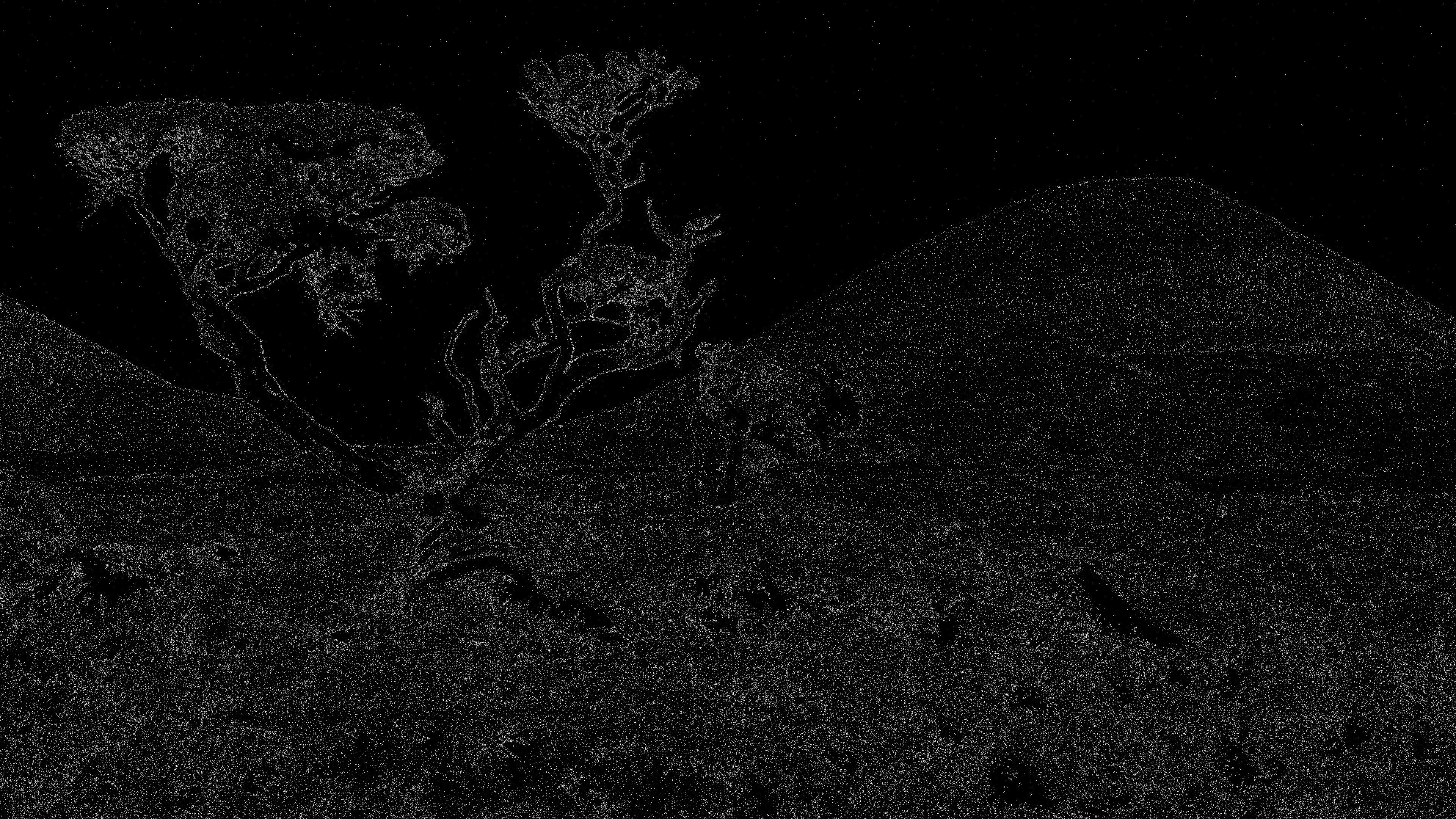}} & 
		\centered{\includegraphics[width=0.3\textwidth]
		{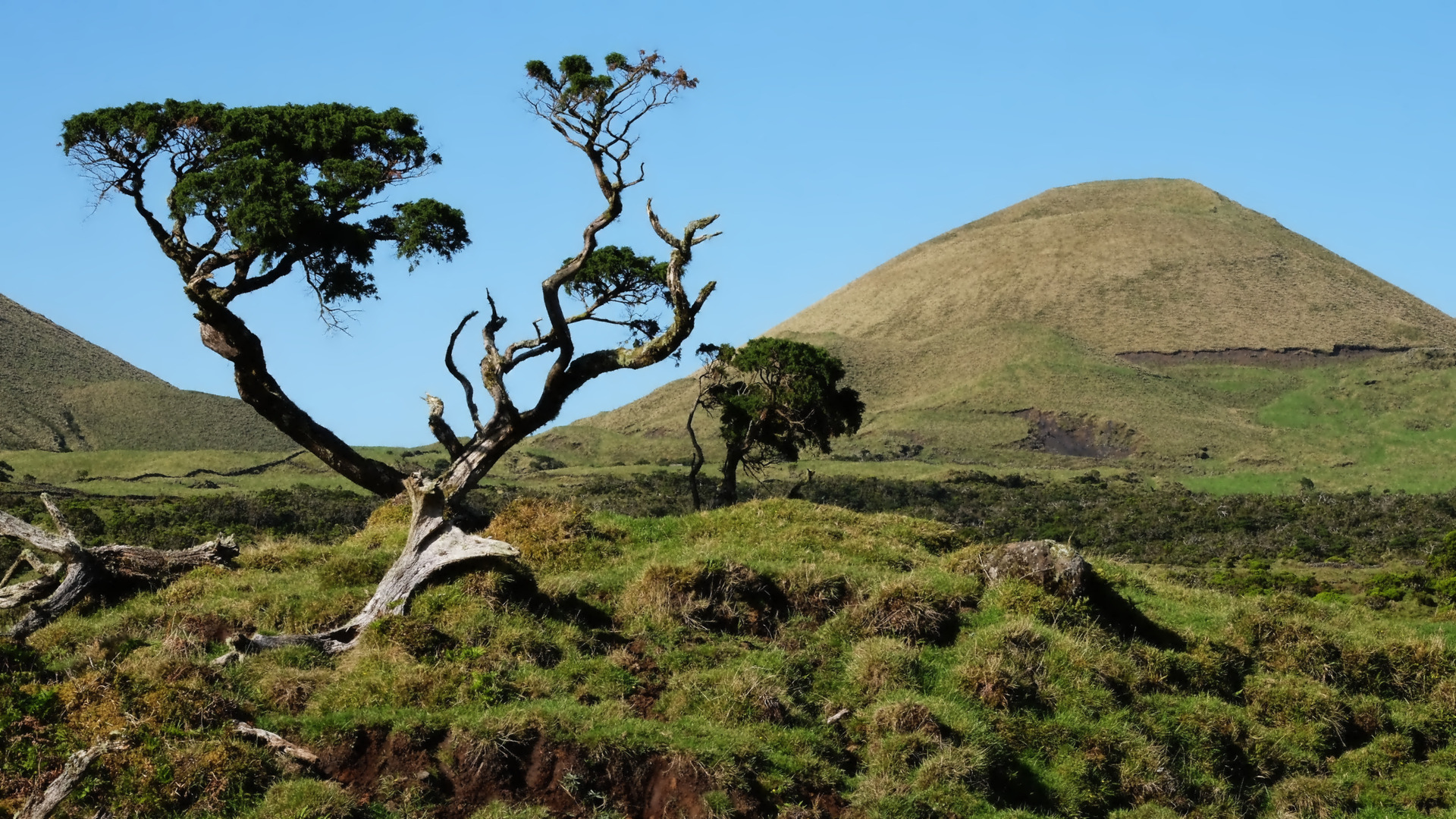}} &
		\centered{\rotatebox{90}{\small PSNR: 27.53}}\vspace{0.2mm}\\ 
		
		\centered{\rotatebox{90}{\small \textit{chairs}}} &
		\centered{\includegraphics[width=0.3\textwidth]
		{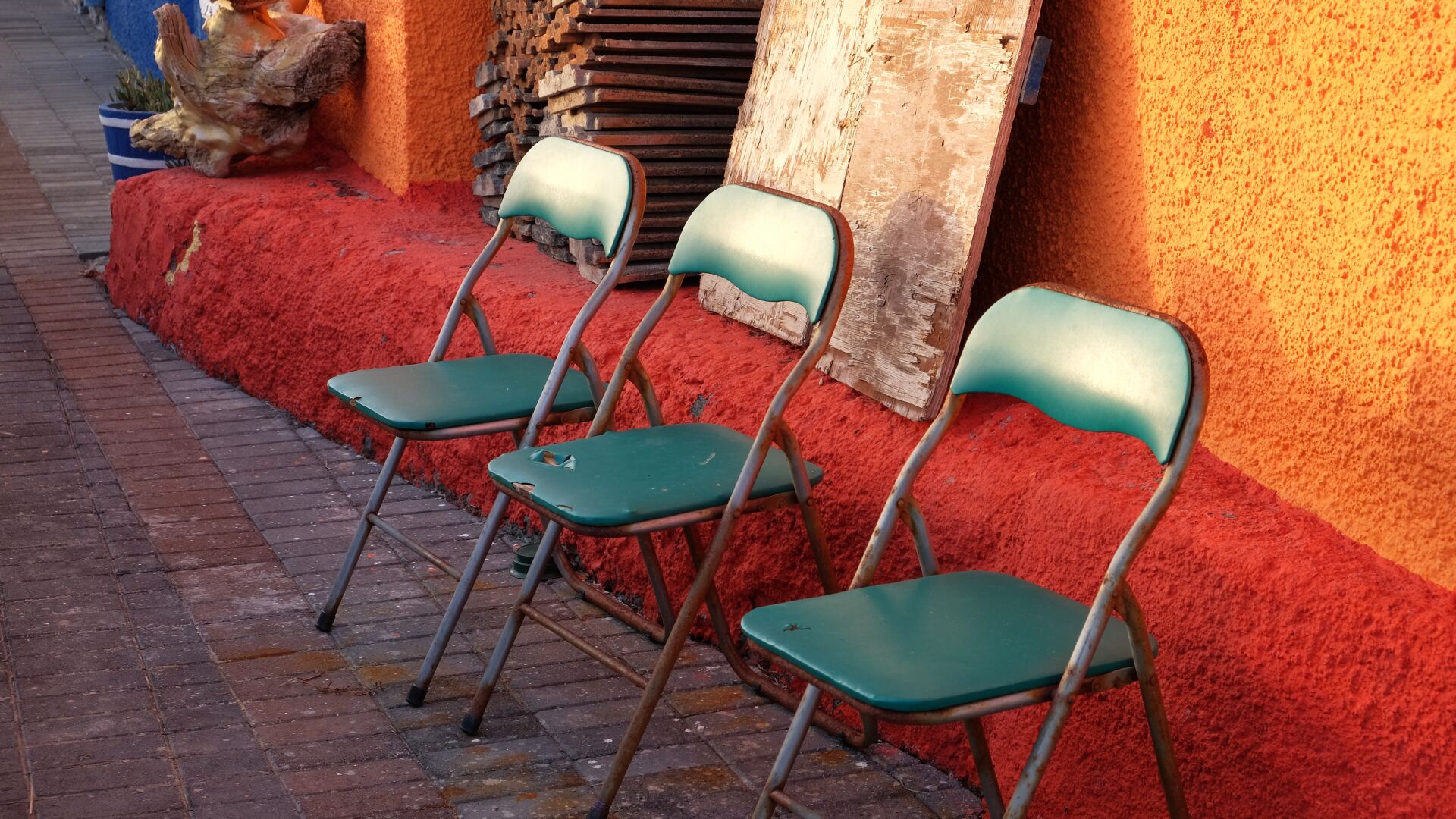}} & 
		\centered{\includegraphics[width=0.3\textwidth]
		{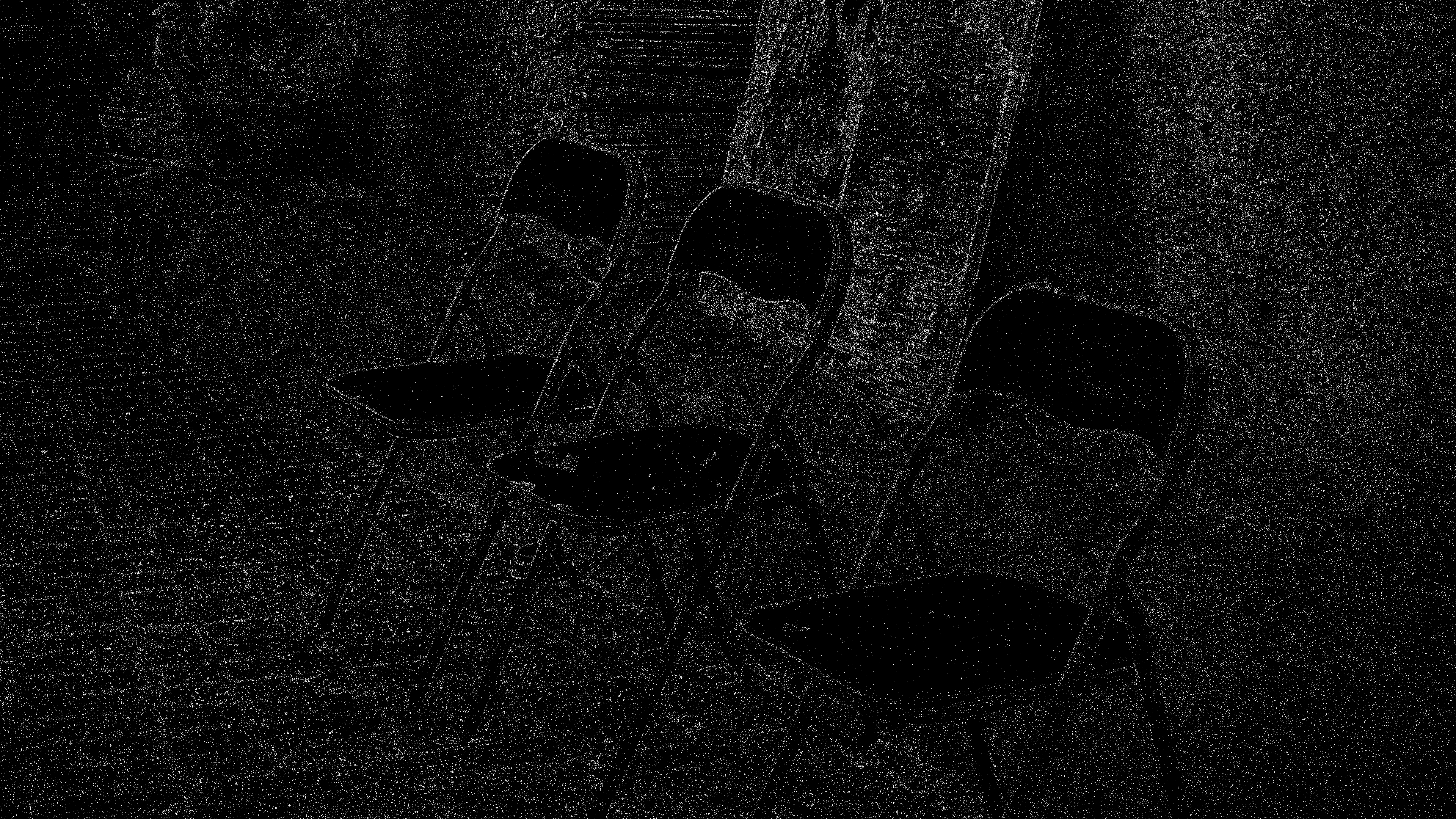}} & 
		\centered{\includegraphics[width=0.3\textwidth]
		{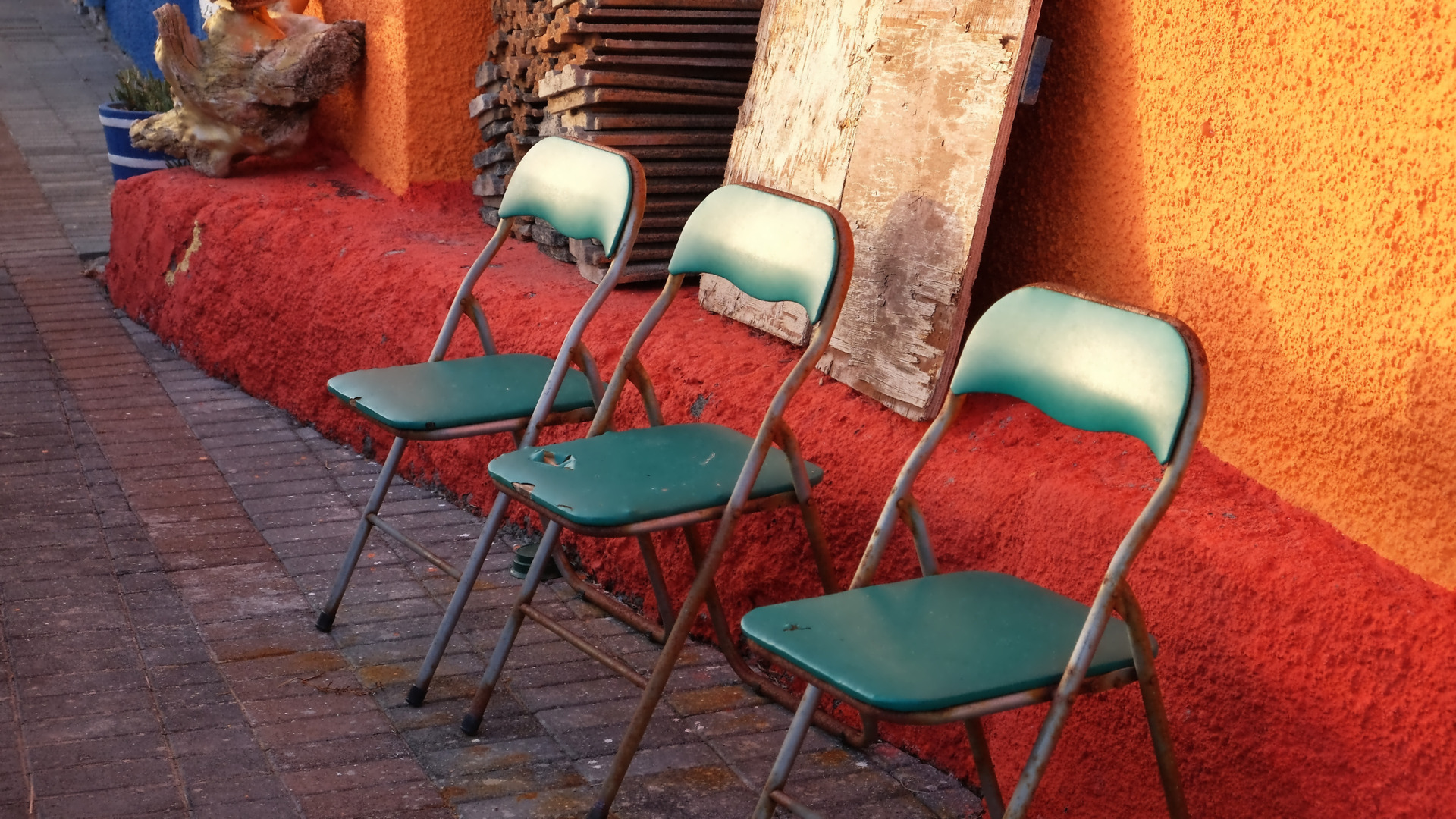}} &
		\centered{\rotatebox{90}{\small PSNR: 32.61}}\vspace{0.2mm}\\
		
		\centered{\rotatebox{90}{\small \textit{butterfly}}} &
		\centered{\includegraphics[width=0.3\textwidth]
		{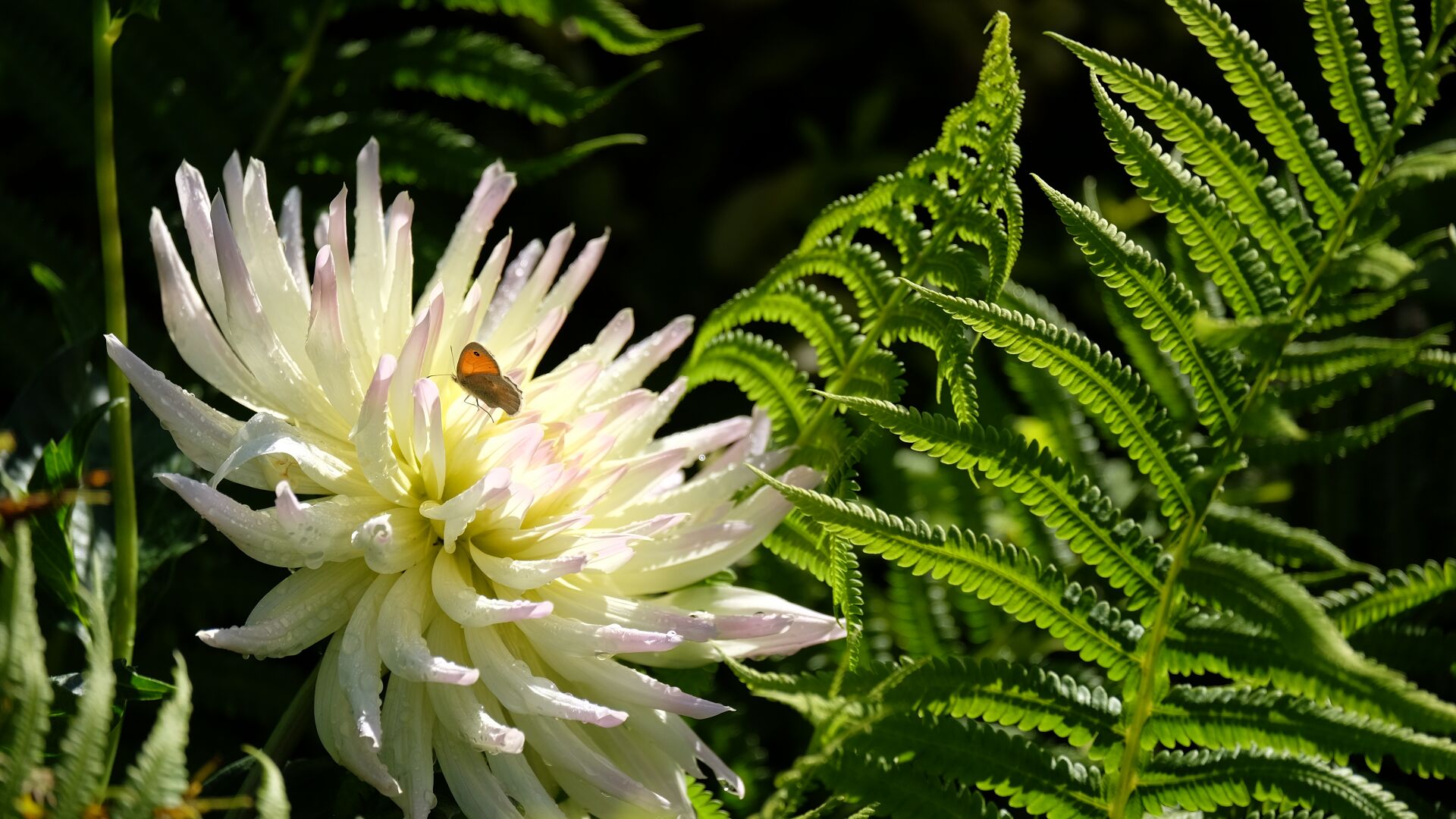}} & 
		\centered{\includegraphics[width=0.3\textwidth]
		{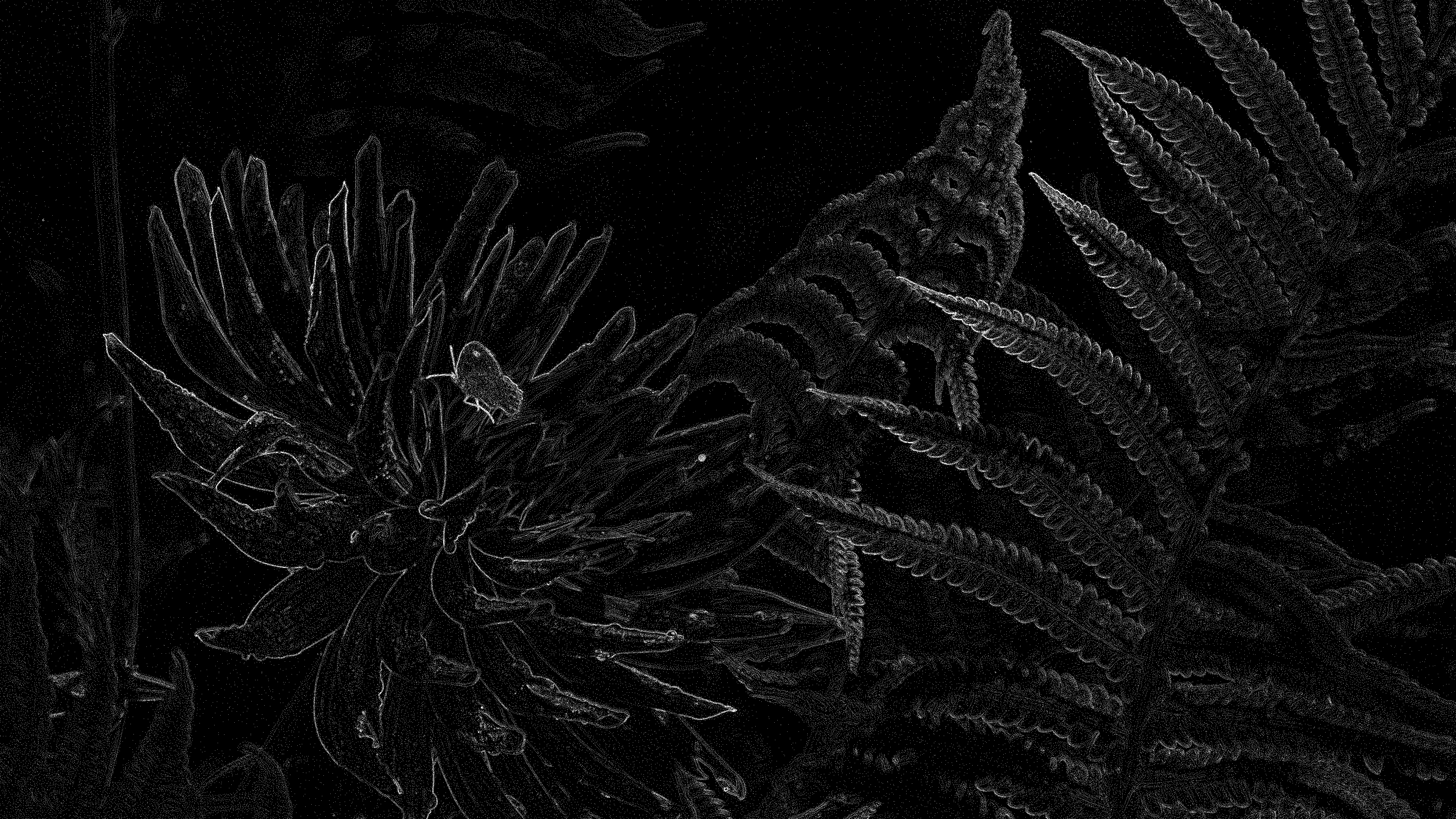}} & 
		\centered{\includegraphics[width=0.3\textwidth]
		{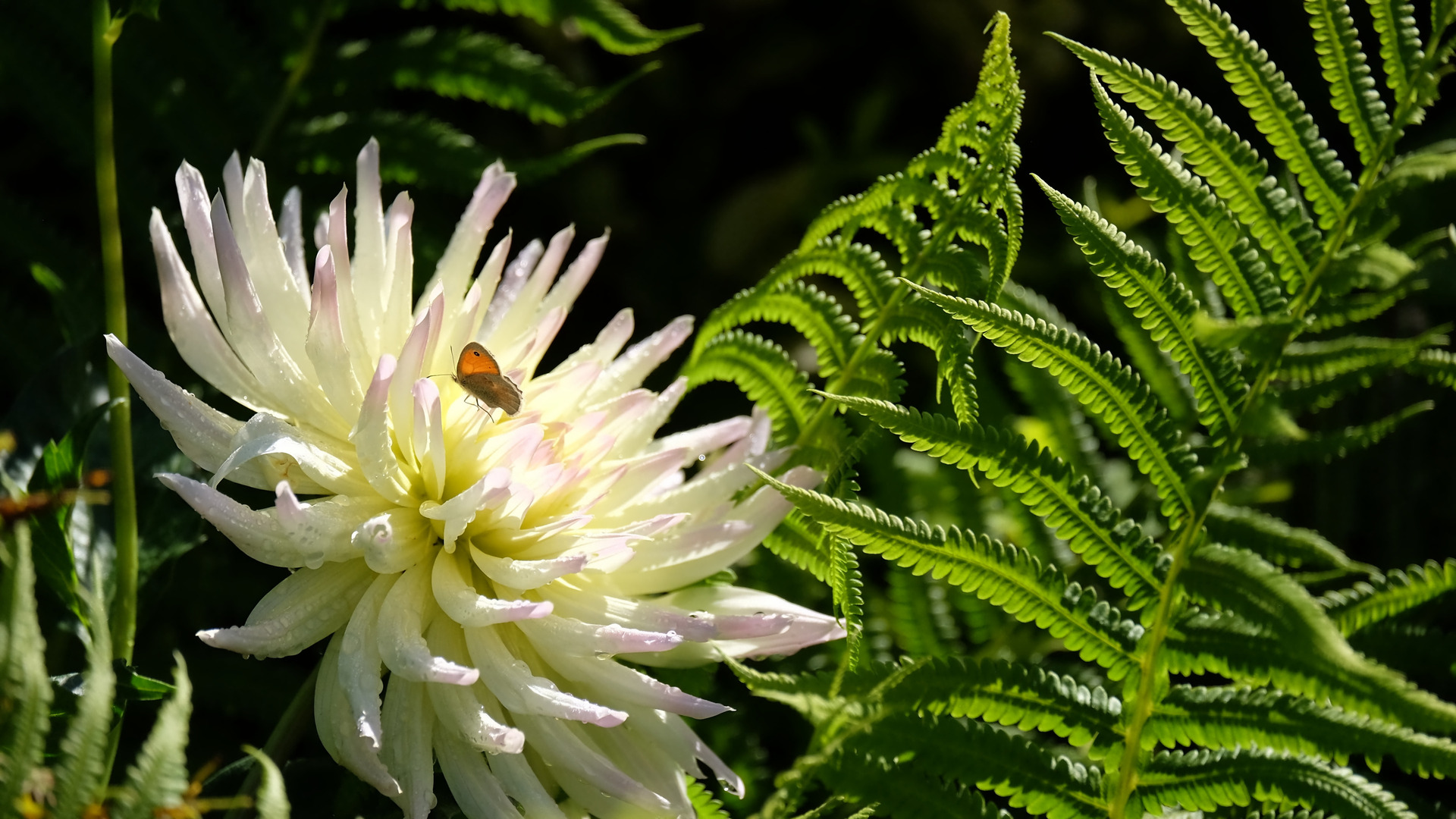}} &
		\centered{\rotatebox{90}{\small PSNR: 41.39}}\vspace{0.2mm}\\ 
		
		\centered{\rotatebox{90}{\small \textit{moon}}} &
		\centered{\includegraphics[width=0.3\textwidth]
		{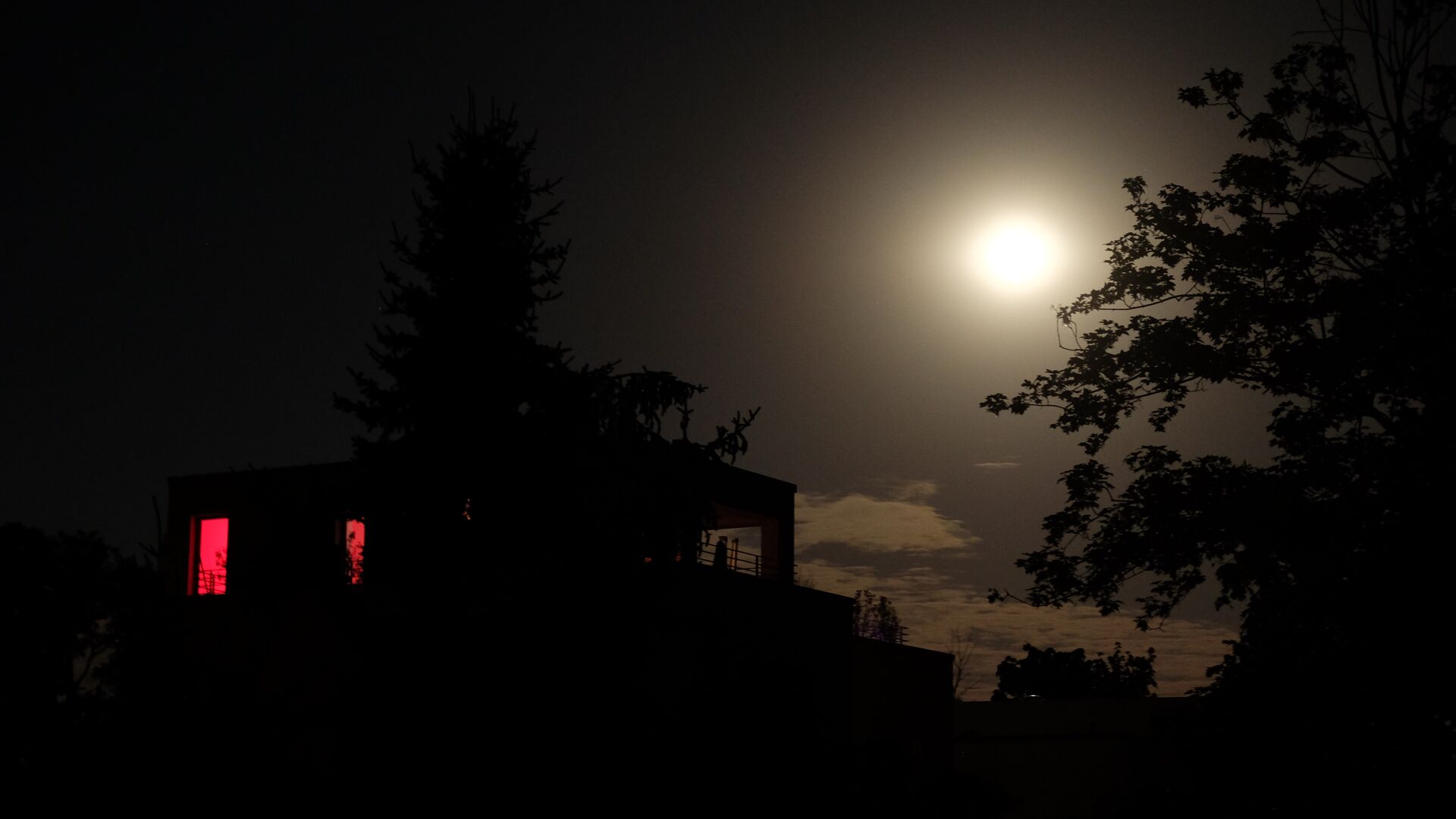}} & 
		\centered{\includegraphics[width=0.3\textwidth]
		{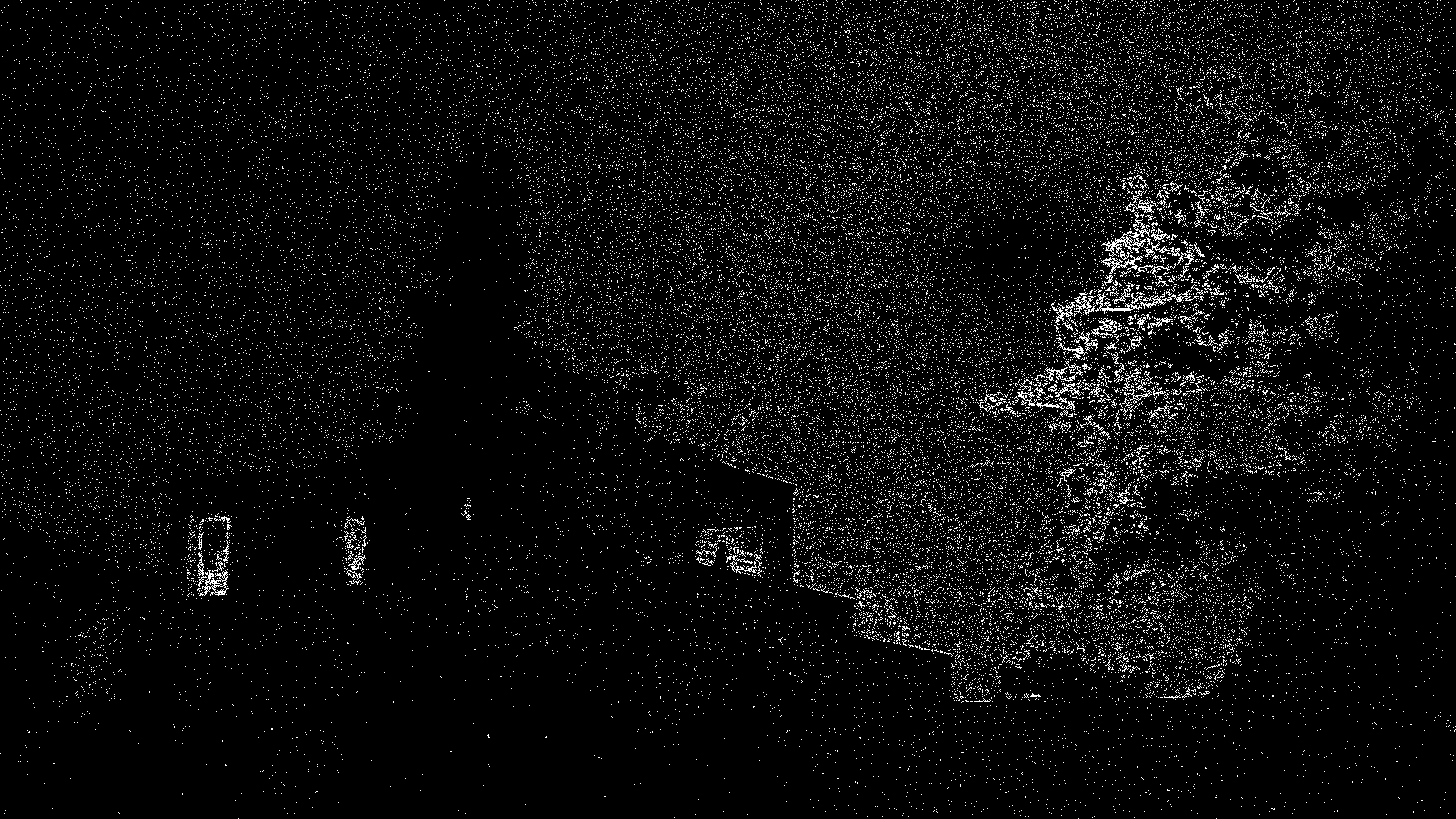}} & 
		\centered{\includegraphics[width=0.3\textwidth]
		{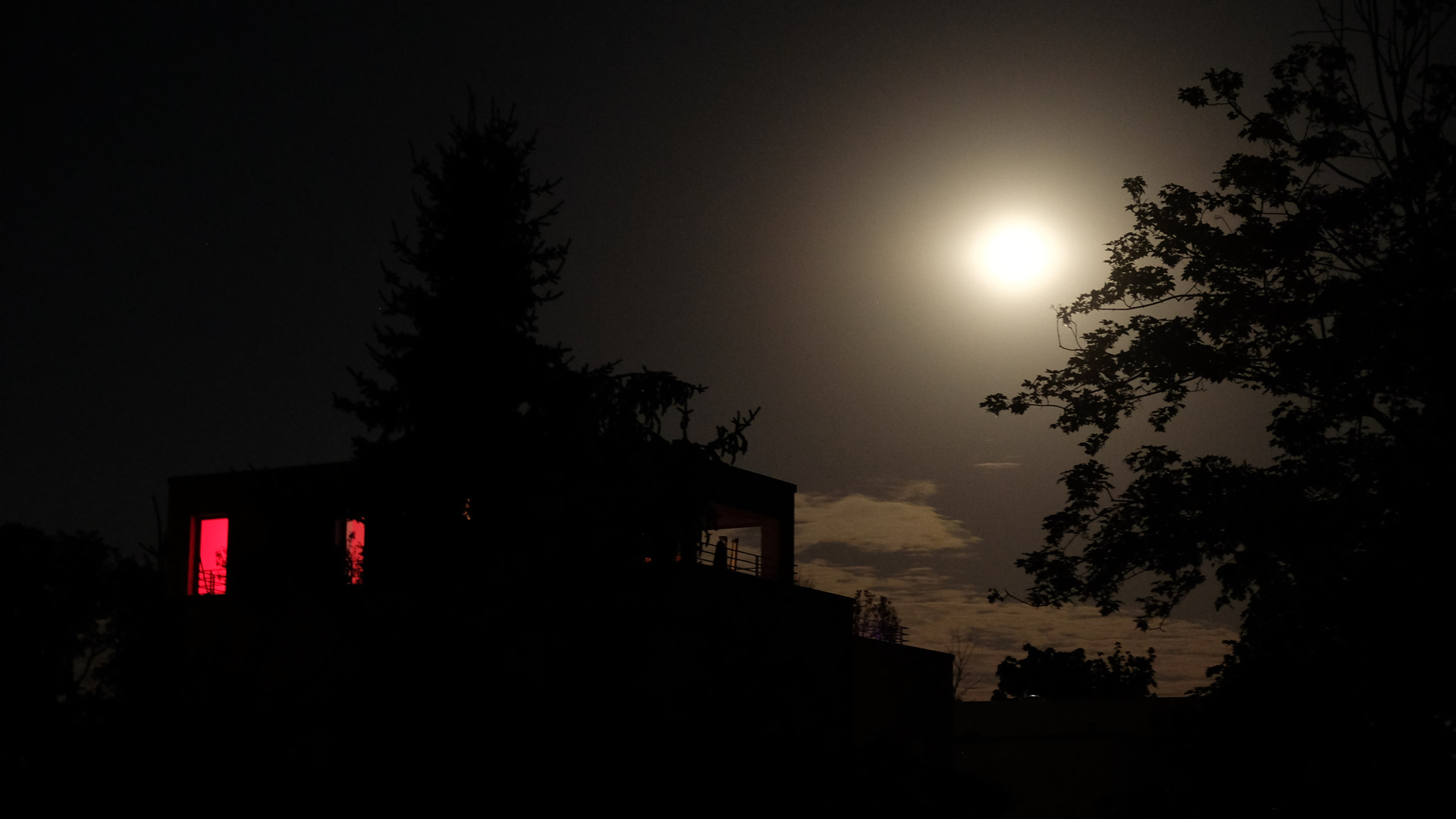}} &
		\centered{\rotatebox{90}{\small PSNR: 48.43}}\vspace{0.2mm}\\
				
	\end{tabular}
	
	\caption{\textbf{Sparse Inpainting with 5 \% Known Data}. Images 7 to 12 of our test dataset of size $3840 \times 2160$ with an 5\% optimized 
	inpainting mask and the corresponding inpainting. Photos by J. Weickert.}
	\label{fig:ds-inpainting-2}
\end{figure}

After mentioning the technical details of our experimental setup in 
\cref{sec:exp_setup}, we show the improvements of our full multigrid ORAS 
inpainting in comparison to our previous ORAS multilevel 
inpainting~\cite{KW22} in \cref{sec:exp_inpainting}.

\subsection{Experimental Setup}
\label{sec:exp_setup}

For the experimental evaluation of our full multigrid ORAS inpainting, we 
compare against the multilevel ORAS approach from our conference contribution 
\cite{KW22} and CG, as these methods are suitable for parallelization and 
can be efficiently implemented on a GPU. To allow for a fair comparison, we
also implement multilevel and multigrid variants of CG that use the same 
framework as their ORAS-based counterparts, but with CG as a pre- and 
post-smoothing operator.
We excluded other inpainting solvers such as the Green's functions 
inpainting~\cite{AWA19, HPW15} from our comparisons, since they become 
infeasible for large images due to runtime and memory restrictions.

We conducted our experiments on an \textit{AMD Ryzen 5900X@3.7GHz} and 
an \textit{Nvidia GeForce GTX 1080 Ti} GPU. We used state-of-the-art 
optimized inpainting masks, obtained with a Delaunay densification strategy, 
proposed in the second part of this companion paper~\cite{KCW23a}. 
To enable a representative evaluation of our inpainting method, we performed 
our experiments on twelve 4K photographs that contain a varying amount of 
texture and coarse and fine structures 
(see \cref{fig:ds-inpainting-1,fig:ds-inpainting-2}). 
\cref{fig:ds-inpainting-1,fig:ds-inpainting-2} also show the optimized 
inpainting masks for a density of 5\% and the corresponding inpainting 
results obtained with our multigrid ORAS solver. We achieve an average 
runtime of 14.3 milliseconds, which corresponds to \textbf{a frame rate of 
nearly 70 frames per second}.  

\subsection{Inpainting Experiments}
\label{sec:exp_inpainting}

\begin{figure}[tb]
	\begin{subfigure}[b]{0.49\linewidth}
		\centering
		\centerline{\includegraphics[width=1.0\linewidth]
			{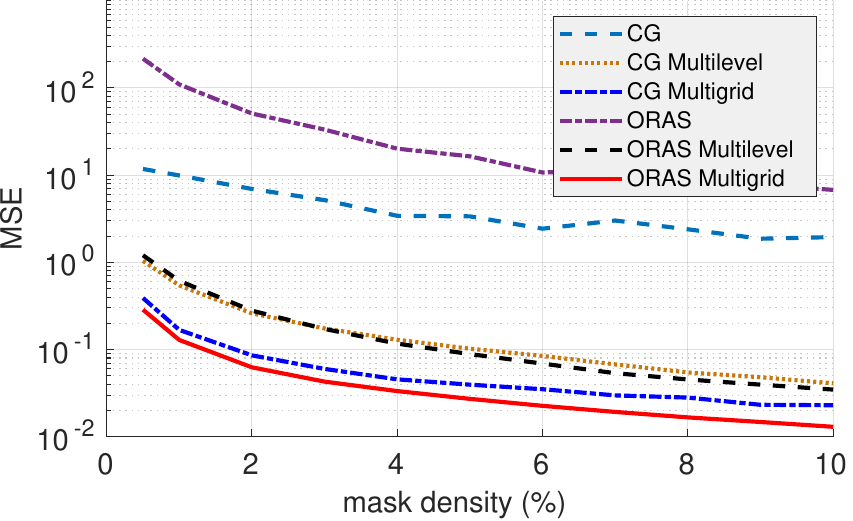}}
		\subcaption{inpainting approximation quality}
		\label{fig:inpainting_approxquality}
	\end{subfigure}
    \hfill
 	\begin{subfigure}[b]{0.49\linewidth}
		\centering
		\centerline{\includegraphics[width=1.0\linewidth]
			{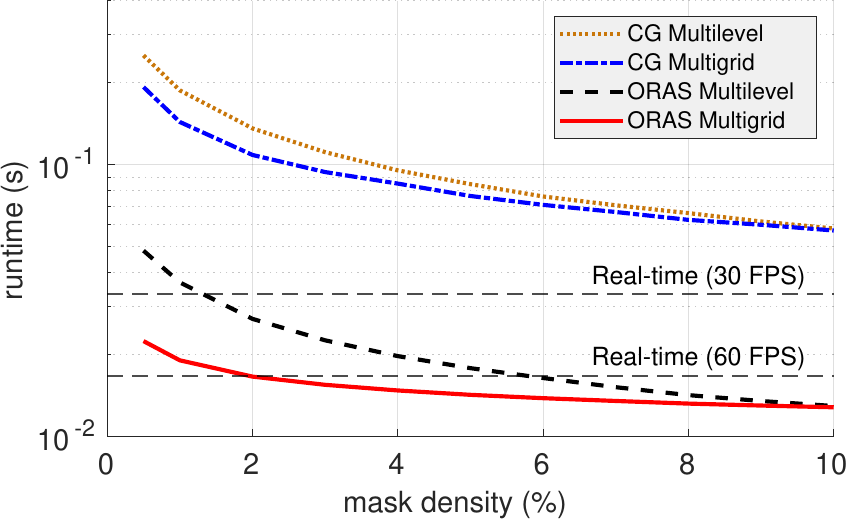}}
		\subcaption{runtime}
		\label{fig:inpainting_densities}
	\end{subfigure}
	\smallskip
	\caption{\textbf{Inpainting Comparison for Different Mask Densities}. 
		\textbf{(a)} MSE between the inpainting approximation and 
            a converged inpainting for a stopping criterion of $10^{-3}$. The 
            multigrid methods are closest to the converged inpainting with our 
            ORAS multigrid performing the best. 
            Both single-level methods are clearly not fully converged and 
            would require a much stricter stopping criterion for a reasonable 
            approximation quality.
  		\textbf{(b)} Runtimes of multilevel and multigrid methods for a 
        stopping criterion of $10^{-3}$.  Our ORAS-based methods are more than 
		4 times faster than their corresponding CG counterparts. Only 
		the ORAS multigrid methods achieve real-time performance for all 
		densities.
        }
	\label{fig:inpainting}
\end{figure}

In \cref{fig:inpainting} we compare our multilevel and multigrid ORAS 
inpainting algorithms with their corresponding CG counterparts on different 
densities from 0.5\% to 10\%, averaged over all twelve 4K images.

\paragraph{Stopping Criterion}
As a stopping criterion we use the relative decay of the Euclidean 
norm of the residual. We note that this alone is not a good 
indicator of the true reconstruction error, which can be seen in 
\cref{fig:inpainting_approxquality}. It shows the approximation 
quality with respect to a fully converged inpainting result for a relative 
residual norm of $10^{-3}$. 
For multilevel and multigrid methods, we obtain reasonably good approximation 
qualities for all mask densities.
Our ORAS multigrid method even outperforms all other methods. The results for 
both single level methods are, however, clearly not fully converged and would 
require a much stricter stopping criterion for a reasonable approximation 
quality.
Due to this, we only consider multilevel and multigrid methods from this 
point on, as they are clearly superior. We use a relative residual norm of 
$10^{-3}$ as a stopping criterion, as we have seen that this is sufficient 
for the multilevel and multigrid methods.

\paragraph{Mask Density Runtime Scaling}
\cref{fig:inpainting_densities} shows the corresponding inpainting runtimes.
We observe that the domain decomposition methods clearly outperform their 
corresponding CG-based solvers by more than a factor of $4$. This demonstrates 
the capabilities of domain decomposition methods over simpler algorithms.
While the performance of the ORAS multigrid and the ORAS multilevel method 
is similar for a mask density of 10\%, our multigrid method clearly 
outperforms the multilevel approach for lower mask densities. This is because 
even though both methods use a coarse-to-fine initial guess, only multigrid 
performs V-cycles that efficiently reduce lower frequency error modes. Since 
for sparser masks the mask pixel contributions are lower frequency, those 
masks benefit much more from the multigrid approach.
This allows us to perform real-time inpainting with 30 frames per second 
for all tested densities and even 60 frames per second for mask densities 
higher than 2\%.

\begin{figure}[tb]
		\centering
		\centerline{\includegraphics[width=0.49\linewidth]
			{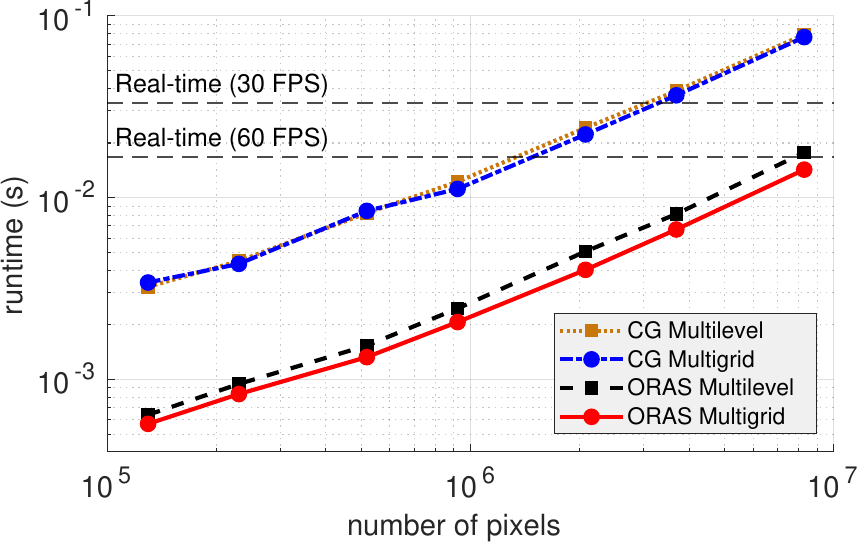}}
	\smallskip
	\caption{\textbf{Comparison on Different Resolution 
                Levels for a 5\% Mask Density}. 
                The horizontal dashed lines represent real-time inpainting 
                with 30 and 60 frames per second.
                Our ORAS multigrid method achieves more than 60 frames 
                per second for a 4K resolution (last data point). With 
                CG-based methods we achieve real-time performance only 
                up to a FullHD resolution (third data point from the right) 
                and only with approximately 45 frames per second.}
	\label{fig:inpainting_scaling}
\end{figure}

\paragraph{Resolution Runtime Scaling}
To evaluate our inpainting methods over different image sizes, we conduct  
experiments over resolutions ranging from $480 \times 270$ to 
$3840 \times 2160$ with 5\% masks. The results are shown in 
\cref{fig:inpainting_scaling}. We can see that similar to the results from 
\cref{fig:inpainting_densities}, our ORAS-based inpainting methods are over 
four times faster than their CG-based counterparts on all image resolutions. 
While both of our domain decomposition methods are able to inpaint 4K images 
in real-time, with at least 30 frames per second, the CG-based methods achieve 
this only for FullHD resolutions.
\cref{fig:inpainting_scaling} also reveals that, at least for higher image 
resolutions, all four methods show a nearly linear behavior in the double 
logarithmic plot. As the slope is approximately $1$ for all methods, we 
observe an ideal linear scaling behavior.

\section{Conclusions and Outlook}
\label{sec:conclusions}

One of the biggest challenges in inpainting-based compression has been the 
high computational complexity associated with inpainting algorithms. 
While qualitatively those are a viable alternative to transform-based 
approaches, standard solvers result in long inpainting runtimes for larger 
images. 
Our work substantially alleviates this issue. By adapting and fusing two 
of the most efficient concepts of numerical analysis and taking advantage of 
the computing power of modern parallel hardware, we have obtained a very 
efficient solver for homogeneous diffusion inpainting. This enabled us, 
for the first time ever, to achieve homogeneous diffusion inpainting 
at 60 frames per second for 4K color images on a contemporary 
GPU (in actuality four generations old).  This suggests that even 8K 
resolution in real-time would be possible with the latest generation
of GPUs.
Furthermore, most state-of-the-art methods for data optimization on the 
encoding side rely on multiple inpaintings. Thus, our work not only benefits 
the decoding side but is also able to improve the runtime of encoding 
methods.
This demonstrates that inpainting-based compression has left its infancy to 
become a serious alternative to classical transform-based codecs not only in 
terms of quality but also for real-time applications.

Our experience with homogeneous diffusion inpainting suggests that approaches 
like ours, which judiciously adapt multiple numerical methods and exploit 
modern parallel architectures, could also be successfully transferred to 
numerous other image processing tasks. 


Last but not least, while our work mainly focuses on mathematically sound 
algorithms, we have nevertheless extracted interesting insights about the 
structure of homogeneous diffusion inpainting and its interplay with numerical 
methods. 
We have seen that devising methods that attempt to preserve continuous 
properties such as the localization of the Dirichlet boundary's influence on 
coarser levels pays off. We have also observed the relationship between mask 
density, multilevel, and multigrid methods. The latter are able to achieve 
much lower runtimes for low densities, since the basis functions at low 
densities are inherently smoother. 
Finally, we noted that the widely used stopping criterion based on the 
relative residual can be highly misleading, especially in the context of 
image processing while using different classes of solvers.

In our ongoing research, we are working on extending our domain 
decomposition approach to more sophisticated inpainting operators, such as 
anisotropic nonlinear diffusion. It has been shown that those can offer 
an improved reconstruction quality compared to the simple homogeneous 
diffusion inpainting. In the companion paper to this one we tackle 
one of the main challenges in the encoding phase: data optimization.


\bibliographystyle{siamplain}
\bibliography{myrefs,additional_refs}
	
\end{document}